\documentclass[hyper,10pt,letterpaper]{JHEP3}
\usepackage{amsmath}

\newcommand{\ep}{\epsilon}

\title{Dissipative superfluid dynamics from  gravity}
\author{Jyotirmoy Bhattacharya$^a$, Sayantani Bhattacharyya$^b$ and Shiraz Minwalla$^a$\\
$^a$Dept. of Theoretical Physics, Tata Institute of Fundamental Research, Homi Bhabha Rd, \\
Mumbai 400005, India. \\
$^b$Harish-Chandra Research Institute, Chhatnag Road, Jhunsi, Allahabad-211019.\\
Email:\ \ {\bf jyotirmoy@theory.tifr.res.in, sayanta@hri.res.in, minwalla@theory.tifr.res.in}}

\abstract{

Charged asymptotically $AdS_5$ black branes are sometimes unstable to the 
condensation of charged scalar fields. For fields of infinite charge and 
squared mass $-4$ Herzog was able to analytically determine the phase 
transition temperature and compute the endpoint of this instability
in the neighborhood of the phase transition. We generalize Herzog's 
construction by perturbing away from infinite charge in an expansion in 
inverse charge and use the solutions so obtained as input for the fluid 
gravity map. Our tube wise construction of patched up locally hairy black 
brane solutions yields a one to one map from the space of solutions of 
superfluid dynamics to the long wavelength solutions of the Einstein Maxwell
system.  We obtain explicit expressions for the metric, gauge field and 
scalar field dual to an arbitrary superfluid flow at first order in the 
derivative expansion. Our construction allows us to read off the 
the leading dissipative corrections to the perfect superfluid 
stress tensor, current and Josephson equations. A general framework for 
dissipative superfluid dynamics was worked out by Landau and Lifshitz 
for zero superfluid velocity and generalized to nonzero fluid velocity by 
Clark and Putterman. Our gravitational results do not fit into the 13 parameter 
Clark-Putterman framework. Purely within fluid dynamics we present 
a consistent new  generalization of Clark and Putterman's equations to a set 
of superfluid  equations parameterized by 14 dissipative 
parameters. The results of our gravitational calculation fit perfectly into 
this enlarged framework. In particular we compute all the dissipative constants 
for the gravitational superfluid.}

\begin{document}

\section{Introduction and Summary}

It was pointed out by Gubser \cite{Gubser:2008px} that charged asymptotically 
$AdS_5$ black 
branes are sometimes unstable in the presence of charged scalar fields. 
The endpoint of this instability is a hairy black brane: a black brane 
immersed in a charged scalar condensate. The AdS/CFT correspondence maps 
the hairy black brane to a phase in which a global $U(1)$ 
symmetry is spontaneously broken by the vacuum expectation value of a charged 
scalar operator. In condensed matter physics a phase with a spontaneously 
broken global $U(1)$ symmetry is referred to as a superfluid. 

In this paper we study aspects of the fluid dynamical description 
of superfluids using the AdS/CFT correspondence. Our work builds on a large
body of earlier studies (see e.g. 
\cite{Hartnoll:2008vx,Hartnoll:2008kx,Herzog:2009md,
Yarom:2009uq,Herzog:2010vz,Sonner:2010yx, Policastro:2001yc,Basu:2008st, 
Herzog:2008he,Amado:2009ts,Karch:2008fa,Herzog:2009ci,Arean:2010xd,
Horowitz:2008bn,Gauntlett:2009bh,Franco:2009yz,Gauntlett:2009dn,
Son:2000ht,Horowitz:2009ij,Mannarelli:2009ia,Arean:2010wu}) but differs from these works in 
its emphasis on the study of dissipative terms in super fluid dynamics.

As we describe in detail in the next section, the variables of relativistic 
superfluid dynamics consist of two velocity fields; the normal fluid velocity 
$u^\mu$ and a superfluid velocity field $u^\mu_s$;  together with a temperature
and chemical potential field. The superfluid velocity is the unit vector 
in the direction of $-\xi_{\mu}$ where $\xi_\mu$ is 
 the gradient of the phase of the scalar 
condensate. Conservation of the stress 
tensor and charge current together with the assertion that 
$\xi_\mu$ is curl free constitute the equations of superfluid 
dynamics. These equations constitute a closed dynamical system once 
they are supplemented with constitutive relations that express the 
the stress tensor, charge current and the component of $\xi_\mu$ 
along the normal velocity as functions of the fluid dynamical variables. 
As superfluid dynamics is a long distance effective field theory, 
it is natural to specify
the relevant constitutive relations in an expansion in derivatives. 

Over fifty years ago Landau and Tiza \cite{Landau:1941,Tisza:1947zz} presented a simple and elegant proposal 
for the structure of superfluid constitutive relations at leading (zero)
order in the derivative expansion. Landau and Tiza (see \S \ref{review} 
below) proposed a form for the constitutive relations that is entirely 
determined by a single thermodynamical `Free Energy' $P(T, \mu, \xi).$
An obvious first question in the study of holographic superfluids is 
the following: do the perfect fluid constitutive relations of holographic
superfluids respect the Landau-Tiza ansatz? This question was largely answered
in the affirmative in a beautiful recent paper by Sonner and Withers
\cite{Sonner:2010yx}. 
These authors used Einstein's equations and the holographic dictionary 
to demonstrate that the stress tensor 
and charge current of a homogeneous stationary holographic superfluid flow 
takes the form predicted by Landau and Tiza, with the free energy or pressure
interpreted as the value of Einstein's action for the relevant bulk solutions.
We re-derive and slightly extend the results of Sonner and Withers in 
\S \ref{eq} and Appendix \ref{SgravThermo} below. As we review in 
Appendix \ref{SgravThermo}, a beautiful 
feature of the gravitational derivation of the Landau-Tiza model 
is its abstract nature. The results of Sonner and Withers 
are derived on general grounds, and do not use the explicit form of the 
gravitational solution dual to homogeneous stationary superfluid flows. 
This is fortunate as no completely explicit analytic
solutions are known for static hairy black branes ( i.e. 
holographic superfluids at rest) much less for hairy black branes 
in motion.  

There also exists a large literature on the subject of dissipative 
corrections to the equations of superfluid dynamics, accurate to first 
order in the derivative expansion (see e.g.  
\cite{Clark,Putterman,LLvol6,Carter:1992a,Pujol:2002na,Son:1999pa,
Valle:2007xx,Lane,Khalatnikov,LebedevKhalatnikov,CarterKhalatnikov,Israel1,
Israel2,Mannarelli:2009ia}). In particular, a general first order theory of dissipative fluid 
dynamics was presented by Landau and Lifshitz \cite{LLvol6} (under the assumption of small 
superfluid velocities) and was generalized to apply to flows with finite 
superfluid velocities in \cite{Clark,Putterman}.
It turns out that the 
most general one derivative corrections to the equations of perfect 
superfluid dynamics are parameterized by 36 dissipative corrections (assuming 
parity invariance). 
Like Landau and Lifshitz, in \cite{Clark,Putterman} the authors assumed that the entropy current of 
superfluid dynamics takes a natural `canonical' form at one derivative order.
Once again, like Landau and Lifshitz, the authors in \cite{Clark,Putterman} used the requirement of 
local positivity of entropy production, together with the `Onsager Reciprocity
relations' to cut the parameter space for physically permissable 
equations of superfluids down to a 13 parameter set (see Appendix $VI$ in \cite{Putterman}). To our knowledge this Clark-Putterman formalism 
\cite{Clark,Putterman} has not been tested by any first principles dynamical 
calculations within a quantum field theory (e.g. has not been derived 
from a field theoretically motivated set of Boltzmann equations). The AdS/CFT 
correspondence offers us the opportunity to test this abstract formulation of viscous superfluid dynamics. 

In this paper we perform such a test in a very particular holographic 
super fluid. The system we study is a small perturbation of 
the `analytic superconductor' studied by Herzog in \cite{Herzog:2010vz}.
In very broad terms we use the fluid gravity map 
\cite{Bhattacharyya:2008jc,Bhattacharyya:2008xc, Bhattacharyya:2008ji, 
Banerjee:2008th, Bhattacharyya:2008mz,Erdmenger:2008rm,Haack:2008cp,
VanRaamsdonk:2008fp} to derive the equations of superfluid
dynamics from Einstein equations to first order in the derivative expansion.
We then read off the constitutive relations of the stress tensor, charge 
current and Josephson equation at first order in the derivative expansion.
In the rest of this introduction we describe our calculations, their results
and their interpretation in more detail.

As we have mentioned we work with a generalization of the  model studied by  
Herzog \cite{Herzog:2010vz}. Herzog's model consists of the Einstein Maxwell 
system interacting with  a charged scalar field of $m^2=-4$ in the so called 
probe limit of infinite charge $e$. Herzog demonstrated that 
this system undergoes a second order phase transition towards 
superfluidity whenever $|\frac{\mu}{T}| \geq 2$. The stable gravitational 
solution, for $|\frac{\mu}{T}|$ just larger that $2$, has a background 
scalar vev. 
Let $\epsilon$ denote the value of this vev. 
In \cite{Herzog:2010vz} Herzog analytically determined the relevant bulk 
solutions perturbatively in $\epsilon$ and separately in the 
difference between superfluid and normal velocities. 

To start with we  generalize Herzog's infinite charge solutions beyond the 
strict probe approximation,  to first nontrivial order in the $\frac{1}{e^2}$.
This generalization is necessary in order to allow for the study of the 
response of the normal velocity and temperature fields to the dynamics of 
the superfluid velocity and chemical potential fields. As a check on our 
algebra we explicitly verify that our solutions obey all the predictions 
of the Landau Tiza model, to the order that we are able to compute, in 
accordance with the results of Sonner and Withers and of Appendix 
\ref{SgravThermo}. We then proceed to use these solutions as raw 
ingredients for the fluid gravity correspondence
\cite{Bhattacharyya:2008jc,Bhattacharyya:2008xc, Bhattacharyya:2008ji, 
Banerjee:2008th, Bhattacharyya:2008mz,Erdmenger:2008rm,Haack:2008cp,
VanRaamsdonk:2008fp}.

Following the procedure of the fluid gravity correspondence, we search for 
solutions of the Einstein Maxwell scalar system that tube wise approximate 
the stationary solutions described in the previous paragraph. More explicitly, 
we study a perturbative expansion to the solutions of Einstein's equations 
whose first term is given by the stationary solutions of the previous 
paragraph -written in ingoing Eddington Finklestein coordinates - with the 
eight 
parameters of equilibrium superfluid flows replaced by slowly varying 
functions of spacetime. The configuration described in this paragraph 
does not obey the bulk equations; however it may sometimes 
be systematically corrected, 
order by order in boundary derivatives, to yield a solution to these equations.
This procedure works if and only if our eight fields are chosen such 
that $\xi^\mu(x)$ is curl free, and such that the energy momentum and 
charge current built out of these fields is conserved. The constitutive 
relations that allow us to express the stress tensor and charge current in
terms of fluid dynamical fields is generated by the perturbative procedure
itself. In other words the output of our perturbative procedure is a set 
of gravitational solutions that are in one to one correspondence with 
the solutions of superfluid dynamics, with superfluid 
constitutive relations that are determined by the bulk gravitational equations. 

Note that the construction described in the previous paragraph is 
carried out in a triple expansion. We follow Herzog to 
expand our equilibrium solutions in a power series in the deviations from 
criticality (let us denote the relevant parameter by $\epsilon$) 
, and further expand these solutions in a power series in
$\frac{1}{e^2}$. We then go on to use the solutions as ingredients in a 
spacetime derivative expansion. We would like to emphasize that this 
procedure is sensible only if the derivatives times mean free path 
 are assumed to be parametrically smaller than $\epsilon$. The 
physical reason for this is that precisely at $\epsilon=0$ we have a new 
massless mode, corresponding to fluctuations of the vev of the charged 
scalar field. When $\epsilon$ is nonzero this mode is no longer massless, 
but it is light at small $\epsilon$. The fluid dynamical description ignores
the dynamics of this light mode. This is justified only when derivatives are 
all much smaller than the mass of this mode, i.e. when derivatives are 
parametrically small compared to $\epsilon$. The fluid dynamical expansion 
breaks down if derivatives are held fixed as $\epsilon$ is taken to zero. 
This fact formally shows up in the blow up of several dissipative fluid 
coefficients at small $\epsilon$ in the equations presented in this paper.

We have implemented the perturbative procedure that determines the 
gravitational solution dual to superfluid dynamics  to first 
order in the derivative expansion, and thereby determined the stress tensor, 
charge current and Josephson equation of our holographic superfluid to 
first order in the derivative expansion. Our results pass several 
consistency checks and also have several attractive features. To start 
with our results respect Weyl invariance. This fact, while 
necessarily true of the results of any  gravitational calculation in an 
asymptotically $AdS$ space, is not algebraically automatic, and so yields 
consistency check on the rather involved algebraic manipulations of our paper. 

More importantly, we are also able to compute the natural gravitational 
entropy current for our gravitational fluid flow. As explained in 
\cite{Bhattacharyya:2008xc}, a gravitational construction of fluid dynamics
is always automatically accompanied a local entropy current that 
is guaranteed, by the area increase theorems of general relativity, to 
be of positive divergence. We demonstrate by explicit computation in our 
particular solution that the gravitational entropy current of 
\cite{Bhattacharyya:2008xc} agrees precisely with the `canonical'
entropy current (see \S \ref{CEntrop} for details) 
of Clark and Putterman (and Landau Lifshitz) \cite{Clark,Putterman,LLvol6}. We regard this 
agreement as a nontrivial check of one of the central physical assumptions 
of the Landau Lifshitz and Clark Putterman formulations of dissipative fluid 
dynamics. \footnote{We should, however, emphasize that we have checked this 
agreement only to leading nontrivial order in the $\epsilon$ expansion. 
It would certainly be useful to verify this agreement - and all the other 
results of our paper - at higher orders in the $\epsilon$ expansion, but we 
leave that for future work.} 
Despite this fact, however, the results we obtain for the 
dissipative coefficients for the gravitational superfluid described 
in this paper do not fit into Clark-Putterman's 13 parameter framework for 
dissipative fluid dynamics \footnote{In an earlier version of this 
draft we compared our results to those of Landau and Lifshitz but not to 
those of Clark and Putterman, as we were unaware of their work. We reported 
that our results did not agree with the predictions of Landau and Lifshitz; 
while this is true, the comparison itself is inappropriate, as Landau 
and Lifshitz apaprently intended their analysis to apply only to the 
limit of zero superfluid velocity, while in this paper we work at
arbitrary superfluid velocity. In this version of our paper we have 
instead compared our results to the predictions of Clark and Putterman who 
explicitly work at finite superfluid velocity.
We once again find disagreement with their predictions, and have motivated
us to construct a slight generalization of the Clark-Putterman formalism, 
with which our gravitational results now agree. We thank C. Herzog and 
A. Yarom for making us aware of Clark and Putterman's work.}.

Given the fact that the gravitational entropy current agrees with the 
current employed by Clark and Putterman \cite{Clark,Putterman}, it is puzzling that the gravitational 
results do not fit into Clark and Putterman's framework. We believe that the 
resolution to this puzzle is that Clark and Putterman missed a parameter  
in their analysis. In \S \ref{dissrev} below we demonstrate that the 
most general equations of super fluid dynamics consistent with 
positivity of the canonical entropy current is parameterized by 21 dissipative
coefficients rather than 20 as mistakenly asserted by Clark and Putterman. Imposition 
of the 7 Onsager relations then leaves us with a 14 (rather than 13) parameter
set of consistent equations of first order dissipative super fluid dynamics.
We present this generalized 14 parameter set of equations for dissipative
fluid dynamics in \S \ref{dissrev} below. It turns out that our gravitational 
results fit perfectly into this enlarged 14 parameter framework \footnote{
Imposing conformal invariance (which is relevant for the gravitational calculation)
sets 4 out of these 14 parameters to zero (see \S \ref{weyl} for more details).}. In 
\S \ref{resinDifframe} below we explicitly list the dissipative parameters for our 
gravitational superfluid.  

Our paper is organized as follows. In \S \ref{review}, \S \ref{dissrev} 
and \S \ref{llcp} we present a general 
framework for relativistic superfluid dynamics including dissipative effects at first order in the 
derivative expansion. These sections are purely fluid dynamical and makes 
no reference to the equations of gravity or the AdS/CFT correspondence. 
The chief new result of \S \ref{dissrev} is the presentation of the 14 parameter 
set of equations that generalize Clark and Putterman's 13 parameter equations \cite{Clark,Putterman}. 
We also demonstrate that the equations of Weyl invariant superfluid dynamics 
are parameterized by 10 dissipative parameters, and present the 
most general form for these equations. In \S \ref{eq} below we review 
and slightly generalize Sonner and Wither's gravitational derivation of the 
Landau Tiza 2 fluid model from gravity. In \S \ref{static} we perturb Herzog's
construction of hairy black branes away from the strict large charge or 
probe limit. In \S \ref{1ordderexp} we use the results of \S \ref{static} as 
an input into 
the fluid gravity map to generate the gravitational solutions dual to 
superfluid flows. We compute the dissipative part of the stress tensor, 
charge current and entropy current dual to superfluid flows and also 
verify the Weyl invariance of our results. In \S \ref{resinDifframe} we transform 
our gravitational stress tensor and charge current to a frame convenient
for the study of fluid dynamics, and demonstrate that our gravitational 
results are a special case of the 14 parameter set of dissipative fluid 
dynamical equations derived in \S \ref{dissrev}. We also explicitly list 
the values of all dissipative parameters of our gravitational superfluid.
In appendix \ref{entropy} we extend the discussion of the canonical 
entropy current in \S \ref{CEntrop} to demonstrate that it is independent of 
the choice of frame and also compute its divergence. In appendix \ref{SgravThermo} we present 
an abstract geometrical derivation of the analogue of Gibbs Duhem relations for 
superfluids in thermodynamic equilibrium directly from gravity. In appendix \ref{stab} we 
show the existence of an upper bound for the superfluid velocity beyond which
the superfluid phase is unstable. Finally, in appendix \ref{rot} we complete
the discussion of \S \ref{App:zetazero} by manifestly recovering complete $SO(3)$ 
rotational invariance in the zero superfluid velocity limit.

{\it Note Added}: Our paper has substantial overlap with \cite{Herzog:2011ec}, which 
was posted on the ArXiv simultaneously with the first version of this paper.

\section{Review of relativistic superfluid dynamics}\label{review}

\subsection{Relativistic superfluids in equilibrium}\label{fleq}

Consider a relativistic quantum field theory with a conserved $U(1)$ charge. 
In the sequel we denote the conserved $U(1)$ current of this theory by $J^\mu$ 
and the conserved stress tensor by $T^{\mu\nu}$. 
Consider this system at a finite temperature $T$ and finite chemical 
potential $\mu$. It is conceivable that a charged operator $O$ (charged under 
the $U(1)$ charge described above) develops a nonzero vev over a certain 
range of temperatures and chemical potentials. Whenever this happens 
our system is said to display superfluidity. Superfluidity, then, is 
associated with the spontaneous breakdown of a global $U(1)$ symmetry. 
One of the most striking facts about a superfluid is that it admits more 
stationary homogeneous solutions to the equations of motion, 
on $R^{3,1}$, that might naively have been supposed. Any system with 
a $U(1)$ charge admits at least a 5 parameter set of homogeneous solutions
on $R^{3,1}$; these solutions represent the system in equilibrium at 
temperature $T$ and chemical potential $\mu$, moving at a uniform 
four velocity $u^\mu$. However superfluids actually admit an 8 parameter
set of homogeneous stationary solutions, as we will now explain. 

Let $\psi$ denote the phase of the condensed operator $O$. $\psi$ is 
effectively a massless scalar field in the superfluid phase. A massless
scalar field admits solutions of the form $\psi(x)= e \xi_\mu x^\mu$ for 
arbitrary constant values of $\xi_\mu$ (here $e$ is the charge of 
the operator $O$). It turns out that a superfluid 
admits homogeneous stationary solutions at every constant value of $T$, $u^\mu$ and 
$\xi_\mu$.  

There is another way to think of the 8 parameter set of solutions listed above.
We can `gauge' the global symmetry of the field theory, and so couple 
the theory to a non dynamical gauge field $A_\mu$. The solutions described 
above may all, equivalently, be thought of as solutions with constant 
values of the phase $\psi$, but with the non dynamical gauge fields 
$$A_\mu=\xi_\mu $$
where $e$ is the charge of the field 
The equivalence of these two ways of describing these solutions follows from 
the fact that a phase $e \xi_\mu x^\mu$ of a field of charge $e$  is 
gauge equivalent to the gauge field listed above.

Finally some definitions. We define 
$$\mu =\xi_\mu u^\mu .$$ 
$\mu$ is referred to as the chemical potential of the system (or 
sometimes as the chemical potential of the normal part of the system). 
This terminology is reasonable as $\mu$ is equal to the asymptotic value of the time component 
of the gauge field in the frame in which the normal fluid is at rest (i.e. 
in the frame in which $u^\mu=(1,0,0,0)$). Also let 
$$\xi=\sqrt{-\xi_\mu \xi^\mu}.$$ 
We define the normalized 4 vector 
$$u_s^\mu =-\frac{\xi^\mu}{\xi}$$
and will refer to $u_s^\mu$ as the `superfluid velocity' in the solution. 
Roughly speaking, $u_s^\mu$ is the four velocity at which the stuff associated
with the condensate, $\langle O \rangle$ moves.

With all this terminology in place, we can reword the central assertion of 
the previous paragraph as follows: the stationary non dissipative solutions 
of a superfluid are labeled by the three components of $u^\mu$, the `velocity'
of the normal component of the fluid, the three components of $u_s^\mu$, 
the velocity of the superfluid (or condensate) of the fluid, the temperature
$T$ and the chemical potential $\mu$. Most importantly we have exactly 
stationary solutions in which the `normal' and `superfluid' parts of the 
system move at arbitrary speeds with respect to one another. 

We now turn to the quantitative characterization of the stationary homogeneous
flows described above. We would first like quantitative expressions for the 
stress tensor and the charged current of the superfluid, as a function of 
the eight parameters characterizing flow. These eight parameters have 
three associated scalars, namely $T, \mu, \xi$. Symmetry considerations 
immediately allow us to write expressions for the stress tensor and current
in terms of a set of arbitrary functions of these scalar fields. 
\begin{equation} \label{ltconstperf}
\begin{split}
T^{\mu\nu}&=(\rho_n+P) u^\mu u^\nu + P \eta^{\mu\nu} + 
\frac{\rho_s}{\xi^2} \xi^\mu \xi^\nu \\
&=(\rho_n+P) u^\mu u^\nu + P \eta^{\mu\nu} + 
\rho_s u_s^\mu u_s^\nu\\
J^\mu &=q_n u^\mu - q_s \frac{\xi^{\mu}}{\xi}\\
&=q_n u^\mu + q_s u_s^\mu\\
u^\mu \xi_\mu&=\mu \\
\end{split}
\end{equation}
The third equation in \eqref{ltconstperf} is simply a statement of 
definitions.

The reader may wonder why we have not allowed a 
$u^\mu \xi^\nu$ cross term in the first of \eqref{ltconstperf}. 
The reason is as follows. Were such a cross term to infact appear in the 
expansion, we could get rid of it by a redefinition of $u^\mu$. 
The assertion that no such cross term appears in the expansion of the stress 
tensor  constitutes our definition of the normal velocity $u^\mu$.

The quantities $\rho_n$, $P$, $\rho_s$, $q_n$, $q_s$ are all as yet 
arbitrary functions of the three scalar quantities $(T, \mu, \xi)$. 
The claim of the Landau Tiza two fluid theory is that all these quantities 
may be derived from a single function, the pressure of these solutions, 
 $P=P(T, \mu,  \xi)$ via the formulas 
\begin{equation} \label{fconst} 
\begin{split} 
\rho_n+P&=q_n \mu + Ts\\
\rho_s&=\mu_s q_s\\
\mu_s&= \xi = \xi_\mu u_s^\mu\\
dP&=sdT +q_sd \mu_s +q_n d \mu\\
  &=sdT +q_sd \xi +q_n d \mu
\end{split}
\end{equation}

\subsection{Relativistic Superfluid Dynamics} \label{dyn}

At long distances (compared to an effective mean free path) a superfluid 
admits a fluid dynamical description. In this subsection we will present 
what appears to us to be the simplest and most logically satisfactory 
formulation of the theory of dissipative super fluid dynamics. Our 
presentation differs by a field redefinition from the more `traditional' 
presentations of, for instance, Clark, Putterman and Landau Lifshitz \cite{Clark, Putterman, LLvol6}. 
We will later describe the field redefinitions that allow us to
transform our formulation to the more traditional one. 

In the relativistic context we choose the variables of 
superfluid dynamics simply to be $u^\mu(x)$, the four 
velocity of the `normal' fluid, $T(x)$ the local temperature of the fluid, 
and $\xi^\mu(x)$, the local value of the gradient of the superfluid phase. 
More specifically, $\xi^\mu$ is given in terms of the superfluid phase 
$\psi(x)$ by  
\begin{equation}\label{phase}
\xi^\mu(x)=-\partial^\mu \psi(x)
\end{equation} 
We have a total of eight fluid dynamical fields. We will also sometimes use 
the terminology
\begin{equation}\label{mde}
\mu(x)=u(x).\xi(x).
\end{equation}
We emphasize, though, that within the formulation presented in this subsection, 
$\mu(x)$ is not an independent dynamical field, but merely terminology for 
the projection of $\xi(x)$ on the normal velocity field $u(x)$.

The equations of super fluid dynamics consist of the following relations 
\begin{equation}\label{eomsfd} \begin{split}
\partial_\mu T^{\mu\nu}&=0\\
\partial_\mu J^\mu&=0\\
\partial_\mu \xi_\nu -\partial_\nu \xi_\mu&=0\\
\end{split}
\end{equation}
The first two of these equations are simply the statement of the conservation
of the stress tensor and the conservation of the charge current; these 
equations are true in any field theory. The last of these equations asserts
that $\xi_\nu$ is the gradient of a scalar. In order to see that we have 
as many independent equations as variables, we note that 
\eqref{phase} is a local solution of the last equation in \eqref{eomsfd}. 
This leaves us with 5 variables $u^\mu(x)$, $T(x)$ and  $\psi(x)$ 
subject to the 5 remaining equations in the first two lines of \eqref{eomsfd}. 

While $\xi_\mu(x)$  may be traded for $\psi(x)$ for counting purposes, 
the equations of fluid dynamics are more conveniently formulated directly 
in terms of the variables $\xi_\mu(x)$ rather than the phase field $\psi(x)$. 
The reason for this is that 
gradients of $\psi(x)$ are not necessarily small in the regime of validity of 
superfluid dynamics, while gradients of $\xi^\mu(x)$ necessarily are. 
The introduction of the variables $\xi_\mu(x)$ allows us to formulate the 
equations of superfluid dynamics in a systematic derivative expansion of 
all its participating fields. 

The equations of superfluid dynamics constitute a closed dynamical system 
once they are combined with constitutive relations that determine the 
stress tensor and charge current in terms of the variables of fluid 
dynamics $u^\mu, T, \xi^{\mu}$. The constitutive relations take the 
form 
\begin{equation} \label{ltconst}
\begin{split}
T^{\mu\nu}&=(\rho_n+P) u^\mu u^\nu + P \eta^{\mu\nu} + 
\frac{\rho_s}{\xi^2} \xi^\mu \xi^\nu + \pi^{\mu\nu} \\
J^\mu &=q_n u^\mu - q_s \frac{\xi^{\mu}}{\xi} + J^\mu_{diss}\\
\end{split}
\end{equation}
where $\pi^{\mu\nu}$ and $J_{diss}^\mu$ are respectively 
tensors and vectors that are first or higher order in 
an expansion in derivatives (of the fluid dynamical fields) and all other 
quantities were defined in the previous subsection. 

\subsubsection{A canonical fluid frame}\label{varframe}

The equations of superfluid dynamics change their detailed form under field
redefinitions. The quantity $\xi_\mu(x)$ has a microscopic definition. In the 
rest of this paper $\xi_\mu(x)$ will always refer to this microscopically
unambiguous value, and we will not permit arbitrary 
field redefinitions of $\xi_\mu(x)$. 
 
On the other hand the fluid variables $T(x)$ and $u^\mu(x)$ are only really 
defined in thermodynamical equilibrium, and so allow for possible field 
redefinitions at derivative order (as derivatives parameterize departures from 
thermodynamical equilibrium).

In order to completely specify the equations of superfluid dynamics we need 
to specify unambiguous definitions of the thermodynamical fields $T(x)$, 
and $u^\mu(x)$. We specify these definitions by prescribing certain 
conditions on the dissipative terms $\pi^{\mu\nu}$ and $J^\mu_{diss}$. 
We must choose these conditions so that they are not automatic, but 
can always be reached by an appropriate field redefinition, and completely 
fix field redefinition ambiguity. For instance we could work with the 
superfluid analogue of the `Landau Frame' of ordinary fluid dynamics
\begin{equation}\label{defvar}
\pi_{\mu \nu} u^\mu=0
\end{equation}
These 4  conditions  serve to provide unambiguous definitions for the 
velocity, chemical potential \footnote{Note that here $\mu(x)$ is not 
an independent field and is given by\eqref{mde}.} and temperature fields. For 
reasons that will become clearer below, we will refer to this choice of fluid 
field variables as the $\mu_{diss}=0$ frame. 

The equation \eqref{defvar} may equivalently be written as 
\begin{equation}\label{defvarn}
T^{\mu \nu} u_\nu= -\rho_n u^\mu + (u.u_s) \rho_s u_s^\mu
\end{equation} 
where it is important that the functions $\rho_n$ and $\rho_s$ on the RHS of 
\eqref{defvarn} are not independent functions, but are related to each 
other by the thermodynamical relations \eqref{fconst} \footnote{
In our presentation 
above we have used the temperature field as one of the independent fields
of the superfluid dynamical description. Of course this choice is arbitrary; 
we could as well use any other thermodynamical field (for instance the 
energy density $\rho_n$ instead of the temperature, where $\rho_n$ is
defined as the thermodynamical function of $T$, $u. \xi$ and $\xi$) in 
place of the temperature. As such thermodynamical reparameterizations are 
ultralocal (i.e. do not involve derivatives) they do not affect the split of 
the stress tensor and current into perfect fluid and dissipative parts, and 
so do not constitute a change of frame.}.

\subsection{`Fluid Frames'}\label{other}

While the frame presented in the previous subsection seems 
to us to be rather natural from several points of view, it turns 
out not to be the fluid frame most commonly employed in analysis of 
superfluid dynamics. In this subsection we will describe a generalized 
framework for superfluid dynamics. In the next subsection 
we will describe how to transform between fluid frames. 

Conventional descriptions of superfluid dynamics are presented in terms of 
9 fields subject to a single constraint rather than 8 independent fields. 
In this subsection we will take these 9 fields to be $T(x)$, $\mu(x)$, 
$u^\mu(x)$ and $\xi^\mu(x)$. 

While in the the previous subsection the field $\mu(x)$ was simply
convenient notation for $u^\mu(x) \xi_\mu(x)$, in this subsection 
$\mu(x)$ is an independent field variable; the relation between $\mu(x)$ 
and $u(x).\xi(x)$ is taken to be given by the so called Josephson equation
\begin{equation} \label{josephson}
u(x).\xi(x) = \mu(x) +\mu_{diss}(x)
\end{equation}
where $\mu_{diss}(x)$ is a function of derivatives of fluid variables (i.e. 
it vanishes when all fluid variables are constants in spacetime). The 
quantity $\mu_{diss}$ will be chosen in order to ensure that another
condition we specify below is satisfied. Comparing 
\eqref{mde} with \eqref{josephson} explains why we referred to the frame of the 
previous section as the $\mu_{diss}=0$ frame. 

The stress tensor and charge current continue to be given by the form 
\eqref{ltconst}, with the understanding, however, that all thermodynmical 
functions in those formulas are to be regarded as functions of the fields 
$(T(x), \mu(x), \xi(x))$ rather than the fields $(T(x), u.\xi(x), \xi(x))$ 
(these two choices were identical in the previous subsection). As 
$\mu(x)$ is ${\it not}$ equal to $u.\xi(x)$ for a general fluid flow, it 
follows that the perfect fluid part of the fluid stress tensor (at a given 
spacetime point) is, in general, not equal to stress tensor for any 
equilibrium flow (this is a complication that was absent in 
our formulation of the previous subsection). 

In the current formulation we have 5 thermodynamical fields - $T(x)$, 
$u^\mu(x)$ and $\mu(x)$ - that require precise definitions. This requires 
us to specify five equations (generalizing the four conditions in, e.g. 
\eqref{defvar}) to give meaning to these fields. One natural choice \cite{Pujol:2002na}
for these equations is 
\begin{equation}\label{defvarnn}
\begin{split}
\pi_{\mu \nu} u^\mu&=0\\
J^\mu_{diss} u_\mu&=0.\\
\end{split}
\end{equation}
We refer to the fluid frame defined by these equations as the 
Transverse Frame. These equations effectively determine the previously 
undetermined quantity $\mu_{diss}$. 

Of course \eqref{defvarnn} is only one set of the possible set of 
five equations we must impose on our thermodynamical variables. 
Many other choices are possible. One other possible choice is 
\begin{equation}\label{defvarnnn}
\begin{split}
\pi_{\mu \nu} u^\mu&=0\\
\mu_{diss}&=0.\\
\end{split}
\end{equation}
which defines the $\mu_{diss}=0$ frame of the previous subsection. It follows 
that the formulation of superfluid dynamics, described in this subsection, 
is a generalization of the formulation of the previous subsection, and 
includes the later as a special case.

We end this subsection by emphasizing our terminology. We refer to the 
formulation of fluid dynamics, presented in this subsection, as the 
formulation in terms of fluid frames. The key feature of this description
of fluid dynamics is that the full microscopically defined gradient of 
phase, $\xi^\mu$, is taken as one of the variables of description. 
The perfect fluid part of the stress tensor and charge current is written 
in terms of $\xi^\mu$ and thermodynamical functions of $\xi$. Superfluid
dynamics in a fluid frame is to be contrasted with super fluid dynamics 
in a modified phase frames, introduced in \S \ref{mpf}.

\subsection{Transforming between fluid frames}\label{transf}

In this subsection we supply the equations that allow us to 
transform between fluid frames. Let us suppose we want to 
make a change of variables that will 
take us from a completely unspecified fluid frame labeled `our' 
to another frame labeled `there', where the frame `there' is a 
well defined fluid frame (e.g. the transverse or the $\mu_{diss}=0$ frame). 
We set 
\begin{equation} \begin{split}\label{fo}
u^\mu_{there}&= u^\mu_{our} + \delta u^\mu\\
T_{there}&=T_{our} + \delta T\\
\mu_{there}&=\mu_{our}+ \delta \mu
\end{split}
\end{equation}
The quantities $\delta u^\mu$, $\delta T$ and $\delta \mu$ are necessarily 
of first or higher order in derivatives. As we work only to first order 
in derivatives in this section, we will effectively work only to linear 
order in these variations. 

Note that in order that $u^\mu_{there}$ and $u^{\mu}_{our}$ both be unit 
normalized, it is necessary that 
$$\delta u_\mu u^\mu=0$$
(at the order 
at which we work $\delta u_\mu u^\mu_{our}=\delta u_\mu u^\mu_{there}$ as 
the two differ at quadratic order in $\delta$. Whenever an equation is 
true both of - say - $u^\mu_{our}$ and $u^\mu_{there}$ we will simply 
omit the subscript in this subsection). 

With infinitesimal variations restricted as above we find 
\begin{equation} \begin{split} \label{channge}
\delta \pi^{\mu\nu} & = 
(u^\mu \delta u^\nu + u^\nu \delta u^\mu) (P+\rho_n) + 
u^\mu u^\nu d(P+\rho_n) + \frac{\xi^\mu \xi^\nu}{\xi^2} d(\rho_s)
+ \eta^{\mu\nu} dP\\
\delta J^{\mu}_{diss}&=q_n \delta u^\mu + d q_n u^\mu - d q_s 
\frac{\xi^\mu}{\xi} \\
\delta{\mu_{dis}} &= -\delta u^{\mu}\xi_{\mu}+\delta \mu.
\end{split}
\end{equation}
where by  
$$\delta \pi^{\mu \nu}=\pi^{\mu \nu}_{our}- \pi^{\mu \nu}_{there}$$
$$\delta J^{\mu}_{diss} = (J^{\mu}_{diss})_{our}-
(J^{\mu}_{diss})_{there}$$
and
$$\delta \mu_{diss} = (\mu_{diss})_{our}- (\mu_{diss})_{there}$$
and in these equations the symbol $d f(\mu, \xi, T)$ represents the change in 
the function $f$ under the first order variable change \eqref{fo}. 
These equations may then be used to obtain the 5 equations that determine 
the five unknowns $\delta \mu$, $\delta T$ and $\delta u^\mu$. 
For instance, if we are interested in transforming to the transverse 
frame (i.e. we want to set `there' to be the transverse frame) 
we would require that 
$\pi^{\mu\nu}_{there}$ and $J^\mu_{there}$ be orthogonal to $u^\mu_{there}$, 
giving the equations
\begin{equation} \label{fc}
\begin{split}
&d \rho_n + \frac{\mu^2}{\xi^2} d \rho_s - \pi_{our}^{\mu\nu} u_\mu u_\nu=0\\
&(P+\rho_n) \delta u^\mu =\frac{\mu}{\xi^2}  d \rho_s \xi^\mu
-u^\mu d \rho_n - \pi_{our}^{\mu \nu} u_\nu \\
&d q_n  +d q_s\frac{\mu}{\xi} +(J^\mu_{diss})_{our} u_\mu=0  \\
 \end{split}
\end{equation}

The 5 equations \eqref{fc} determine $\delta u^\mu$, $\delta \mu$, 
$\delta T$, and so the change in the dissipative part of the stress tensor 
and current etc from \eqref{channge}. A similar procedure may be employed 
to transform into the $\mu_{diss}=0$ frame or any other frame of interest.

\subsection{A canonical Entropy Current}\label{CEntrop}

In this section we will define a `canonical' entropy current in any 
fluid frame. Our definition is
\begin{equation}\label{entcur} 
J^\mu_s= s u^\mu - \frac{\mu}{T} J^\mu_{diss} - \frac{u_\nu \pi^{\mu\nu}}{T}
\end{equation}
where $s$ is the thermodynamical entropy density of our fluid. 

Although this is not obvious, we have shown in Appendix \ref{entropy} that 
this current is frame invariant. This means, for instance, that
\eqref{entcur} defines the same vector field in the transverse as well as 
the $\mu_{diss}=0$ frames. 

In the same Appendix we 
have also demonstrated that the divergence of this current is given, in 
any fluid frame,  by  
\begin{equation}\label{diventcur}
\partial_\mu  J^\mu_s = -\partial_\mu \left[\frac{u_\nu}{T} \right]\pi^{\mu\nu}  
- \partial_\mu\left[\frac{\mu}{T} \right]J_{diss}^{\mu} 
+\frac{\mu_{diss}}{T}\partial_\mu\left(\frac{\rho_s}{\xi^2}\xi^\mu\right)
\end{equation}

\subsection{Modified Phase frames}\label{mpf}

As we have emphasized above, the theory of superfluid dynamics formulated 
in any fluid frame in which $\mu_{diss} \neq 0$ (like the transverse frame) 
has slightly hybrid features. The perfect fluid part of the charge current
and stress tensor involves thermodynamical functions of $\mu$ not $u.\xi$, 
 but is written (in index structure) in terms of the full field $\xi^\mu$. 
This ensures that, for a general fluid flow, the prefect fluid stress tensor
and charge current, at any given point, does not equal the stress tensor
and charge current for {\it any} equilibrium flow. 

Fluid dynamics formulated in a modified phase frame eliminates this slightly
unpleasant feature by working with a modified gradient of phase field, 
$\xi^\mu_0(x)$ defined as 
\begin{equation} \begin{split} \label{xinotdef}
\xi_0^\mu &= -\mu u^\mu +w^\mu\\
\xi_0 &= \sqrt{\mu^2 -w^2}\\
\end{split}
\end{equation}
where $w^\mu$ is defined as the part of $\xi^\mu$ projected orthogonal 
to $u^\mu$. 

Note that $\xi_0^\mu$ is not equal to the phase field $\xi^\mu$ (because 
$\mu \neq u_\mu\xi^\mu$ ). Instead the relationship between these two fields
is given by 
\begin{equation} \begin{split} \label {xirel}
\xi^\mu &= \xi_0^\mu - \mu_{diss}u^\mu\\
\xi & = \xi_0+ \mu_{diss}\frac{\mu}{\xi_0}\\
\end{split}
\end{equation}
Note also that, by construction, $\xi_0.u= \mu$.

The fluid stress tensor, charge current, in a modified phase frame, 
are assumed to take the form  
\begin{equation} \label{ltconst2}
\begin{split}
T^{\mu\nu}&=(\rho_n+P) u^\mu u^\nu + P \eta^{\mu\nu} + 
\frac{\rho_s}{\xi_0^2} \xi_0^\mu \xi_0^\nu + {\tilde \pi}^{\mu\nu} \\
J^\mu &=q_n u^\mu - q_s \frac{\xi^{\mu}_0}{\xi_0} + {\tilde J}^\mu_{diss}\\
\end{split}
\end{equation}
where all thermodynamical functions are taken to be functions of $(T, \mu, 
\xi_0)$. 

As in the case of fluid frames, the precise definition of any given modified 
phase frame requires the specification of 5 additional conditions (to give 
precise meaning to the fields $T(x)$, $u^\mu(x)$ and $\mu(x)$). 
The `Landau-Lifshitz-Clark-Putterman' frame is a modified phase frame in which 
the additional conditions are taken to be
$${\tilde J}^\mu_{diss}= u_\mu u_\nu {\tilde \pi}^{\mu\nu}=0$$
As the name suggests, this is the frame employed by Landau Lifshitz, Clark and 
Putterman \cite{Clark, Putterman, LLvol6}(in a non relativistic context) in their analysis of superfluid 
dynamics. 

It is not difficult to generalize the analysis of \S \ref{transf} to 
describe the transformation from a modified phase frame to a fluid 
frame or vice versa. We will implement such a frame change in the next 
subsection.

\subsection{The canonical entropy current in modified phase frames}

The most general modified phase frame may be obtained starting from 
the most general fluid frame and then reexpressing all quantities in terms 
of $\xi_0^\mu$ and $\xi_0$ rather than $\xi^\mu$ and $\xi$. Let the 
fluid frame we start with be characterized by $\pi^{\mu\nu}$ and $J^\mu_{diss}$
and $\mu_{diss}$. The generalized phase frame, obtained from this fluid frame 
by reexpressing $\xi^\mu$ as a function of $\xi^\mu_0$ has 
\begin{equation} \label{genf} \begin{split}
{\tilde \pi}^{\mu\nu}&= \pi^{\mu\nu} -  d(\rho_n + P) u^\mu u^\nu - dP\eta^{\mu\nu}
-df \xi_0^\mu\xi_0^\nu -f ~\mu_{diss} (u^\mu\xi_0^\nu + u^\nu \xi_0^\mu) \\
{\tilde J}^\mu_{diss}&= J^\mu_{diss}- dq_n u^\mu + d f \xi_0^\mu 
+f~ \mu_{diss} u^\mu \\
\xi_0.u&=\mu\\
\xi.u&= \mu +\mu_{diss}
\end{split}\
\end{equation}
where $f = \frac{q_s}{\xi_0} = \frac{\rho_s}{\xi_0^2}$
Further, any thermodynamical function $A$ in the fluid frame is related 
to the corresponding thermodynamical function ${\tilde A}$ 
in the modified phase 
frame by 

$$dA = \tilde A - A =
A(\xi_0) - A(\xi) = -\mu_{diss}\left(\frac{\mu}{\xi_0}\right)\frac{\partial A}{\partial \xi}$$

As we have explained above, the canonical entropy current in an arbitrary
fluid frame is given by \eqref{entcur}. 
Applying the transformation formulas described above, this entropy current
may be expressed in terms of the modified phase thermodynamical and dissipative
parameters as (see Appendix \ref{cc})
\begin{equation}\label{mfent}
\tilde J^\mu_s= s(\xi_0) u^\mu - \frac{\mu}{T} \tilde J^\mu_{diss} - \frac{u_\nu \tilde \pi^{\mu\nu}}{T}
+\frac{f}{T}\mu_{diss} w^\mu
\end{equation}
In obtaining \eqref{mfent} we have used the thermodynamical identity \eqref{fconst}.

It is also possible to transform the equation for the divergence 
of the entropy current, \eqref{diventcur}, into the modified phase frame 
(or simply rederive this expression directly in the modified 
phase frame).  We find (see Appendix \ref{cb})
\begin{equation}\label{divergentfpa}
 \begin{split}
\partial_\mu\tilde J^\mu_S
 =& -\partial_\mu\left[\frac{u_\nu}{T} \right]\tilde \pi^{\mu\nu} 
 - \partial_\mu\left[\frac{\mu}{T} \right]\tilde J_{diss}^{\mu} 
+\mu_{diss}P^{\mu\nu}\partial_\mu\left(\frac{fw_\nu}{T}\right)
 \end{split}
\end{equation}

\section{A theory of first order dissipative superfluid dynamics 
in fluid frames} \label{dissrev}

In this subsection we will present a `theory' of dissipative superfluid 
dynamics to first order in the derivative expansion. By this we mean that 
we will parameterize the allowed forms of $\pi^{\mu\nu}$ and $J^\mu_{diss}$ 
at first order in the derivative expansion. Our parameterization will be 
in terms of a certain number of undetermined functions of 
$T$ $\mu$ and $\xi$. One of these functions is the viscosity of the 
normal part of the superfluid. Following standard (but slightly misleading) 
usage, we will refer to these functions as dissipative parameters of the 
superfluid. 

\subsection{Summary of arguments and results}

As the analysis of this section will be rather lengthy, we 
first present a summary of our logic and our procedure. To start with 
we simply classify all onshell inequivalent one derivative 
contributions to $\pi^{\mu \nu}$, $J^\mu_{diss}$ and $\mu_{diss}$. 
It is not difficult to establish that, in any given frame (e.g. the 
transverse frame or $\mu_{diss}=0$ frame or the Landau-Lifshitz-Clark-Putterman
frame) there exists a 36 parameter space 
of inequivalent first derivative corrections to the equations of superfluid 
dynamics (assuming parity invariance). 

In order to cut down the set of possibilities we then
follow Landau Lifshitz, Clark and Putterman \cite{Clark, Putterman, LLvol6} to make the central assumption 
of this subsection. We assume that the entropy current takes the canonical 
form described in \S \ref{CEntrop} \footnote{This assumption is not 
universally valid. In particular it seems 
certain to fail in situations in which the $U(1)$ symmetry in question has 
a $U(1)^3$ anomaly. It may also fail in other circumstances. We leave the 
investigation of the validity of this assumption to future work.}. A local form 
of the second law of thermodynamics then asserts that the equations 
of superfluid dynamics should be geared to ensure that the divergence 
of this entropy current is positive. Using the expression \eqref{diventcur}
we find that this requirement cuts down the 36 parameter space of possible 
one derivative corrections to the entropy current to a 21 parameter space 
of possibilities. The coefficients in this 21 parameter space are further 
constrained by a complicated set of inequalities that ensure positivity 
of the entropy production. (One of these inequalities, for instance, 
asserts the positivity of the normal viscosity). Finally, 
the Onsager reciprocity relations relate 7 of the remaining parameters to 
each other, leaving us with a 14 parameter space of dissipative 
coefficients. As mentioned above, these 14 parameters (each of which is 
a function of $T$, $\mu$ and $\xi$) are further constrained to obey a set 
of inequalities. As far as we are aware, there are no further 
restrictions on this 14 parameter space from general principles. 

To end this summary we explain how the framework presented 
in this subsection relates to previous work. The programme outlined in 
the paragraph above was implemented by Landau and Lifshitz \cite{LLvol6} for the 
special case of flows with zero (or negligibly small) superfluid velocities. 
Landau and Lifshitz  found a set of equations with 5 first order dissipative
parameters. Our 14 parameter set of equations indeed reduce to the Landau 
Lifshitz form upon setting the superfluid velocity to equal the normal 
velocity; consequently our framework agrees with that of Landau and Lifshitz
within its domain of validity. 

Clark and Putterman \cite{Clark, Putterman} extended the Landau Lifshitz programme to the case of nonzero 
superfluid velocities. The end result of their analysis was a thirteen parameter
set of equations. We believe that Clark and Putterman overlooked one 
allowed parameter \footnote{Specifically, the traceless symmetric 3 index 
tensor listed in equation (A VI-$9$) of Putterman's book is not unique. Another 
such tensor is given by 
$$Y_{ijk} =  w_i ( w_j w_k - (1/3) w^2 \delta_{j k} ).$$}

Reinstating that parameter yields our 14 parameter set of 
equations. Thus Clark and Putterman's equations are a special case of our 14 parameter 
equations with one parameter set to zero. 

As we have explained in the previous section, one of the complicating features 
of superfluid dynamics is that one can work in many different frames. 
In this section we work out the `theory' of dissipative superfluid dynamics 
in the two natural fluid frames described in the previous subsection, namely 
the transverse frame and the $\mu_{diss}=0$ frame. In the next section 
we present the equivalent analysis for in the Landau-Lifshitz-Clark-Putterman
frame.

\subsection{Counting of parameters}

At a given spacetime point, a superfluid flow has two independent velocities;
the normal velocity $u^\mu$ and the superfluid velocity $\xi^\mu_s$. These
two velocities break the local Lorentz rotation group $SO(3,1)$ down to 
$SO(2)$, the group of spatial rotations in the plane orthogonal to both 
velocities.

In \S \ref{consteq} we enumerate the onshell inequivalent first 
derivatives of all fluid dynamical fields at a point. We found it convenient
to classify these derivatives as scalars (spin 0), vectors (spin $\pm 1$) 
and tensors (spin $\pm 2$) under the unbroken $SO(2)$ described in the 
previous paragraph. As explained in \S \ref{consteq} it turns out 
that there are 6 onshell inequivalent parity even scalars, 
5 onshell inequivalent parity even vector and 2 onshell inequivalent parity 
even tensor first derivatives of fluid dynamical fields.
\footnote{In addition we have one additional parity odd scalar field. 
Further, every vector $V_\mu$ can be transformed to a psuedo vector 
${\tilde V}_\mu$ according to the formula ${\tilde V}_\mu=
\epsilon_{\mu\nu\alpha \beta} V^\nu n^\alpha u^\beta$.}

In order to be specific we will assume in the rest of this subsection that 
we are working in the transverse frame. Very similar arguments can be 
made in a fluid or modified phase frame, and give identical results. 

In the transverse frame, $\pi^{\mu \nu}$ has two inequivalent scalar 
components  $\xi^\mu \xi^\nu \pi_{\mu\nu}$ and $\pi^\mu_\mu$, one vector 
component $\tilde P^\mu_\alpha\pi^{\alpha\nu}\xi_\nu$ and a single tensor component
${\tilde P}_{\alpha \mu} {\tilde P}_{\beta \nu} \pi^{\alpha \beta}$, 
where ${\tilde P}_{\alpha \beta}$ is the projector orthogonal to the 
$u^\mu$, $\xi^\mu$ plane. On the other 
hand $J^\mu_{diss}$ has one scalar component $J^\mu_{diss}\xi_\mu$ and 
one vector component ${\tilde P}_{\alpha \mu} J^\alpha_{diss}$. 
Finally $\mu_{diss}$ has one scalar component. It follows 
that the total number of undetermined parameters in the 
arbitrary expansion of $\pi^{\mu\nu}$, $J^\mu_{diss}$ and $\mu_{diss}$
   in terms of 
first derivatives of the fluid dynamical fields (assuming parity invariance) 
is given by 
$$4 \times 6  +2 \times 5+ 2=36$$
where the three terms above originate in the scalar, vector and tensor 
sector respectively. \footnote{Dropping the assumption of parity invariance 
we have $4 \times 7+ 2 \times 10 + 2=50$ independent dissipative coefficients.}

\subsection{Constraints from positivity of entropy production and 
Onsager relations in the transverse frame}\label{ep}

In this subsection we will explore the constraints on dissipative 
coefficients from the physical requirements of positivity of entropy 
production and the Onsager reciprocity relations. We will find these 
requirements cut down the 36 parameter set of possible dissipative 
coefficients (assuming parity invariance) to a 14 parameter set of coefficients that are further 
constrained by positivity requirements.

For concreteness we present our analysis in the transverse frame. We will 
record the results of similar analysis in other fluid frames in 
later subsections.

\subsubsection{Constraints from positivity of entropy production}\label{cep}

The divergence of the `canonical' entropy current, 
given by \eqref{diventcur}, involves only terms proportional to 
$\partial_\mu u_\nu \pi^{\mu \nu}$, $\partial_\nu (\mu/T) J^\nu_{diss}$ 
and $\mu_{diss}\partial_\mu(q_s \xi^\mu/\xi)$. Let us examine these terms 
one by one. In the transverse frame 
$$\partial_\mu u_\nu \pi^{\mu \nu}=\sigma_{\mu \nu} \pi^{\mu\nu}
 + \left(\frac{\partial_\mu u^\mu}{3}\right) \pi^\theta_\theta$$ 
where $\sigma_{\mu\nu}$ is the traceless symmetric part of $\partial_\mu u_\nu$, projected in the direction
perpendicular to $u^\mu$.

\begin{equation}\label{sigmadef}
 \begin{split}
&  \sigma_{\mu\nu} = P^\alpha_\mu P^\beta_\nu
 \left(\frac{\partial_\alpha u_\beta +\partial_\beta u_\alpha}{2}
 - \eta_{\alpha\beta} \frac{[\partial.u]}{3}\right)\\
&\text{and}\\
&P_{\mu\nu} = \text{The projector}= \eta_{\mu\nu} + u_\mu u_\nu
 \end{split}
\end{equation}

Now the field 
$\sigma_{\mu\nu}$ has one scalar piece of data 
\footnote{In the terminology of \S \ref{consteq} below, $S_w=\frac{2 S_4 - S_6}{3}$}
$$S_{w}= n^\mu n^\nu\sigma_{\mu\nu}$$ 

one vector piece of data
\footnote{In the terminology of \S \ref{consteq} below, $[V_b]_\mu=\frac{1}{2}[V_5]_\mu$}
$$[V_b]_\mu = \tilde P_\mu^\nu n^\alpha\sigma_{\nu\alpha} $$
and a tensor piece of data 
\footnote{In the terminology of \S \ref{consteq} below, ${\cal T}_{\mu\nu}=\frac{1}{2}[T_1]_{\mu\nu}$}
$${\cal T}_{\mu\nu} =\tilde P_\mu^\alpha \tilde P_\nu^\beta\sigma_{\alpha\beta}$$
The trace of $\pi^{\mu\nu}$ couples to another scalar piece of data
\footnote{In the terminology of \S \ref{consteq} below, $S_{w'}=S_4 + S_6$.}
$$S_{w'}=\partial_\mu u^\mu.$$

In the expressions above $n^\mu$ is the unique normal vector in the plane 
spanned by $u^\mu$ and 
$\xi^\mu$ that is orthogonal to $u^\mu$ and is given by
$$ n_{\mu} \equiv \frac{w_{\mu}}{w},$$
with $w$ being the norm of the $w_{\mu}$ vector.

Similarly, in the transverse gauge 
$$\partial_\nu (\mu/T) J^\nu_{diss}= 
P^\nu_\alpha \partial_\nu (\mu/T) J^\alpha_{diss},$$
where $P_{\mu \nu}$ is the projection operator (defined in \eqref{sigmadef})
 that projects orthogonal to $u_{\mu}$
only .
The quantity $P_\alpha^\nu \partial_\nu (\mu/T)$ has one 
scalar piece of data 
$$ S_b = (n^\mu\partial_\mu)\left(\mu/T \right) $$
and one vector piece of data 
$$[V_a]_\mu = \tilde P_\mu^\sigma\partial_\sigma \left(\mu/T \right) $$

Finally
$$ S_a= \frac{\partial_\mu(q_s \xi^\mu/\xi)}{T^3}$$
 is itself a scalar piece of data. 

In other words we conclude that the expression for the divergence 
of the entropy current, \eqref{diventcur}, depends explicitly (i.e. 
apart from the dependence of $\pi^{\mu\nu}$ $J^\mu_{diss}$ and 
$\mu_{diss}$ on these terms)  only on 
4 scalar expressions, 2 vector 
expressions and one tensor expression. Let us choose these 4 vectors 
scalars $S_a$, $S_b$, $S_{w}$ and $S_{w'}$, supplemented by 3 other arbitrarily 
chosen scalar expressions $S^I_{m}$ ($m = 1 \ldots 3$) as our 7 
independent scalar expressions. Similarly we choose the 2 vectors $[V_a]_\mu$ 
and $[V_b]_\mu$ supplemented by 3 other arbitrarily chosen 
expressions $[{V^I_m}]_\mu$ ($m = 1 \ldots 3$) as our four independent 
vector expressions. We also choose 
${\cal T}_{\mu\nu}$ as one of our two independent tensor expressions
\footnote{We could now go ahead and use the perfect fluid equations to 
solve for  for the dependent data in terms of independent data; however 
we will not need the explicit form of this 
solution in this subsection}. We proceed to express 
$\pi^{\mu\nu}$, $J^\mu_{diss}$ and $\mu_{diss}$ as the most general 
linear combinations of all combinations of independent expressions 
allowed by symmetry

\begin{equation}\label{nfstcrf}
\begin{split}
 \pi^{\mu\nu}&= T^3\bigg[\left( P_a S_a + P_b S_b + P_w S_w
+P_{w'} S_{w'} 
+ \sum_{m=1}^3 P^I_m S^I_m \right)
\left( n_{\mu} n_{\nu} -\frac{P_{\mu\nu}}{3}\right)\\
&~~~~~~ ~~+ \left( T_a S_a + 
T_b S_b + T_w S_w + T_{w'} S_{w'} 
+ \sum_{m=1}^3 T^I_m S^I_m \right)  P^{\mu\nu}\\
 &~~~~~~~~+E_a \left(V_a^\mu n^\nu + V_a^\nu n^\mu\right) 
+E_b \left(V_b^\mu n^\nu + V_b^\nu n^\mu\right)
+\sum_{m=1}^3 E^I_{m} \left([V_m^I]^\mu n^\nu + [V^I_m]^\nu n^\mu\right) 
 \\
&~~~~~~~~+\tau {\cal T}^{\mu\nu} +\tau_2 T_2^{\mu\nu}\bigg]\\
J^\mu_{diss}&=
T^2 \bigg[\left( R_a S_a + R_b S_b + R_w S_w + R_{w'} S_{w'}+
\sum_{m=1}^3 R^I_{m} S^I_m \right)n^\mu\\
&~~~~~~~~~+ C_a V_a^\mu  + C_b V_b^\mu
+ \sum_{m=1}^3 C^I_m [V^I_m]^\mu \bigg]\\
\mu_{diss}& = -\left[ Q_a S_a + Q_b S_b + Q_w S_w + Q_{w'} S_{w'}
+ \sum_{m=1}^3 Q^I_{m} S^I_m \right] 
\end{split}
 \end{equation}

Plugging \eqref{nfstcr} into \eqref{diventcur} we now obtain an explicit 
expression for the divergence of the entropy current as a quadratic form 
in first derivative independent data. We wish to enforce the condition 
that this quadratic form is positive definite. Now the quadratic form 
from \eqref{diventcur} clearly has no terms proportional to $(S^I_m)^2$. 
It does, however, have terms of the form (for instance) $S_a S^I_m$, 
and also terms proportional to $S_a^2$. Now it follows from a moments 
consideration that no quadratic form of this general structure can be 
positive unless the coefficient of the $S_aS^I_m$ term vanishes. 
\footnote{For instance the quadratic form $x^2 + c xy$ (where $c$ is a constant)
can be made negative by 
taking $\frac{y}{x}$ to either positive or negative infinity (depending 
on the sign of $c$) unless $c=0$.} Using similar reasoning we can immediately
conclude that the positive definiteness of 
\eqref{diventcur} requires that 
\begin{equation} \label{zer}
P^I_m=T^I_m=E^I_{m}= C^I_m= R^I_{m}=\tau_2=0.
\end{equation}
\eqref{zer} is the most important conclusion of this subsubsection. It tells us 
that a 21 parameter set of first derivative corrections to the constitutive
relations are consistent with the positivity of the canonical entropy 
current. 

Of course the remaining 21 parameters are not themselves arbitrary, but 
are constrained to obey inequalities in order to ensure positivity. In 
order to derive these conditions we plug \eqref{zer} into \eqref{nfstcrf} and use 
\eqref{diventcur} so that the divergence of the entropy current 
is the linear sum of three different quadratic forms (involving the 
tensor terms, vector terms and scalar terms respectively) 
\begin{equation} \label{divcur}
\partial_\mu J^\mu_s=T^2\left(Q_s+ Q_V+Q_T\right)
\end{equation}
where 
$$Q_T=- \tau {\cal T}^2$$
\begin{equation}\label{vecdivergf}
 \begin{split}
  Q_V=&- C_a  V_a^2 - \left(C_b +  E_a\right) V_b V_a 
- E_b V_b^2\\
=& -C_a \left[V_a + \left(\frac{C_b + E_a}{2 C_a}\right) V_b\right]^2
 - \left[E_b - \frac{(C_b + E_a)^2}{4 C_a}\right] V_5^2
\end{split}
\end{equation}
\begin{equation}\label{scaldivergf}
 \begin{split}
  Q_S=
&-P_w S_w^2-T_{w'} S_{w'}^2 - Q_a S_a^2 - R_b S_b^2\\
& - \left(Q_w +P_a\right) S_w S_a - \left(Q_{w'} +T_a\right) S_{w'} S_a
  - \left(R_w +P_b\right) S_w S_b\\
& - \left(R_{w'} +T_b\right) S_{w'} S_b
- \left(R_a + Q_b\right) S_a S_b -\left(T_w +P_{w'}\right) S_w S_{w'} \\
\end{split}
\end{equation}
Positivity of the entropy current clearly requires that $Q_T$ $Q_V$ 
and $Q_S$ 
are separately positive. Let us examine these conditions one at a time. 
For $Q_T$ to be positive it is necessary and sufficient that $\tau\leq 0$.
This is simply the requirement that the normal component of our superfluid 
have a positive viscosity. In order that $Q_V$ be positive, it is necessary 
and sufficient that 
\begin{equation} \label{vecconst}
C_a\leq0,~~ E_b\leq0~~\text{and}~~ 4 E_b C_a \geq (C_b + E_a)^2.
\end{equation}
Note that this expression involves $C_a$ and $E_b$ on the LHS but the 
different quantities $C_b$ and $E_a$ on the RHS; the last inequality 
above is satisfied roughly, when $C_b$ and $E_a$ are larger in modulus 
than $C_a$ and $E_b$. 

Finally $Q_S$, listed in \eqref{scaldivergf},  is a quadratic form in the the 
4 variables $S_a$, $S_b$, $S_w$ and $S_{w'}$. We demand that this scalar 
form be positive. We will not pause here to explicate the precise 
inequalities that this condition imposes on the coefficients. See below, 
however, for the special case of a Weyl invariant fluid.

\subsubsection{Constraints from the Onsager Relations}
\label{onsg}

In the previous subsection we found that first order dissipative 
corrections to the equations of perfect superfluid dynamics take the form 
\begin{equation}\label{fnlstrcurn}
\begin{split}
 \pi^{\mu\nu}&= T^3\bigg[\left( P_a S_a + P_b S_b + P_w S_w
+P_{w'} S_{w'} 
 \right)
\left( n_{\mu} n_{\nu} -\frac{P_{\mu\nu}}{3}\right)\\
&~~~~~~ ~~+ \left( T_a S_a + 
T_b S_b + T_w S_w + T_{w'} S_{w'} 
\right)  P^{\mu\nu}\\
 &~~~~~~~~+E_a \left(V_a^\mu n^\nu + V_a^\nu n^\mu\right) 
+E_b \left(V_b^\mu n^\nu + V_b^\nu n^\mu\right)
 \\
&~~~~~~~~+\tau {\cal T}^{\mu\nu} \bigg]\\
J^\mu_{diss}&=
T^2 \bigg[\left( R_a S_a + R_b S_b + R_w S_w + R_{w'} S_{w'} \right)n^\mu\\
&~~~~~~~~~+ C_a V_a^\mu  + C_b V_b^\mu
 \bigg]\\
\mu_{diss}& = -\left[ Q_a S_a + Q_b S_b + Q_w S_w + Q_{w'} S_{w'}
\right] 
\end{split}
 \end{equation}
where the coefficients in these equations are constrained by the inequalities
listed in the previous subsubsection. The coefficients that appear in these
equations are further constrained by the Onsager reciprocity relations 
(see, for instance, the text book \cite{LLvol6}, for a discussion). 
These relations assert, in the present context, that we should equate 
any two dissipative parameters that multiply the same terms in the formulas
\eqref{vecdivergf} and \eqref{scaldivergf} for entropy production. 
This implies that  
\begin{equation}\label{onsagarrel}
 \begin{split}
  &Q_w = P_a,~~Q_{w'} = T_a,~~R_w = P_b\\
& R_{w'} = T_b,~~R_a = Q_b,~~T_w = P_{w'}\\
&\text{and}\\
&C_b = E_a
 \end{split}
\end{equation}
In summary we are left with a 14 parameter set of equations of first order 
dissipative superfluid dynamics. The requirement of positivity 
constrains further these coefficients to obey the inequalities spelt 
out in the previous subsubsection.

\subsection{Weyl Invariant Superfluid Dynamics in the transverse frame}\label{weyl}

Let us now specialize the results of \S \ref{cep} and \S \ref{onsg} to the case of super fluid
dynamics for a conformal superfluid. The analysis of \S \ref{cep} 
is simplified in this special case by the fact that the trace of the 
stress tensor vanishes in an arbitrary state (and so in the fluid limit) 
of a conformal field theory. This fact reduces the number of 
explicit scalars that appear in \eqref{diventcur} from 4 to 3 (the 
scalar $S_{w'}$ never makes an appearance). It follows that the 
requirement of Weyl invariance forces $P_{w'}=R_{w'}=T_{w'}=Q_{w'}=0$. 
Moreover the requirement that $\pi^{\mu\nu}$ be traceless 
forces $T_a=T_b=T_w=0$. It turns out that there are no 
further constraints from the requirement of Weyl invariance. 
The expansion of the dissipative part of the stress tensor and charge current
for a conformal superfluid is given by 

\begin{equation}\label{weystrcr}
\begin{split}
 \pi^{\mu\nu}&= T^3\bigg[\left( P_a S_a + P_b S_b + P_w S_w
 \right)
\left( n_{\mu} n_{\nu} -\frac{P_{\mu\nu}}{3}\right)\\
 &~~~~~~~~+E_a \left(V_a^\mu n^\nu + V_a^\nu n^\mu\right) 
+E_b \left(V_b^\mu n^\nu + V_b^\nu n^\mu\right)
 \\
&~~~~~~~~+\tau {\cal T}^{\mu\nu} \bigg]\\
J^\mu_{diss}&=
T^2 \bigg[\left( R_a S_a + R_b S_b + R_w S_w  \right)n^\mu\\
&~~~~~~~~~+ C_a V_a^\mu  + C_b V_b^\mu
 \bigg]\\
\mu_{diss}& = -\left[ Q_a S_a + Q_b S_b + Q_w S_w \right] 
\end{split}
 \end{equation}

The entropy production is given by 
\begin{equation} \label{divcur2}
\partial_\mu J^\mu_s=T^2(Q_s+ Q_V+Q_T)
\end{equation}
where 
$$Q_T=- \tau {\cal T}^2$$
\begin{equation}\label{vecdiverg2}
 \begin{split}
  Q_V=&- C_a  V_a^2 - \left(C_b +  E_a\right) V_b V_a 
- E_b V_b^2\\
=& -C_a \left[V_a + \left(\frac{C_b + E_a}{2 C_a}\right) V_b\right]^2
 - \left[E_b - \frac{(C_b + E_a)^2}{4 C_a}\right] V_b^2
\end{split}
\end{equation}
\begin{equation}\label{scaldiverg2}
 \begin{split}
  Q_S=
&-P_{w} S_{w}^2 - Q_a S_a^2 - R_b S_b^2\\
& + \left(Q_{w} +P_a\right) S_{w} S_a
- \left(R_a + Q_b\right) S_a S_b +\left(R_{w} +P_b\right) S_{w} S_b \\
\end{split}
\end{equation}

For the entropy current to be positive it is necessary and sufficient that 
$\tau\leq 0$ and that 
\begin{equation} \label{vecconst}
C_a\leq0,~~ E_b\leq0~~\text{and}~~ 4 E_b C_a \geq (C_b + E_a)^2.
\end{equation}
and that the quadratic form 
\begin{equation}\label{quad}
\begin{split}
Q_S =& ~a_1 x_1 ^2 +  a_2 x_2 ^2 +  a_3 x_3 ^2
+ b_1 x_1 x_2 + b_2 x_2 x_3 + b_3 x_1 x_3\\
=&~ a_1 \left[x_1 + \left(\frac{b_1}{2 a_1} \right) x_2 + \left(\frac{b_3}{2 a_1} \right) x_3\right]^2 \\
&+\left(a_3 -\frac{b_3^2}{4 a_1}\right)
\left[ x_3 + \left(\frac{2 a_1 b_2 - b_1 b_3}{4 a_1 a_3 - b_3^2}\right)x_2\right]^2\\
&+\left[ \frac{(4 a_1 a_2 - b_1^2)(4 a_1 a_3 - b_3^2) - (2 a_1 b_2 - b_1 b_3)^2}{4 a_1 (4 a_1 a_3 - b_3^2)}\right] x_2^2
 \end{split}
\end{equation}
is positive with 
$x_1 = S_{w}$, $x_2 = S_a$ and $x_3 = S_b$ and 
$$ a_1 = - P_{w},~~a_2 = -Q_a,~~a_3 = -R_b,~~b_1 =Q_{w} + P_a,~~b_2 =-(Q_b + R_a)
,~~b_3 =R_{w} + P_b$$
For the last quadratic form to be positive it is necessary and sufficient that 
\begin{equation}\label{simpcond}
 \begin{split}
 &a_1\geq0\\
&4 a_1 a_2>b_1^2\\
&(4 a_1 a_2 - b_1^2)(4 a_1 a_3 - b_3^2)>(2 a_1 b_2 - b_1 b_3)^2
 \end{split}
\end{equation}
By rewriting \eqref{quad} as a sum of squares in a cyclically permuted manner
we can also derive the cyclical permutations of these equations. 

In summary, the most general Weyl invariant fluid dynamics consistent with 
positivity on the entropy current is parameterized by a negative $\tau_1$, 
4 parameters in the vector sector constrained by the inequalities 
\eqref{vecconst} and 9 parameters in the scalar sector, 
subject to the inequalities \eqref{simpcond}. These 14 dissipative parameters
are further constrained by the 4 Onsager relations  
\begin{equation}\label{onsagarrel}
  Q_w = P_a,~~~R_w = P_b,~~~~R_a = Q_b, ~~~~C_b = E_a
\end{equation}
leaving us with a 10 parameter set of final equations.

\subsection{Weyl Invariant Super Fluid dynamics in the 
$\mu_{diss}=0$ frame}\label{mdweyl}

In this section we present constitutive relations of 
super fluid dynamics in the frame that we consider the most natural, namely 
the $\mu_{diss}=0$ fluid frame. We only present our final results, and 
specialize to the case of a Weyl invariant fluid for simplicity. We find 
\begin{equation}\label{nfstcr}
\begin{split}
 \pi^{\mu\nu}&= T^3\bigg[\left[ P_{{\cal S}} {\cal S} + P_b S_b + P_{w}S_w\right]
\left(n^\mu n^\nu - \frac{ P^{\mu\nu}}{3}\right)\\
 &~~~~~~~~+E_a \left(V_a^\mu n^\nu + V_a^\nu n^\mu\right) 
+E_b \left(V_b^\mu n^\nu + V_b^\nu n^\mu\right) \\
&~~~~~~~~+\tau {\cal T}^{\mu\nu}\bigg]\\
\\
J^\mu_{diss}&= T^2\bigg[\left[ Q_{{\cal S}} {\cal S} + Q_b S_b  + Q_{w}S_w\right]
u^\mu\\
&~~~~~~~~~+\left[ R_{\cal S} {\cal S} + R_b S_b + R_{w}S_w\right]n^\mu\\
&~~~~~~~~~+ C_a V_a^\mu + C_b V_b^\mu\bigg]
\end{split}
 \end{equation}
where ${\cal S}= (u.\partial)\left(\frac{\mu}{T}\right)$ 

The equation of entropy production is given by 
\begin{equation} \label{divcurf2}
\partial_\mu J^\mu_s=T^2\left(Q_s+ Q_V+Q_T\right)
\end{equation}
where 
$$Q_T=- \tau{\cal T}^2$$
\begin{equation}\label{vecdivergf2}
 \begin{split}
  Q_V=&- C_a  V_a^2 - \left(C_b +  E_a\right) V_b V_a 
- E_b V_b^2\\
=& -C_a \left[V_a + \left(\frac{C_b + E_a}{2 C_a}\right) V_b\right]^2
 - \left[E_b - \frac{(C_b + E_a)^2}{4 C_a}\right] V_b^2
\end{split}
\end{equation}
and 
\begin{equation}\label{scaldivergf2}
 \begin{split}
  Q_S=
&-P_w S_w^2-Q_{\cal S}^2 {\cal S}^2 - R_b S_b^2\\
&+ \left(Q_w +P_{\cal S}\right) S_w {\cal S} + \left(R_w +P_b\right) S_w S_b
  - \left(R_{\cal S} +Q_b\right) {\cal S} S_b\\
\end{split}
\end{equation}
The positivity of $Q_T$ is equivalent to the requirement that $\tau_1\leq0$. 
The positivity of $Q_V$ is equivalent to the condition  
$$ C_a\leq0,~~ E_b\leq0~~\text{and}~~ 4 E_b C_a \geq (C_b + E_a)^2$$
The implications of the requirement of the positivity of $Q_S$ are 
exactly as in \S \ref{weyl}. 
The Onsagar relations imply that
$$Q_w = P_{\cal S},~~R_w = P_b,~~R_{\cal S} = Q_b,~~C_b = E_a$$
Once again we have a 10 parameter set of dissipative equations of super 
fluid dynamics.

\section{Dissipative Superfluid dynamics 
in the Landau-Lifshitz-Clark-Putterman Frame}\label{llcp}

In the previous section we have presented a `theory' of first 
order dissipative super fluid dynamics in the transverse fluid frame. In 
this subsection we present the equivalent `theory' in the 
Landau-Lifshitz-Clark-Putterman modified phase frame. As the logic of this 
presentation mirrors that of the previous subsection in every way we 
will omit all derivations and present only our final results. After 
we have presented our results we will explicitly take the non relativistic
limit and compare with the analogous expressions in Clark, Putterman and Landau 
and Lifshitz \cite{Clark, Putterman, LLvol6}.

\subsection{Allowed forms for dissipative terms} \label{allowed}

As we have explained above, the Landau-Lifshitz-Clark-Putterman frame 
is a modified phase frame that specified by the relations 
\begin{equation}\label{frameputp}
 u^\mu u^\nu \tilde \pi_{\mu\nu} = 0,~~\tilde J^\mu_{diss} = 0
\end{equation}

We take the stress tensor to be given by 
\begin{equation}\label{strespara1}
\begin{split}
& \tilde \pi^{\mu\nu} =\left( Q^\mu u^\nu + Q^\nu u^\mu\right) + \Pi P^{\mu\nu} + \Pi _{t}^{\mu\nu}\\
&\text{where}\\
&u^\mu Q_\mu = 0,~~u_\nu \Pi_t^{\mu\nu} = 0,~~\left(\Pi_t\right)^\mu_\mu = 0,
~~\tilde \pi^\mu_\mu = 3\Pi
\end{split}
\end{equation}

The divergence of the entropy current in this frame is easily obtained from 
\eqref{divergentfpa} and takes the form 
\begin{equation} \label{enddiv}
 \partial_\mu J^\mu _S =- \frac{\left[  (\partial.u) \Pi+ \Pi_t^{\mu\nu}\sigma_{\mu\nu}\right]}{T}
- \frac{Q^\nu\left[\partial_\nu T + T(u.\partial) u_\nu\right]}{T^2}
 + \mu_{diss}P^{\mu\nu}\left[\partial_\mu \left(\frac{f~w_\nu}{T} \right)\right]
\end{equation}

The first derivative expressions of fluid dynamical fields that appear 
explicitly in \eqref{enddiv} may be written in terms of the following 
scalars
$$\Sigma_1 =n^\nu\frac{\left[\partial_\nu T + T(u.\partial) u_\nu\right]}{T} ,
~~\Sigma_2 = n^\mu \sigma_{\mu\nu} n^\nu,~~\Sigma_3 = \partial_\mu u^\mu,~~
\Sigma_4 =\frac{P^{\mu \nu} \left[\partial_\mu \left(\frac{f~w_\nu}{T} \right)\right]}{T^2}$$
the following vectors 
$${\cal W}_1^\mu =\tilde P^{\mu\nu}\frac{\left[\partial_\nu T + T(u.\partial) u_\nu\right]}{T},~~
{\cal W}_2^\mu =n^\beta \sigma_{\beta\alpha} \Tilde P^{\alpha\mu} $$
and the following tensors
$${\cal T}^{\mu\nu} =\tilde P^{\mu\beta }\sigma_{\beta\alpha} \Tilde P^{\alpha\mu} $$
where
$$n^\mu = \frac{w^\mu}{w},~~\tilde P^{\mu\nu} = P^{\mu\nu }- n^\mu n^\nu$$

In terms of these expressions, positivity of the entropy current requires 
\begin{equation}\label{abarpara}
\begin{split}
\mu_{diss} &=-\left[\mu_1 \Sigma_1+\mu_2 \Sigma_2+\mu_3 \Sigma_3+\mu_4 \Sigma_4\right]
\\
\\
\Pi &=T^3\left[\pi_1 \Sigma_1+\pi_2 \Sigma_2+\pi_3 \Sigma_3+\pi_4 \Sigma_4\right]\\
\\
 Q^\mu &=T^3\left[Q^{(s)}_1 \Sigma_1+Q^{(s)}_2 \Sigma_2+Q^{(s)}_3 \Sigma_3
+Q^{(s)}_4 \Sigma_4\right]n^\mu \\
&+ T^3Q^{(v)}_1{\cal W}_1^\mu+T^3 Q^{(v)}_2{\cal W}_2^\mu\\
\\
\Pi_t^{\mu\nu} &=T^3\bigg[\left(P_1 \Sigma_1+P_2 \Sigma_2+P_3 \Sigma_3+P_4 \Sigma_4 
 \right)
\left(n_{\mu} n_{\nu} - \frac{1}{3}P^{\mu\nu}\right)\\
 &~~~~~~~~+E_1 \left({\cal W}_1^\mu n^\nu + {\cal W}_1^\nu n^\mu\right) 
+E_2 \left({\cal W}_2^\mu n^\nu + {\cal W}_2^\nu n^\mu\right)
 \\
&~~~~~~~~+\tau {\cal T}^{\mu\nu} \bigg]\\
\end{split}
\end{equation}

With this form for the dissipative parameters, the divergence of 
the entropy current is given by 
$$\partial_\mu J^\mu_s =T^2\left( Q_S + Q_v + Q_T\right)$$
where
\begin{equation}\label{nannarar}
 \begin{split}
  Q_S =& - Q^{(s)}_1 \Sigma_1^2 -P_2 \Sigma_2^2 -\pi_3 \Sigma_3^2 -\mu_4\Sigma_4^2\\
&-(Q^{(s)}_4 + \mu_1)\Sigma_1 \Sigma_4 - (P_4 + \mu_2)\Sigma_2 \Sigma_4
-(\pi_4 + \mu_3)\Sigma_3 \Sigma_4\\
&-(Q^{(s)}_2 + P_1)\Sigma_1 \Sigma_2 - (Q^{(s)}_3 + \pi_1)\Sigma_1 \Sigma_3
-(\pi_2 +P_3)\Sigma_3 \Sigma_2\\
\\
 Q_V=&- Q^{(v)}_1  {\cal W}_1^2 - \left(Q^{(v)}_2 +  E_1\right) {\cal W}_1{\cal W}_2
- E_2 {\cal W}_2^2\\
=& -Q^{(v)}_1 \left[{\cal W}_1 
+ \left(\frac{Q^{(v)}_2 +  E_1}{2 Q^{(v)}_1 }\right) {\cal W}_2\right]^2
 - \left[E_2 - \frac{(Q^{(v)}_2 +  E_1)^2}{4  Q^{(v)}_1}\right] {\cal W}_2^2\
\\
Q_T =& - \tau {\cal T}_{\mu\nu}{\cal T}^{\mu\nu}
 \end{split}
\end{equation}
 The positivity of entropy current requires that the each of the three quadratic forms $Q_S$, $Q_v$ and 
$Q_T$ are positive definite.

The Onsager relations take the form
\begin{equation}\label{onsagarp}
 \begin{split}
  &Q^{(s)}_4 = \mu_1,~~P_4 = \mu_2,~~\pi_4 = \mu_3\\
&Q^{(s)}_2 = P_1,~~Q^{(s)}_3 =\pi_1,~~P_3 = \pi_2\\
&Q^{(v)}_2 =  E_1
 \end{split}
\end{equation}
Once again we have 14 indpendent dissipative coefficients, constrained 
by inequalities that follow from the requirement of positivity of the 
entropy current.

\subsection{The Limit of zero superfluid velocity}\label{zero}

In this subsubsection we will examine how the description of the previous 
subsection must simplify in the limit $w \rightarrow 0$. In this limit 
the superfluid and normal velocity are coincident. In this limit the 
vector $n^\mu$ no longer has any significance. The equations for 
the dissipative parts of the stress tensor, current and $\mu_{diss}$ 
must all possess the full $SO(3)$ invariance of rotations that leave $u^\mu$ 
fixed. The expressions for all dissipative quantities must be analytic 
functions of the projected superfluid velocity vector field 
$w^\mu=P^\mu_\nu \xi^\nu =r_c \zeta n^\mu$. This requirement imposes several 
constraints on the $\zeta \to 0$ behaviour of the coefficients in 
dissipative terms, including 
\begin{equation}\label{rellim}
 \begin{split}
&  \lim_{w\rightarrow0}Q^{(s)}_2 =
\lim_{w\rightarrow0}Q^{(s)}_3 =\lim_{w\rightarrow0}Q^{(s)}_4
 =\lim_{w\rightarrow0}Q^{(v)}_2 =0\\
&  \lim_{w\rightarrow0}P_1 =
\lim_{w\rightarrow0}P_3 =\lim_{w\rightarrow0}P_4
 =\lim_{w\rightarrow0}E_2 =0\\
&  \lim_{w\rightarrow0}\mu_1 =
\lim_{w\rightarrow0}\mu_2 =\lim_{w\rightarrow0}\pi_1
 =\lim_{w\rightarrow0}\pi_2 =0\\
&\lim_{w\rightarrow0}P_2 =
\lim_{w\rightarrow0}E_1 =\lim_{w\rightarrow0}\tau
 =\eta\\
&\lim_{w\rightarrow0}Q^{(s)}_1 =
\lim_{w\rightarrow0}Q^{(v)}_1 =\kappa
 \end{split}
\end{equation} 
\footnote{To see where these conditions come from, consider, for example, 
a constitutive relationship for a quantity $v_1$ which we know is a full 
$SO(3)$ vector when ${\vec \zeta}$ vanishes, but is simply the sum of 
an $SO(2)$ vector and an $SO(2)$ scalar at nonzero ${\vec \zeta}$. 
At arbitrary values of ${\vec \zeta}$, the scalar and vector parts of $v_1$ 
can be expanded as arbitrary linear combinations of all independent 
SO(2) scalars and vectors. However as ${\vec \zeta} \to 0$, the scalar 
component can be expanded in only those scalars that arise from the 
decomposition of an $SO(3)$ vector. As all $SO(2)$ scalars are not of this
form, this implies the vanishing of a number of coefficients in the 
expansion. Moreover the coefficient of these 
scalars has to equal the coefficient of the corresponding SO(2) vectors 
in the vector part of $v_1$. This requires that two otherwise independent 
coefficients are equal.  Similar remarks hold for the expansion 
of vector fields.}

It follows that, in this limit the 21 independent dissipative coefficients 
of the previous subsubsection (before imposing Onsager relations) reduce to 
only 6 non-zero coefficients $$\eta,~\kappa,~\mu_3,~\mu_4,~\pi_3,~\pi_4.$$
The relations in \eqref{abarpara} simplify to  
\begin{equation}\label{zerosima}
\begin{split}
\mu_{diss} &=\mu_3 \Sigma_3+\mu_4 \Sigma_4
\\
\\
\Pi &=T^3\left[\pi_3 \Sigma_3+\pi_4 \Sigma_4\right]\\
\\
 Q^\mu &=T^3\kappa ~P^{\mu\nu}\left(\frac{\partial_\nu T}{T} + (u.\partial)u_\nu\right)\\
\\
\Pi_t^{\mu\nu} &=T^3\eta~\sigma^{\mu\nu}\\
\end{split}
\end{equation}
The Onsagar relation set $\mu_3 = \pi_4$, finally leaving us with 
5 dissipative coefficients, as predicted by Landau and Lifshitz \cite{LLvol6}.

\subsection{The Non Relativistic Limit}

In order to make contact with the results of Clark and Putterman \cite{Clark, Putterman}(and 
with those of Landau and Lifshitz \cite{LLvol6} in the limit of zero superfluid 
velocity) we will now explicitly take the non relativistic limit of the
formulas presented in \S \ref{allowed}. 

The non relativistic limit is taken as follows. We set 
$$u_\mu = \{1,\vec v \},~~w_\mu = \{0,\vec w\},~~
n^\mu = \{0,\vec n\}= \left\{0,\frac{\vec w}{w}\right\}$$
(we ignore all terms of quadratic or highe order in ${\vec v}$ and 
${\vec w}$) 
and 
$$P_{\mu\nu} = \delta_{ij},~~\tilde P_{\mu\nu} = \tilde P_{ij}= \delta_{ij} - n_i n_j$$
Note in particular that all vectors and tensor projected orthogonal to 
$u^\mu$, in this limit, are purely spatial. 

The various scalar, vector and the tensor terms listed in \S \ref{allowed} 
reduce to the following expressions in the non relativistic limit:

\begin{equation}\label{nonrel1}
 \begin{split}
 & \Sigma_1 = \frac{(\vec n.\vec\partial)T}{T},~~\Sigma_2 = n_i n_j \sigma_{ij}
,~~\Sigma_3 = \vec \partial.\vec v,~~
\Sigma_4 = \frac{1}{T^2}\vec\partial \left(\frac{f~\vec w}{T}\right)\\
\\
&\left[{\cal W}_1\right]_i = \frac{\partial_i T}{T} - n_i \Sigma_1,
~~\left[{\cal W}_2\right]_i = n_j\sigma_{ji} - n_i \Sigma_2\\
\\
&{\cal T}_{ij} =\left(\delta_{ik} - n_i n_k\right)\sigma_{km}\left(\delta_{mj} - n_m n_j\right)
 \end{split}
\end{equation}
(note that all vectors and tensors are purely spatial, and we have only 
listed their spatial components). It follows that 
\begin{equation} \label{pook} \begin{split}
\tilde \pi^{00}&=0\\
\tilde \pi^{0i}&=Q^i\\
\tilde \pi^{ij}&= \Pi_t^{ij} + \Pi \delta^{ij}\\
\end{split}
\end{equation}
where all the quantities in the equation above are given by plugging 
\eqref{nonrel1} into \eqref{abarpara}. The resultant equations are 
very close to equation AVI-11 - AVI-14 in Putterman's book \cite{Putterman}. The main difference between 
our equations and those of Putterman is that our equations have 21 
free parameters, in contrast with Putterman's 20 free parameters. 
The equations of this subsection reduce to (A VI -11 to A VI-14) 
in \cite{Putterman} if we set  $Q^{(s)}_2 = 0$.


The Onsager relations of \S \ref{allowed} carry 
over unchanged to the non relativisitc limit. These relations are 
identical to the Onasager relations imposed by Putterman \cite{Putterman} and reduce the 
number of dissipative coefficients to 14 in our formulation, but to 13 
according to Putterman \cite{Putterman}. 

The non relativistic version of the ${\vec w} \to 0$ 
constitutive relations \eqref{zerosima} is simply given by by plugging
\eqref{nonrel1} into \eqref{zerosima}; the resulting 6 parameter set of 
constitutive (cut down to a 5 parameter set by the Onsager relations) 
agree with those presented by Landau and Lifshitz \cite{LLvol6}.

\section{Equilibrium Superfluid Thermodynamics from Gravity}\label{eq}

In this section we explain that the equilibrium superfluid thermodynamics, 
reviewed in \S \ref{fleq}, may be derived from gravity very simply. 
This section, which is abstract in nature, is a generalization and 
reworking of the beautiful paper of Sonner and Withers \cite{Sonner:2010yx} 
on the same subject. 

Consider a general gravitational system governed by the action
\begin{equation}\label{sysg}
\begin{split}
 {\cal S} &= \frac{1}{16 \pi G} \int \sqrt{g}\bigg( R +12 - \frac{1}{e^2}\big( V_1(\phi \phi^*) F_{ab}F^{ab} 
+ V_2(\phi\phi^*) D_a\phi \bar{D}^a \phi^* + V_3(\phi\phi^*) \big)\bigg)
\end{split}
\end{equation}
where $D_a = \nabla - i A_a$ and $\bar D_a = \nabla + i A_a$ and all the potentials 
$V_1,~V_2,~V_3$ are real. We assume that this system admits 
a homogeneous stationary asymptotically $AdS$ family of solutions - dual 
to homogeneous stationary superfluid flows - that take the form  
\begin{equation}\label{metanza}
\begin{split}
\text{Metric}:~ ds^2 &= -2 g(\frac{r}{r_c}) u_{\mu} dx^{\mu} dr 
- r_c^2 f(\frac{r}{r_c}) u_{\mu} u_{\nu} dx^{\mu} dx^{\nu}
+r_c^2 ~k(\frac{r}{r_c}) n_{\mu} n_{\nu} dx^{\mu} dx^{\nu}\\
&+r_c^2 ~j(\frac{r}{r_c}) \left( n_{\mu} u_{\nu} + u_{\mu} n_{\nu}\right) dx^{\mu} dx^{\nu}
+ r^2 {\tilde P}_{\mu \nu} dx^{\mu} dx^{\nu},\\
\text{Gauge field}:~r_c A & = H(\frac{r}{r_c})~u^\mu \partial_\mu + L(\frac{r}{r_c})~ n^\mu \partial_\mu \\
\text{Bulk scalar field} &= \phi\left(\frac{r}{r_c}\right)
\end{split}
\end{equation}
where
\begin{equation}
{\tilde P}_{\mu\nu}=\eta_{\mu\nu} +u_\mu u_\nu - n_\mu n_\nu.
\end{equation}
Here $u_{\mu}$ and $n_{\mu}$ are two arbitrary constant vectors obeying
\begin{equation}\label{orthog}
 u^{\mu} n_{\mu} = 0;~~  u^{\mu} u_{\mu} = -1;~~ n^{\mu} n_{\mu} = 1.
\end{equation}

We work in a gauge such that the scalar field is real $\phi^* = \phi$
(this implies $A^r = 0$), so that the boundary value of the gauge field 
gives the superfluid velocity. 
We choose the constant vector $u_\mu$ so as to ensure that the 
killing vector coincides with the generators of the event horizon of our 
solution. $n^\mu$ is then uniquely determined by \eqref{orthog} together 
with the requirement that $A_\mu$ at infinity (i.e. $\xi_\mu$) can be written
as a linear combination of $u^\mu$ and $n^\mu$.

We now choose coordinates so that the killing vector $\partial_v$ points 
along the direction of the $u^{\mu}$ and the vector $\partial_x$ 
points in the direction of $n^{\mu}$. 
Our solution retains rotational invariance 
in the remaining two spatial directions. We also work in the 
rescaled variables $\frac{r}{r_c}$ and $r_c x^\mu$ in terms of which  
\eqref{metanza} reduces to
\begin{equation}\label{metgaganza}
 \begin{split}
    &ds^2 = 2g(r)~dv~ dr  - f(r)dv^2 -2j(r)~dv~dx +k(r)~ dx^2
+r^2 \left(\sum dy_i^2\right)\\
&A^r = 0,~~A^v = H(r),~~A^x = L(r),~~A^y = 0,~~A^z = 0\\
&\text{Bulk scalar field} = \phi(r)
 \end{split}
\end{equation}

The solutions above are translationally invariant in the field theory 
spacetime directions. They are characterized by an onshell action density 
(action \eqref{sysg} per unit volume), a stress tensor and a charge current. 
Merely from symmetry, and with an appropriate definition of $u^\mu$, the 
stress tensor and the charge current of our system necessarily 
take the form \eqref{ltconstperf}. In Appendix \ref{SgravThermo} below we  demonstrate that the functions that appear in \eqref{ltconstperf} obey all the 
thermodynamical relations of the Landau Tiza two fluid model described above.

While the details of our demonstration are presented in Appendix 
\ref{SgravThermo}, we very briefly review the main ideas here. 

Our system has three inequivalent translationally invariant killing 
vectors: $u^\mu \partial_\mu=\partial_v$, $u_s^\mu\partial_\mu$ and 
$k^\mu \partial_\mu$ where $k^\mu$ is transverse to the normal as well as 
the superfluid velocity. As is standard in derivations of black hole 
thermodynamics, we demonstrate using the equations of motion that 
$R^{\mu\nu} \theta_\nu$ (where $\theta^\mu$ is any of the vectors above) 
takes a particularly simple form. The resultant identities allow us 
to demonstrate the onshell vanishing of three total derivatives built out of 
the functions in the metric and gauge field \eqref{metanza}.
Integrating these expressions allows us to prove the 
Smarr-Gibbs-Duhem relations listed in the first two of equations \eqref{fconst}.
We are also able to show that the killing vector that reduces to the 
null generator of the horizon of our solution also defines the 
normal velocity of our fluid. Recall the later is  defined to ensure the 
absence of $u^\mu u_s^\nu$ cross terms in the stress tensor.

Next we follow \cite{Sonner:2010yx} to rewrite the action of an equilibrium 
configuration of the sort studied in this section as the integral 
of a total derivative. Performing the integral we find that the action 
evaluates to the difference of two terms, one at infinity and the second at the 
horizon. The term at the horizon vanishes while the term at infinity 
evaluates to the negative of the pressure of the solution. 

Finally, the action density (hence the pressure) of this system may be 
regarded as a function of its temperature $T$, chemical potential $\mu$ and 
superfluid chemical potential $\mu_s$. It is well known in classical mechanics
that the onshell action as a function of initial and final coordinates, 
$q^i_0$ and $q^i_f$ obeys the relation $dS= p^i_0 dq^i_0-p^i_f dq^i_f$. 
Sonner and Withers \cite{Sonner:2010yx} have demonstrated that the application 
of  this result to our system (with $r$ being regarded as the time) 
gives 
\begin{equation}
dP= q_n d \mu + q_s d\mu_s + s dT
\end{equation}
where $s$ is the entropy density of the solution given by the Hawking-
Beckenstein formula. 

\section{An analytically tractable limit of hairy black branes}\label{static}

As we have explained in the previous section, general gravitational 
methods are powerful enough to tightly constrain the gravitational 
solutions dual to fluid dynamics, in particular to demonstrate that 
the thermodynamics of these solutions is given by the Landau-Tiza two 
fluid model. In order to explicitly determine the thermodynamics of 
any particular gravitational system, however, we need to explicitly 
determine the solutions dual to uniform superfluid flows. Unfortunately, 
the ordinary differential equations that arise in this attempt have 
proved so complicated that it has not proved possible to analytically 
solve for hairy black branes (the gravitational duals to superfluids) 
in any reasonable gravitational system \footnote{Of course much attention has 
been focused on the numerical solutions of the relevant equations in 
several models.}. The only analytic results that we are aware of, for 
hairy black brane solutions, are those of Herzog \cite{Herzog:2010vz}. Herzog considered 
a very special model, the model of a charged scalar field of $m^2=-4$ and 
infinite charge $e$ (i.e. a model in the so called probe limit). He 
demonstrated that this model displays a second order phase transition towards 
superfluidity whenever $|\frac{\mu}{T}| \geq 2$. 

When $|\frac{\mu}{T}|$ is just larger that $2$, the stable gravitational 
solutions develop a scalar vev. Let $\epsilon$ denote the value of this vev. 
Herzog was able to generate the relevant gravitational solutions 
perturbatively in $\epsilon$ and also perturbatively in the difference 
between superfluid and normal velocities. 

In this paper we will be interested in probing the structure of viscous 
superfluid dynamics from gravity. In the infinite charge or probe limit of 
\cite{Herzog:2010vz} 
scalar and gauge dynamics do not back react on spacetime. In order 
to probe the dynamics of the interaction between the stress tensor and the 
charge current we need to go beyond the infinite charge probe limit. In this
section we generalize Herzog's perturbative construction of gravitational 
solutions to go beyond the probe limit. In other words we generalize 
Herzog's infinite $e$ solutions to retain the first nontrivial 
correction in a $(\frac{1}{e})$ expansion. 

In the next section we will use the results of this section as an input
into the fluid gravity map, in order to generate gravitational solutions
dual to viscous superfluid flows. 

\subsection{The bulk system and the equations of motion}

Following Herzog \cite{Herzog:2010vz} we consider the system
\begin{equation}\label{sys}
 \mathcal{L}= \frac{1}{16 \pi G} \int d^5x \sqrt{-g}\left(\mathcal{R} + 12 + \frac{1}{e^2} 
              \left(- \frac{1}{4} F_{\mu \nu} F^{\mu \nu}
              -\frac{1}{2} |D_{\mu} \phi |^2+ 2 |\phi|^2. \right) \right),
\end{equation}
Where $D_{\mu}  = \nabla_{\mu}  - i A_{\mu}  $, and $\nabla_{\mu}$ is the       
gravitational covariant derivative. Note that this system is a special case of
the system \eqref{sysg} (where we have chosen particular simple 
forms of the functions $V_1$,$V_2$ and $V_3$).

The equation of motion for the scalar field and the gauge field that follows from \eqref{sys} 
are respectively
\begin{equation}\label{scaleq}
 D_{\mu}D^{\mu} \phi + 4 \phi = 0,
\end{equation}
and
\begin{equation}\label{maxeq}
 D_{\mu} F^{\mu \nu} =\frac{1}{2} J^{\nu},
\end{equation}
where the current $J_{\mu}= i \left( \phi^{*} D{\mu} \phi - \phi (D_{\mu} \phi)^{*}\right)$.
The Einstein Equation that follows from \eqref{sys} is
\begin{equation}\label{eineq}
 G_{\mu \nu} - 6 g_{\mu \nu} =\frac{1}{e^2}\big( (T_{\text{max}})_{\mu \nu} + (T_{\text{mat}})_{\mu \nu} \big),
\end{equation}
where 
\begin{equation}
 \begin{split}\label{tmattmax}
  (T_{\text{max}})_{\mu \nu}& = -\frac{1}{2} \left( F_{\mu \beta} F^{\beta}_{~~\nu} 
              -\frac{1}{4} g_{\mu \nu} F_{\sigma \beta}F^{\beta \sigma} \right),\\
  (T_{\text{mat}})_{\mu \nu} &= \frac{1}{4} \left(D_{\mu}\phi D_{\nu} \phi^* +D_{\nu}\phi D_{\mu} \phi^* \right)
                       -\frac{1}{4} g_{\mu \nu}\left( |D_{\beta} \phi |^2 -4 |\phi|^2 \right).
 \end{split}
\end{equation}

\subsection{Boundary Conditions and Solutions}

We search for solutions of the form \eqref{metgaganza}.
The 4-vectors $n^\mu$,  defined in the previous section, may be 
computed as follows. Let 
$$r_c \zeta_{\mu}= \left(\eta_{\mu \nu} + u_{\mu} u_{\nu}  \right) \xi^{\mu}.$$ 
It follows that $n_\mu$ is given by  
$$n_{\mu} = \zeta_{\mu}/ |\zeta|.$$ 

As explained in the previous section we choose gauge so that our 
scalar field is real 
$\phi(r)= \phi^*(r)$ by a choice of gauge. This gauge choice, plus the 
static nature of our solution, forces $A^r = 0$ (as can be seen by comparing
the equation of motion of $\phi$ with that of $\phi^*$).

We search for solutions that obey the following large $r$ boundary conditions
\begin{equation} \label{bcinft}
\begin{split}
k(r) & = r^2 + \frac{k_2}{r^2}\\
f(r)&=r^2 + \frac{f_2}{r^2} +{\cal O}\left( \frac{1}{r^3}\right) \\
j(r) & = \frac{j_2}{r^2}+ \dots\\
 L(r)& = \frac{\zeta}{r^2}+ \dots\\
\phi(r)& = \frac{\epsilon}{r^2}+ \dots\\
\end{split}
\end{equation}
It turns out that the conditions above, together with the equations of motion, 
automatically ensure
$$ \lim_{r\rightarrow \infty} g(r) = 1$$
so that this condition, while true, does not have to be additionally imposed.
Also, it turns out that the coefficient of $1/r^2$ term in the asymptotic 
expansion of $H(r)$ is fixed by equations of motion and the requirement that 
$\phi$ be regular at the horizon.

Our functions are also constrained at $r=1$ as follows 
\begin{equation} \label{bchor}
j(1)=f(1)=0
\end{equation}
On the other hand the functions $H(r)$, $k(r)$, $L(r)$ and $\phi(r)$ are 
required only to be regular at $r=1$.

It is possible to argue that there exists an 8 parameter class 
of solutions of the form \eqref{metanza}, to the system \eqref{sys}, 
subject to the boundary conditions listed above. One of these parameters is 
$r_c$ in \eqref{metanza}. The three normal velocity parameters can be set to 
zero by a boost, and rotations can be used to point the superfluid velocity 
in the $x$ direction, as in the previous section.
 This leaves us with a two parameter set of solutions, 
parameterized by $\epsilon$ and $\zeta$. 

\subsection{Perturbative Solutions}

In this subsection we will generalize the work out in 
\cite{Herzog:2010vz} to the hairy black branes of our system, as a function 
of $\epsilon$ and $\zeta$ at small values of those parameters. 
Our starting point is 
Herzog's observation that, at $e=\infty$,  the linearized equations of motion 
about the Reissner Nordstrom black brane at $|\frac{\mu}{T}|=2$ admit a regular
static solution scalar solution proportional to $\frac{\epsilon}{1+r^2}$. 
As was explained in \cite{Herzog:2010vz} this solution can be taken to be the 
starting point for a perturbative expansion of hairy black brane solutions 
in a power series in $\epsilon$. The solutions of \cite{Herzog:2010vz} were 
further generalized to nonzero $\zeta$.

In this subsection we generalize Herzog's solutions away from the infinite 
charge limit, to first order in a power series expansion
in ${\cal O}(\frac{1}{e^2})$, i.e to first order in deviations away from 
the probe approximation. This generalization will prove crucial for generating
the equations of superfluid dynamics including effects of back reaction 
of the superfluid on the normal fluid.

The techniques for obtaining this perturbative expansion are standard. 
We do not pause to explain our computations in detail; in 
the rest of this section we simply present the results of our calculations.
As a function of $\epsilon$ and $\xi$ (with both taken to be small) we find 
that the scalar field is given by 
\begin{equation}\label{adish}
\begin{split}
 \phi(r)=&\bigg\{  \epsilon\left[\frac{1}{r^2+1}+\frac{\zeta ^2 \left(2 \log (r)-\log
   \left(r^2+1\right)\right)}{4 r^2+4}+{\cal O}\left(\zeta ^4\right)\right]\\
&+\epsilon^3\left[\frac{-2 \left(r^2+1\right) \log (r)+\left(r^2+1\right) \log
   \left(r^2+1\right)-2}{48 \left(r^2+1\right)^2}+{\cal O}\left(\zeta
   ^2\right)\right]  + {\cal O}(\epsilon^5)\bigg\} \\
&+ {\cal O}\left(\frac{1}{e^2}\right)
\end{split}
\end{equation}

The functions in the gauge field in \eqref{metanza} are given by
\begin{equation}
 \begin{split}
  H(r)&= 
\left( H_0(r) + H_1(r) \epsilon^2 + 
H_2(r) \epsilon^4 + \mathcal{O}(\epsilon^6) \right) +{\cal O}(1/e^2),\\
  L(r)&= 
\left(L_0(r) + L_1(r) \epsilon^2 + L_2(r) \epsilon^4 + 
\mathcal{O}(\epsilon^6) \right) +{\cal O}(1/e^2)\\
 \end{split}
\end{equation}
where
\begin{equation}
 \begin{split}
 H_0(r) &= \frac{2}{r^2+1}+\frac{\zeta ^2}{2
   \left(r^2+1\right)}-\frac{\zeta ^4 (1-\log (2))}{4
   \left(r^2+1\right)}+O\left(\zeta ^6\right), \\
 H_1(r) &=  \bigg(\frac{r^2-5}{48 \left(r^2+1\right)^2} + \frac{\zeta ^2}{288 \left(r^2-1\right)
   \left(r^2+1\right)^2}
   \bigg(10 r^4+72 r^4 \log (r)-27 r^4 \log (2)+18 r^2+18 r^2
   \log (2)\\ &-36 r^4 \log \left(r^2+1\right)-28+45 \log
   (2)\bigg) \bigg)+O\left(\zeta ^4\right),\\
 H_2(r) &= -\frac{1}{55296
   \left(\left(r^2-1\right)
   \left(r^2+1\right)^3\right)}\bigg(253 r^6+589 r^4-589 r^2+48 \left(r^2+1\right)^2 \log
   (64)\\&-336 \left(r^2-1\right) \left(r^2+1\right)^2 \log (2)+576
   \left(r^6+r^4\right) \log
   \left(\frac{r^2}{r^2+1}\right)-253 \bigg)+O\left(\zeta ^2\right),\\
 \end{split}
\end{equation}
and
\begin{equation}
 \begin{split}
 L_0(r) &= \frac{\zeta}{r^2},\\
 L_1(r) &= -\frac{\zeta}{8 r^2 (1+r^2)} + \mathcal{O}(\zeta^3),\\
 L_2(r) &= \mathcal{O}(\zeta).\\
 \end{split}
\end{equation}

The functions in the metric in \eqref{metanza} are given by 
\begin{equation}
 \begin{split}
  f(r)&=\left(r^2 - \frac{1}{r^2}\right)
 +\frac{1}{e^2} \left( f_0(r) + f_1(r) \epsilon^2 + f_2(r) \epsilon^4 + \mathcal{O}(\epsilon^6)\right)
 +\mathcal{O}\left(\frac{1}{e^4}\right),\\
 g(r)&=1
 +\frac{1}{e^2} \left( g_0(r) + g_1(r) \epsilon^2 + g_2(r) \epsilon^4 + \mathcal{O}(\epsilon^6)\right)
 +\mathcal{O}\left(\frac{1}{e^4}\right),\\
 j(r)&=0
 +\frac{1}{e^2} \left( j_0(r) + j_1(r) \epsilon^2  + \mathcal{O}(\epsilon^4)\right)
 +\mathcal{O}\left(\frac{1}{e^4}\right),\\
 k(r)&=r^2
 +\frac{1}{e^2} \left( k_0(r) + k_1(r) \epsilon^2 + k_2(r) \epsilon^4 + \mathcal{O}(\epsilon^6)\right)
 +\mathcal{O}\left(\frac{1}{e^4}\right).\\
 \end{split}
\end{equation}
where
\begin{equation}
 \begin{split}
  f_0(r) & =-\frac{4 \left(r^2-1\right)}{3 r^4}-\frac{2 \left(r^2-1\right)
   \zeta ^2}{3 r^4}+\frac{\zeta ^4 \left(3 r^2+r^2 (-\log
   (16))-3+\log (16)\right)}{12 r^4}+O\left(\zeta ^6\right),\\
  f_1(r) & = \frac{-7 r^4+12 r^2-5}{36 r^4 \left(r^2+1\right)} \\ &+\frac{\zeta ^2}{432
   r^2 \left(r^2+1\right)}
   \bigg(\frac{54 r^6+r^4 (54 \log (2)-23)-36 r^2 (2+\log
   (2))+41-90 \log (2)}{r^2} \\ & +18 \left(3 r^6+3 r^4-9 r^2-1\right)
   \left(2 \log (r)-\log \left(r^2+1\right)\right)\bigg)+O\left(\zeta ^4\right),\\
  f_2(r) & =\frac{1}{48 r^2}\bigg(-\frac{2 \left(r^6+r^4-2 r^2\right) \left(2 \log (r)-\log
   \left(r^2+1\right)\right)}{3 \left(r^2+1\right)} \\ &-\frac{1}{864
   r^2 \left(r^2+1\right)^3}\bigg(576
   r^{10}+989 r^8+624 r^8 \log (2)-1538 r^6+1248 r^6 \log
   (2)-1044 r^4 \\ &+914 r^2-1248 r^2 \log (2)+103-624 \log (2) \bigg)\bigg),\\
 \end{split}
\end{equation}

\begin{equation}
 \begin{split}
  g_0(r) & = \mathcal{O}\left(\zeta ^6\right),\\
  g_1(r) & =-\frac{1}{6 \left(r^2+1\right)^2}\\ &+\frac{\zeta ^2 \left(54
   \left(r^2+1\right)^3-6 \left(r^2+1\right)^2 \left(-9 r^4-18
   r^2+3\right) \left(2 \log (r)-\log
   \left(r^2+1\right)\right)\right)}{864
   \left(r^2+1\right)^4}+O\left(\zeta ^4\right),\\
  g_2(r) & =\frac{-6 r^6-21 r^4-14 r^2-6 \left(r^2+1\right)^2 \left(r^4+2
   r^2\right) \left(2 \log (r)-\log
   \left(r^2+1\right)\right)+4}{864 \left(r^2+1\right)^4}+O\left(\zeta ^2\right),\\
 \end{split}
\end{equation}

\begin{equation}
 \begin{split}
  j_0(r) & = \mathcal{O}\left(\zeta ^6\right),\\
  j_1(r) & = \frac{\left(r^2-1\right) \zeta }{8 \left(r^2+1\right)} +\mathcal{O}\left(\zeta ^3\right),\\
 \end{split}
\end{equation}
and
\begin{equation}
 \begin{split}
  k_0(r) & = \mathcal{O}\left(\zeta ^6\right),\\
  k_1(r) & =\frac{r^2 \zeta ^2 \left(-2 \left(r^2+1\right) \log
   (r)+\left(r^2+1\right) \log \left(r^2+1\right)-1\right)}{8
   \left(r^2+1\right)}+O\left(\zeta ^4\right),\\
  k_2(r) & = \mathcal{O}\left(\zeta ^2\right),\\
 \end{split}
\end{equation}

Upon  setting $\frac{1}{e^2}=0$, our result exactly matches with the 
equations 2.30, 2.31, 2.32 in \cite{Herzog:2010vz},if we replace $u=1/r$ in those equations.

\subsection{The first correction to the phase transition curve}

As we have seen above, Herzog's model undergoes a superfluidity 
nucleation phase transition at $|\frac{\mu}{T}|=2$. It is easy to 
work out how this phase transition curve is corrected at 
${\cal O}(\frac{1}{e^2})$. The phase transition curve is determined by finding
the black brane solution that admits a static regular and 
normalizable solution to the linearized scalar field equations about 
a Reissner Nordstrom black brane. Now the black brane solution, at 
fixed temperature and chemical potential, depends on the parameter $e$. 
In the probe limit the metric of the charged black brane is simply 
independent of the chemical potential and equal to that of the 
uncharged black brane of the same temperature. This 
metric (and the accompanying gauge field) are slightly deformed at 
${\cal O}(\frac{1}{e^2})$. The condition for the existence of a normalizable
and regular zero mode to the scalar equation is also, consequently, deformed. 

Let the restriction of the gauge field at infinity be given by
$$\xi_\mu = r_c (-\nu u_\mu)$$
where $r_c$ is the location of the horizon. 
It turns out that a normalizable and regular zero mode to the 
linearized scalar equation occurs at 
$$\nu_c = -2 + \frac{1}{e^2}\left(4 - \frac{16 \log(2)}{3}\right) + {\cal O}\left(\frac{1}{e^2}\right)$$
The temperature of this solution (taking the back reaction of the gauge field 
into account) is given by 
$$T =\frac{r_c}{\pi}\left[ 1- \frac{2}{3 e^2} + {\cal O}\left(\frac{1}{e^4}\right)\right]$$
It follows that $\frac{\mu}{T}$ at the superfluidity phase 
transition is given by 
$$|\frac{\mu_c}{T}|= \frac{r_c \nu_c}{T}=\pi\left[2 +\frac{8}{3e^2}\left( 2\log(2)-1\right) 
+ {\cal O}\left(\frac{1}{e^2}\right)\right]
$$

\subsection{Boundary Thermodynamics}

Using the solution obtained in the previous section we evaluate the 
boundary stress tensor charge current. For this purpose we use 
the standard AdS/CFT formulas \cite{Balasubramanian:1999re,Henningson:1998gx}
\begin{equation}\label{fromst}
\begin{split}
\text{Boundary stress tensor} &= T^{\mu}_{\nu} =\frac{1}{16 \pi G} 
  \lim_{r \rightarrow \infty} r^4 \bigg(2 \left(\delta^\mu_\nu K_{\alpha \beta} \gamma^{\alpha \beta}-
 K^{\mu}_{\nu} \right) - 6 \delta^{\mu}_{\nu}+\frac{ \phi^* \phi}{e^2}\delta^\mu_\nu\bigg)\\
\text{Boundary charge current} &= j^{\mu} =\frac{1}{16 \pi G~e^2} \lim_{r \rightarrow \infty} r^3 F^{\mu r}\\
\text{Entropy density} &= s = \frac{\sqrt{k(1)}}{4 G},\\
 \text{Temperature} & =T = \frac{f'(1)}{4 \pi g(1)}.
\end{split}
\end{equation}
where $\gamma_{\alpha \beta}$ and $K_{\alpha\beta}$ are respectively the induced metric and
extrinsic curvature of a constant $r$ surface.

The result for the stress tensor and current can be parameterized as the form 
\begin{equation}\label{fromstcc}
\begin{split}
 T^{\mu\nu}& =\frac{1}{16 \pi G}\bigg[A u^\mu u^\nu + B n^\mu n^\nu  
+ C \left( n^\mu u^\nu +  u^\mu n^\nu \right)
+\left( \frac{A-B}{2}\right) {\tilde P}^{\mu\nu}\bigg]\\
 j^{\mu}& =\frac{1}{16 \pi G }\left[Q_1 u^\mu + Q_2 n^\mu\right] \\
\end{split}
\end{equation}

$A,~B$ and $C$  are given by the following expressions.
\begin{equation}\label{abc}
 \begin{split}
A &=3 r_c^4 +\frac{r_c^4}{e^2}\bigg\{\left[4+2 \zeta ^2+\zeta ^4 \left(\log
   (2)-\frac{3}{4}\right)+{\cal O}\left(\zeta ^6\right)\right]\\
& +\epsilon^2\left[\frac{7}{12}+\zeta ^2 \left(\frac{59}{144}-\frac{3 \log
   (2)}{8}\right)+{\cal O}\left(\zeta ^4\right)\right]\\
&+\epsilon^4\left[\frac{624 \log (2)-451}{13824}+{\cal O}
\left(\zeta ^2\right)\right] +{\cal O}(\epsilon^6) \bigg\}
+ {\cal O}\left(\frac{1}{e^4}\right)\\
\\
B &= r_c^4 +\frac{r_c^4}{e^2}\bigg\{\left[\frac{4}{3}+\frac{2 \zeta ^2}{3}+\zeta ^4 \left(\frac{\log
   (2)}{3}-\frac{1}{4}\right)+{\cal O}\left(\zeta ^6\right)\right]\\
& +\epsilon^2\left[\frac{7}{36}+\zeta ^2 \left(\frac{131}{432}-\frac{\log
   (2)}{8}\right)+{\cal O}\left(\zeta ^4\right)\right]\\
&+\epsilon^4\left[\frac{624 \log (2)-451}{41472}+{\cal O}\left(\zeta ^2\right)\right] + {\cal O}(\epsilon^6) \bigg\}
+ {\cal O}\left(\frac{1}{e^4}\right)\\
\\
C &= \frac{r_c^4}{e^2}\bigg[\epsilon^2\left[\frac{\zeta}{2} + {\cal O}\left(\zeta^3\right)\right]
+{\cal O}(\epsilon^4) \bigg]+ {\cal O}\left(\frac{1}{e^4}\right)
\end{split}
\end{equation}

While $Q_1$ and $Q_2$ are given by the following expressions

\begin{equation}\label{q1q2}
 \begin{split}
  Q_1 &=-\frac{r_c^3}{e^2} \bigg\{\bigg[ 4+\zeta ^2+\frac{1}{2} \zeta ^4 (\log (2)-1)+{\cal O}\left(\zeta^6\right)\bigg] 
+ \epsilon^2 \bigg[ \frac{7}{24}+\zeta ^2 \left(\frac{7}{36}-\frac{5 \log
   (2)}{16}\right)+{\cal O}\left(\zeta ^4\right)\bigg] 
\\ &+ \epsilon^4 \bigg[ \left(\frac{13 \log
   (2)}{576}-\frac{493}{27648}\right)+{\cal O}\left(\zeta ^2\right)\bigg]+ {\cal O}(\epsilon^6) \bigg\}
+ {\cal O}\left(\frac{1}{e^4}\right)\\
\\
  Q_2 &=  \frac{r_c^3}{e^2}\bigg\{\epsilon^2\left[-\frac{ \zeta}{4}+   {\cal O}(\zeta^3)\right]
+   {\cal O}(\epsilon^4)\bigg\}
+ {\cal O}\left( \frac{1}{e^4}\right)\\
 \end{split}
\end{equation}

Using the above expressions we can compute the coefficients in \eqref{ltconst}. We find
\begin{equation}\label{fexp2}
 \begin{split}
 16 \pi G\left( \rho_n\right)&= 3 r_c^4
+ \frac{r_c^4}{e^2}\bigg\{\left[4+2\zeta^2 + {\cal O}(\zeta^4) \right]
+\epsilon^2\left[-\frac{5}{12} + {\cal O}(\zeta^2)\right] + {\cal O}(\epsilon^4) \bigg\}
+ {\cal O}\left(\frac{1}{e^4}\right)\\
\\
16 \pi G (\rho_s) &=  \frac{r_c^4}{e^2}\bigg\{ \left[{\cal O}(\zeta^4) \right]
+\epsilon^2\left[1 + {\cal O}(\zeta^2)\right] + {\cal O}(\epsilon^4) \bigg\}
+ {\cal O}\left(\frac{1}{e^4}\right)\\
\\
16 \pi G (P) &= r_c^4+  \frac{r_c^4}{e^2}\bigg\{\left[ \frac{4}{3} + \frac{2}{3}\zeta^2
+{\cal O}(\zeta^4) \right]
+\epsilon^2\left[\frac{7}{36}+ {\cal O}(\zeta^2)\right] + {\cal O}(\epsilon^4) \bigg\}
+ {\cal O}\left(\frac{1}{e^4}\right)\\
\\
16 \pi G  (q_n) &= -\frac{r_c^3}{e^2}\bigg\{\left[4 + \zeta^2 + {\cal O}(\zeta^4) \right]
+ \epsilon^2\left[ - \frac{5}{24} + {\cal O}(\zeta^2) \right]+ {\cal O}(\epsilon^4)\bigg\}
+{\cal O}\left( \frac{1}{e^4}\right) \\
\\
16 \pi G (q_s) &=\frac{r_c^3}{e^2}\bigg\{\left[ {\cal O}(\zeta^4) \right]
+ \epsilon^2\left[ \frac{1}{2} + {\cal O}(\zeta^2) \right]+ {\cal O}(\epsilon^4)\bigg\}
+{\cal O}\left( \frac{1}{e^4}\right) \\
 \end{split}
\end{equation}

Further the chemical potential and $\mu_s$ of our solution is given by 
\begin{equation}\label{mudef}
 \begin{split}
 \mu =u^{\mu} \xi_{\mu}=&~
r_c \bigg\{ \bigg[-2-\frac{\zeta ^2}{2}+\zeta ^4 \left(\frac{1}{4}-\frac{\log
   (2)}{4}\right)+{\cal O}\left(\zeta ^6\right)\bigg)]\\
&+ \epsilon^2 \bigg[ -\frac{1}{48}+\zeta ^2 \left(\frac{3 \log
   (2)}{32}-\frac{5}{144}\right)+{\cal O}\left(\zeta ^4\right)\bigg]
\\ &+ \epsilon^4 \bigg[ \left(\frac{253}{55296}-\frac{7 \log
   (2)}{1152}\right)+O\left(\zeta ^2\right)\bigg)+ {\cal O}(\epsilon^6) \bigg]\bigg\}
+ {\cal O}\left(\frac{1}{e^2}\right)\\
\\
\mu_s = u^\mu_s \xi_\mu=&~r_c \bigg\{ \bigg[2+\frac{\zeta ^2}{4}
+\zeta ^4 \left(-\frac{13}{64}+\frac{\log
   (2)}{4}\right)+{\cal O}\left(\zeta ^6\right)\bigg)]\\
&+ \epsilon^2 \bigg[ +\frac{1}{48}+\zeta ^2 \left(-\frac{3 \log
   (2)}{32}+\frac{43}{1152}\right)+{\cal O}\left(\zeta ^4\right)\bigg]
\\ &+ \epsilon^4 \bigg[ \left(-\frac{253}{55296}+\frac{7 \log
   (2)}{1152}\right)+O\left(\zeta ^2\right)\bigg)+ {\cal O}(\epsilon^6) \bigg]\bigg\}
+ {\cal O}\left(\frac{1}{e^2}\right)\\
\end{split}
\end{equation}

Moreover we find 
\begin{equation}
 s = \frac{r_c^3}{4 G}\bigg[1+
\frac{1}{e^2}
\bigg\{  \epsilon^2\left[ \frac{\log (4)-1}{32}\zeta^2
+ {\cal O}(\zeta^4) \right]+{\cal O}(\epsilon^4) \bigg\} + {\cal O}\left( \frac{1}{e^4}\right)\bigg]
\end{equation}

\begin{equation}
\begin{split}
 T &=  \frac{r_c}{\pi} +\frac{r_c}{4\pi e^2} \bigg\{ \bigg[ -\frac{8}{3}-\frac{4 \zeta ^2}{3}+\zeta ^4
   \left(\frac{1}{2}-\frac{2 \log (2)}{3}\right)+{\cal O}\left(\zeta
   ^6\right)\bigg]
\\ &+  \epsilon^2 \bigg[ \frac{1}{9}+\zeta ^2 \left(\frac{\log
   (2)}{4}-\frac{23}{216}\right)+{\cal O}\left(\zeta ^4\right)\bigg]
\\ &+\epsilon^4  \bigg[  \left(\frac{91}{20736}-\frac{\log (2)}{108}\right) 
+ {\cal O}\left(\zeta^2\right)\bigg]
+{\cal O}(\epsilon^6) \bigg\} + {\cal O}\left( \frac{1}{e^4}\right) 
\end{split}
\end{equation}

Using these expressions and the quantities obtained in \eqref{fexp2} we 
have verified all the relations \eqref{fconst} to the order to which 
we have evaluated our solution. 

\subsection{Stability} \label{stability}

In Appendix \ref{stab} we have demonstrated that the solutions we have 
computed above are unstable whenever $\frac{\zeta}{\epsilon} \geq 
\frac{1}{4} \sqrt{\frac{7}{3}}$ (working at leading order in $\frac{1}{e^2}$
and first nontrivial order in $\zeta$ and $\epsilon$ with both 
quantities taken to be of the same order).

It follows that superfluid flows with $\zeta \gg \epsilon$ are unstable. 
For this reason in the rest of this paper we will specialize to the case 
that $\zeta \leq  {\cal O}(\epsilon)$. 

It is of some interest to characterize the superfluid phase on a phase 
diagram with axes labeled by $\frac{\mu}{T}$ and $\zeta$ (by conformal 
invariance the phase diagram can only depend on the ratio $\frac{\mu}{T}$). 
By restricting \eqref{mudef} to lowest order, and noting that $\epsilon 
\geq 0$, we conclude that the superfluid phase exists only when 
\begin{equation} \label{existance}
|\frac{\mu}{T}|-2 \geq  \frac{\zeta^2}{2}
\end{equation}
On the other hand this phase is unstable when 
\begin{equation} \label{existance}
|\frac{\mu}{T}|-2 \geq  \frac{9 \zeta^2}{14}
\end{equation}
As $\frac{9}{14}>\frac{1}{2}$, it follows that as we increase $\zeta$ at fixed 
$|\frac{\mu}{T}|$, the superfluid phase first goes unstable and then stops 
existing. It would be interesting to investigate whether this qualitative 
behaviour persists at finite values of $\zeta$.

\section{Superfluid dynamics to first order in the derivative expansion} \label{1ordderexp}

In the previous section we have determined the equilibrium solutions 
for hairy black branes, perturbatively in $\epsilon$ and the superfluid 
velocity, and separately in an expansion in $\frac{1}{e^2}$. In this section 
we use the results of the previous subsection as an input into the fluid 
gravity map. 

The basic idea here is a simple generalization of the ideas spelt out in 
\cite{Bhattacharyya:2008jc,Bhattacharyya:2008xc, Bhattacharyya:2008ji, 
Banerjee:2008th, Bhattacharyya:2008mz,Erdmenger:2008rm,Haack:2008cp,
VanRaamsdonk:2008fp}. 
We search for gravitational solutions 
that tube wise approximate the 8 parameter hairy black brane solutions 
described in the previous section,
with values of the temperature, the chemical potential, 
$\zeta^\mu$ and $u^\mu$ varying in space and time. The tubes in question 
run along null ingoing geodesics, and foliate our spacetime. Technically, 
this programme is implemented by working in ingoing Eddington Finklestein 
coordinates (as we have been through this paper) but promoting the parameters
of our solutions to fields that vary in spacetime. 

The fluid gravity map generates the gravitational solutions dual to fluid flows
perturbatively in a boundary derivative expansion. The zero order ansatz 
for such a solution is simply the solution \eqref{metanza} with $\epsilon$, $r_c$, 
$\zeta^\mu$ and $u^\mu$ promoted to arbitrary slowly varying functions of 
spacetime. This ansatz of course solves the equations of motion \eqref{scaleq}, \eqref{maxeq} and \eqref{eineq},
when all parameters are constant, but does not solve these equations 
when these parameter vary in spacetime. As in 
\cite{Bhattacharyya:2008jc, Banerjee:2008th, Bhattacharyya:2008mz,Erdmenger:2008rm} this  ansatz may 
be corrected to obtain a true solution ( systematically in a derivative 
expansion) provided the eight fluid fields that parameterize our ansatz 
obey certain constraint equations. These constraint equations are simply 
the fluid equations \eqref{eomsfd} with holographically generated constitutive
relations for the stress tensor, the charge current, and a holographically 
generated correction to the Josephson equation.  

In this section we implement this programme to first order in the derivative
expansion. 

\subsection{The method}\label{method}
As we have explained above, we will search for gravity solutions that tube 
wise approximate the equilibrium solutions of the previous section. In principle
our solutions could be labeled by a temperature and a chemical potential 
field in addition to the normal and superfluid velocity fields. However, 
for calculation purposes we will find it convenient to trade chemical 
potential for $\epsilon(x)$, the local expectation value of the operator $O$, 
and a temperature like variable $r_c(x)$, together with $u^\mu(x)$ and 
$\zeta^\mu(x)$. The precise definitions of our field variables is given
by the equations 
\begin{equation}\label{gravdef}
 \begin{split}
&\phi(r,x) = \frac{r_c^2(x)\ep(x)}{r^2} + {\cal O}\left(\frac{1}{r^3}\right)\\
  &u^\mu T_\mu^\nu (x)= -\rho_n(x)u^\nu 
+ \frac{\rho_s(x)}{\mu(x)^2 - r_c^2(x)\zeta(x)^2} \left[-\mu(x) u^\nu + 
r_c(x) \zeta^\nu(x)\right]\\
 & P^{\mu\nu}\xi_\nu(x) = r_c(x) \zeta^\mu(x),~~\zeta = \sqrt{\zeta^\mu\zeta_\mu}\\
 \end{split}
\end{equation}
where $\phi(r,x)$ is the slowly varying bulk scalar field and 
$T^{\mu\nu}(x)$ is  the boundary stress tensor. The functions
$\rho_s(x)$, $\rho_n(x)$ and $\mu(x)$ are given in terms of  
$r_c(x)$, $\epsilon(x)$ and $\zeta(x)$ determined by thermodynamics (i.e. from 
previous sections). As usual,  $P^{\mu\nu} = \eta^{\mu\nu} + u^\mu u^\nu$. 
Note that the second 
equation in \eqref{gravdef} would have agreed precisely with \eqref{defvar} 
were we to replace the function $\mu(x)$ in \eqref{gravdef} with 
$u^\mu \xi_\mu$. As it stands the velocity field 
defined by \eqref{gravdef} is close to but distinct from the velocity field 
defined by \eqref{defvar}.

The fluid gravity map is generated by solving Einstein's equations tube wise, 
point by point on the boundary. At any given boundary point we can always 
boost and rotate coordinates so that
 $$u_\mu = (-1,0,0,0),~~~~n_\mu = (0,1,0,0) $$
In the neighborhood of our special point, however, 
\begin{equation}
 \begin{split}
  u_\mu &= \gamma_u (-1,\beta_1,\beta_2,\beta_3),~~n_\mu = \gamma_n  (-n_v, 1, n_2,n_3),
~~n_v = \beta_1 + n_2\beta_2 + n_3\beta_3 \\
\gamma_u &=\frac{1}{ \sqrt{1 - \beta_1^2 - \beta_2^2 - \beta_3^2 }},~~
 \gamma_n  = \frac{1}{ \sqrt{n_v^2 - 1 - n_2^2 - n_3^2 }}
 \end{split}
\end{equation}
where $\beta_i$ and $n_i$ are of first or higher order in 
derivatives of fluid fields at the special point. 

In this paper we will work only to first order in the derivative expansion. 
At this order we are sensitive only to first derivatives of 
$\beta_1$ $\beta_2$, $\beta_3$, $n_2$ and $n_3$ along with 
the first derivatives of $\xi$, $r_c$ and $\epsilon$. 

The solution at 
our special point preserves an $SO(2)$ symmetry (of rotations in a plane 
perpendicular to $u_\mu$ and $n_\mu$; the $yz$ plane in our coordinates). 
This symmetry will help us organize our calculation. To start with it will 
prove useful to organize first derivative `fluid data', i.e. all the first 
derivatives of the fluid fields at our special point, in terms of their 
$SO(2)$ transformation properties. We list our results 

\begin{itemize}\label{derdata}
\item First order derivative excitations with spin $0$ (scalars):\\
 $ S_1 =\frac{1}{\epsilon} \partial_1\zeta,~S_2  = \frac{1}{\epsilon}\partial_1\epsilon,
~S_3 =\partial_0\beta_1,~S_4 = \partial_1\beta_1,~
S_5 = \partial_i n_i ,$ 

$S_6  = \partial_i \beta_i ,~ S_7 =\epsilon_{ij} \partial_i n_j , $

$S_8 = \partial_0 r_c,~S_9 = \partial_1r_c,~S_{10}= \frac{1}{\epsilon}\partial_0\zeta,~,S_{11} =\frac{1}{\epsilon} \partial_0\epsilon,
~S_{12}=\epsilon_{ij} \partial_i \beta_j$
\item First order derivative excitations with spin $\pm 1$ (vectors):\\
$\left[V_1\right]_i =\frac{1}{\epsilon} \partial_i\epsilon,~\left[V_2\right]_i = \frac{1}{\epsilon}\partial_i\zeta,
~\left[V_3\right]_i = \partial_1 n_i,
~\left[V_4\right]_i =\partial_0\beta_i, $

$~\left[V_5\right]_i = \partial_i\beta_1 + \partial_1\beta_i~\left[V_6\right]_i = \partial_i r_c,~\left[V_7\right]_i = \partial_0 n_i, ~ \left[V_8\right]_i =\partial_i\beta_1 -\partial_1\beta_i$
\item First order derivative excitations with spin $\pm 2$ (traceless 
symmetric tensors):\\
$\left[T_1\right]_{ij} = \partial_i \beta_j + \partial_j \beta_i - (\partial_k\beta_k)\delta_{ij},
~~\left[T_2\right]_{ij} = \partial_i n_j + \partial_j n_i - (\partial_k n_k)\delta_{ij}$

Here $\{i,j\} = \{2,3\}$.
\end{itemize}

Following the methods of 
\cite{Bhattacharyya:2008jc, Banerjee:2008th, Bhattacharyya:2008mz}, in order
to derive the metric dual to a fluid flow we need to solve the equations 
of motion, order by order, in the derivative expansion. That is we set 
the metric $g$ of our solution to $g_0+ \epsilon g_1 \ldots$ (and similarly
for the gauge fields and the scalars) and solve the bulk equations of motion 
at first order in $\epsilon$. As explained in 
\cite{Bhattacharyya:2008jc, Banerjee:2008th, Bhattacharyya:2008mz}, the 
resulting equations are of two sorts. The Einstein and Maxwell constraint 
equations reduce simply to the equations of energy momentum and current 
conservation, and do not involve the unknown fields $g_1$ etc. These 
equations relate some of the independent derivatives listed above
to others. On the other hand the dynamical Einstein and Maxwell equations 
allow you us compute the unknown fields $g_1$ etc in terms of the constrained 
derivative data listed above. 

\subsection{The constraint equations} \label{consteq}

We will now first describe the solution 
of the constraint equations, before turning to the dynamical equations. 

In addition to the conservation equations described above, there is 
one additional source of constraints on the derivative data
given in \S \ref{method}. Our demand that our solution be  asymptotically
$AdS$ requires, in particular that the boundary field strength vanishes, 
implying
$\partial_\mu \xi_\nu-\partial_\nu \xi_\nu$ vanishes. We must add this 
equation to the list of equations that constrain independent data. 

It is convenient to decompose the constraint equations according to the 
its quantum numbers under the preserved $SO(2)$. We now perform the relevant decompositions,
and state which pieces of data we use these constraints to solve for.
\begin{itemize}
 \item {\it Current conservation:}
It is a spin-0 constraint. Using this we shall solve for $S_{11}$.
\item {\it Stress-tensor conservation:}
It is effectively four equations. Among them
two are spin-0 constraints and one spin-1 constraint. Using this we 
shall solve for $S_8$, $S_9$ and $V_6$,.
\item {\it Curl-free condition on $\xi_\mu$:}
This imposes a set of 6 equations.

 Two of them transform in spin-0 ($[\partial_0\xi_1 - \partial_1\xi_0]$ and $\epsilon_{ij}\partial_i\xi_j$). 
Using these we solve for $S_{10}$ and $S_{12}$ respectively.

Four of them transform in two separate spin-1 ($[\partial_i\xi_1 - \partial_1\xi_i]$ and $[\partial_i\xi_0 - \partial_0\xi_i]$).
Using these we solve for $V_7$ and $V_8$.
\end{itemize}

After solving for dependent data
\footnote{As we have indicated above, we solve for some first derivatives
of fluid fields in terms of other derivatives. The relevant equations are 
linear and easy to solve; the solutions are explicit but lengthy and we 
do not present them here.}, the remaining independent one derivative 
pieces of data are given as follows. We have seven spin-0 ($S_1, \cdots, S_7$), five spin-1 ($V_1, \cdots, V_5$) and two spin-2 ($T_1,  T_2$) boundary data.

For later use we will find it useful to list covariant expressions for 
the independent data. These expressions are most usefully written in terms 
of the projector normal to the velocity/ superfluid velocity frame
$$\tilde P_{\mu\nu} = u_\mu u_\nu + \eta_{\mu\nu} - n_\mu n_\nu$$
Using this projector one can write the following covariant expressions for 
our choices of independent boundary data as follows:
\\

{\it Spin-0}
\begin{equation}\label{covsc}
 \begin{split}
  S_1 &= \frac{1}{\epsilon}(n^\mu\partial_\mu)\zeta,~~~S_2 = \frac{1}{\epsilon}(n^\mu\partial_\mu)\epsilon,
~~~S_3 = u^\mu n^\nu \partial_\mu u_\nu,
~~~S_4 = n^\mu n^\nu \partial_\mu u_\nu,\\
S_5 &=\tilde P^{\mu\nu}\partial_\mu n_\nu,~~~ S_6 = \tilde P^{\mu\nu}\partial_\mu u_\nu,~~~
S_7=\epsilon^{\mu\nu\rho\sigma} n_\mu u_\mu\partial_\rho n_\sigma\\
 \end{split}
\end{equation}
\\

{\it Spin-1}
\begin{equation}\label{covvv}
 \begin{split}
 [V_1]_\mu &=\frac{1}{\epsilon} \tilde P_\mu^\sigma\partial_\sigma\epsilon,
~~~[V_2]_\mu = \frac{1}{\epsilon}\tilde P_\mu^\sigma\partial_\sigma\zeta,
~~~[V_3]_\mu = \tilde P_\mu^\nu n^\sigma\partial_\sigma n_\nu,\\
[V_4]_\mu &= \tilde P_\mu^\nu u^\sigma\partial_\sigma u_\nu,
~~~[V_5]_\mu = \tilde P_\mu^\nu n^\sigma\left(\partial_\nu u_\sigma + \partial_\sigma u_\nu\right)\\
 \end{split}
\end{equation}
\\

{\it Spin-2}
\begin{equation}\label{covtt}
 \begin{split}
 [T_1]_{\mu\nu} &=\tilde P_\mu^\sigma \tilde P_\nu^\rho\left[\partial_\sigma u_\rho +\partial_\rho u_\sigma\right] - S_6 \tilde P_{\mu\nu},\\
[T_2]_{\mu\nu} &= \tilde P_\mu^\sigma \tilde P_\nu^\rho\left[\partial_\sigma n_\rho +\partial_\rho n_\sigma\right] - S_5 \tilde P_{\mu\nu}
 \end{split}
\end{equation}

\subsection{The dynamical equations}

Following earlier work on the fluid gravity correspondence 
\cite{Bhattacharyya:2008jc, Banerjee:2008th, Bhattacharyya:2008mz,Erdmenger:2008rm} , we work 
in the gravitational gauge $g_{rr}=0$ and $g_{r\mu}=u_\mu$. For the $U(1)$ 
field we continue to demand that the scalar field be real. With derivatives
taken into account this requirement no longer sets $A^r$ to zero, but allows
us to determine $A^r$ rather simply, by demanding the consistency of the 
equations for $\phi$ and $\phi^*$.

We will now solve for the first derivative corrections about the basic 
fluid gravity ansatz. As we have determined the equilibrium solutions, 
in the previous section,  only to 
order $\frac{1}{e^2}$, we can of course compute the metric dual to fluid 
flows only at the same order in $\frac{1}{e^2}$.

We now describe in rough terms how we determine the 
deviations away from the zero order fluid ansatz.
Let us start with the gauge field and scalars. At leading order in 
$\frac{1}{e^2}$ we take derivative corrections to the 
gauge field and the scalar field to have the form ($\delta A^M$ is the 
derivative correction for the gauge field)
\begin{equation}
 \begin{split}\label{abkalan1}
  &\delta A^r =\sum_{i = 1}^7 \delta A_i\left(\frac{r}{r_c}\right) S_i+{\cal O}\left(\frac{1}{e^2}\right) \\
&\delta A^\mu =\frac{1}{r_c^2}\left[ u^\mu\sum_{i = 1}^7 \delta H_i\left(\frac{r}{r_c}\right) S_i ~
 +~ n^\mu\sum_{i = 1}^7 \delta L_i\left(\frac{r}{r_c}\right) S_i
+\sum_{i = 1}^5X_i\left(\frac{r}{r_c}\right)~[V_i]^\mu\right]  + {\cal O}\left(\frac{1}{e^2}\right) \\
\\
&\delta \phi =\frac{1}{r_c}\left[ \sum_{i = 1}^7 \delta \phi_i\left(\frac{r}{r_c}\right) S_i\right]  + {\cal O}\left(\frac{1}{e^2}\right)\\
 \end{split}
\end{equation}

We now describe in structural terms how we have solved for these functions, 
emphasizing boundary conditions
\begin{enumerate}
 \item It turns out that $\delta A_i(r)$ obeys a first order 
differential equation in $r$. The general solution of $\delta A^r$ diverges 
linearly in  $r$ while expanded around $r=\infty$. We fix the 
constant of integration (coefficient of the homogeneous solution in 
this equation)  by setting the coefficient of the linear term in 
$r$ to zero. This choice of boundary conditions is forced on us by the 
requirement that the bulk current goes to zero at the boundary so that 
the boundary current is really conserved.
 \item $\delta H_i(r)$ obeys a second order differential equation (arising 
from the $r$ component of the  Maxwell equation). The two integration 
constants for this equation are fixed as follows. One of the integration constant is determined from the requirement of regularity at the horizon.
 The other integration constant is obtained from the requirement that 
there exist a regular scalar field solution (see below).
\item $\delta L_i(r)$ obeys a second order differential equation given by the $x$ component of the  Maxwell
equation. Here one of the integration constant is determined imposing the regularity of the solution at the horizon. The other integration 
constant is fixed using the fact that according to  equation \eqref{gravdef} $\zeta^\mu$ does not receive any derivative correction.
A generic solution of $\delta L_i(r)$ dies of at infinity like 
$\frac{1}{r^2}$; the coefficient of this $\frac{1}{r^2}$ must be set to zero. 
\item The equation for $X_i(r)$ comes from the $y$ or $z$ component of the Maxwell equation.
 This is also a second order differential equation and its integration constants are determined in a similar way as 
in $\delta L_i(r)$.
\item The equation for the scalar field determines $\delta\phi_i(r)$. 
Normalizability and the definition of $\epsilon(x)$
as given in equation \eqref{gravdef} fixes the two integration constants here.
More specifically, an expansion about infinity of a generic solution to 
the scalar field equation takes the form
$$\delta\phi_i(r) =a_i \frac{\ln (r)}{r^2} + \frac{b_i}{r^2} + {\cal O}\left(\frac{1}{r^2}\right)$$
Our boundary conditions are that both $a_i$ vanishes 
(from the requirement of normalizability) and that $b_i$ vanishes (from 
our definition of $\epsilon$). These two requirements completely fix the 
scalar fluctuation. As described above, the further requirement that the 
scalar fluctuation be regular at the horizon yields a boundary condition on 
$\delta H_i(r)$ (see above).
\end{enumerate}

Let us now turn to the metric field. In the strict limit of 
$\frac{1}{e^2}\rightarrow 0$ the scalar and gauge field do not back react 
on the metric. The derivative expansion of the metric in this 
limit is thus that of uncharged fluid dynamics and was determined in 
\cite{Bhattacharyya:2008jc} to be 
\begin{equation}\label{probemetric}
 \begin{split}
  ds^2 =&- 2 u_\mu dx^\mu dr+
r^2\bigg[   -\left(1- \frac{r_c^4}{r^4}\right)u_\mu u_\nu +
+ P_{\mu\nu}\bigg]dx^\mu dx^\nu\\
&-dx^\mu dx^\nu\bigg[2r u_\mu\left(\left(u.\partial\right)u_\nu  - 
\frac{1}{3}\left(\partial.u\right) u_\nu\right) +r_c F\left(\frac{r}{r_c}\right)\sigma_{\mu\nu}\bigg]
 \end{split}
\end{equation}
  Where $$F(r) = -\frac{r^2 }{2} \left[-\log \left(r^2+1\right)+4 \log (r)-2
   \log (r+1)+2 \tan ^{-1}(r)-\pi \right]$$ and 
$$\sigma_{\mu\nu} = P_\mu^\alpha P_\nu ^\beta\left(\frac{\partial_\alpha u_\beta + \partial_\beta u_\alpha}{2} 
- \frac{\partial.u}{3}\eta_{\alpha\beta}\right)$$

The new results of this paper are for the derivative correction to the metric 
at order ${\cal O}\left(\frac{1}{e^2}\right)$. We parameterize the corrections
to the metric as 
\begin{equation}\label{ccmmgg}
 \begin{split}
&\delta(ds^2)\\
 &= - \frac{2}{e^2}\left[ \sum_{i=1}^7\frac{S_i}{r_c}~\delta g_i\left(\frac{r}{r_c}\right) \right]u_\mu dx^\mu dr\\
&+\frac{r_c}{e^2} dx^\mu dx^\nu\bigg\{ \left[ \sum_{i=1}^7S_i~\delta f_i\left(\frac{r}{r_c}\right) \right]u_\mu u_\nu 
+ \left[ \sum_{i=1}^7S_i~\delta K_i\left(\frac{r}{r_c}\right) \right]n_\mu n_\nu \\
&~~~~~~~~~~~~~+ \left[ \sum_{i=1}^7S_i~\delta J_i\left(\frac{r}{r_c}\right) \right]\left(n_\mu u_\nu + n_\nu u_\mu\right)\bigg\}\\
&+\frac{r_c}{e^2} dx^\mu dx^\nu\bigg\{ \sum_{i = 1}^5\left[Y_i\left(\frac{r}{r_c}\right)\left(u_\mu [V_i]_\nu + u_\nu [V_i]_\mu\right) 
+ W_i\left(\frac{r}{r_c}\right)
\left(n_\mu [V_i]_\nu + n_\nu [V_i]_\mu\right)\right]\\
&~~~~~~~~~~~~~~~~~~~~~~~~+\sum_{i = 1}^2 Z_i\left(\frac{r}{r_c}\right)[T_i]_{\mu\nu}\bigg\}
 + {\cal O}\left(\frac{1}{e^4}\right)\\
\\
 \end{split}
\end{equation}

We now describe, very qualitatively, how we have solved for these functions.
\begin{enumerate}
 \item $\delta g_i(r)$, $\delta f_i(r)$ and $\delta K_i (r)$ are determined solving three coupled equations obtained from the $(rr)$
$(rv)$ and $(xx)$ component of the Einstein equations.
 Once decoupled using appropriate combination of these functions, two of the 
equations become first order and the third one is a second order differential equation. 
Two of the four integration constants are determined 
using the asymptotic AdS condition of the metric. \footnote{
After this condition is 
imposed $\frac{1}{r}$ expansion of the 
functions $\delta g_i(r)$ and $\delta K_i(r)$ take the form
 $$\lim_{r\rightarrow\infty}\delta g_i(r) = {\cal O}\left(\frac{1}{r^4}\right),~~
\lim_{r\rightarrow\infty}\delta K_i(r) = {\cal O}\left(\frac{1}{r^2}\right)$$}
A third integration constant is fixed by demanding the regularity of the 
function $\delta K_i(r)$ at $r= r_c$. The last integration
 constant is fixed to ensure that the $vv$ component of the boundary stress tensor receives no derivative corrections so that the equation \eqref{gravdef} 
is satisfied.

\item The function $\delta J_i(r)$ is determined using the $(rx)$ or $(vx)$ component of the Einstein equation. 
This is a second order differential equation in $r$. The two integration constants are determined using the normalizability
and the definition of boundary stress tensor (according to \eqref{gravdef}
the $vx$ component of the boundary stress tensor should not receive any 
derivative correction). The general solution for $\delta J_i(r)$
has the following expansion around $r= \infty$. 
$$\lim_{r\rightarrow\infty}\delta J_i(r) = j_0~ r^2 + \frac{j_1}{r^2} + {\cal O}\left(\frac{1}{r^2}\right)$$
Our boundary condition is that $j_0$ and $j_1$ both vanish.

\item $Y_i(r)$ is  determined from the $(vy)$ or $(vz)$ component of the Einstein equation .
This is  a second order differential equation in $r$. The two integration constants are determined using normalizability 
of the metric and the definition of boundary stress tensor (the $(vy)$ or $(vz)$ component of the stress tensor
should not receive any derivative corrections). This condition is exactly 
same as that of $\delta J_i(r)$ in terms of
the coefficients of $\frac{1}{r}$ expansion.

\item    $W_i(r)$ is  determined from the $(xy)$ or $(xz)$ component of the Einstein equation .
This is  a second order differential equation in $r$. The two integration constants are determined using normalizability 
and regularity
of the metric respectively. A generic solution of $W_i$ behaves like 
$r^2$ at large  $r$. Our boundary conditions are that the leading coefficient
of this leading $r^2$ piece vanish.
\item    $Z_i(r)$ is  determined from the $(yz)$ component of the Einstein equation .
This is  a second order differential equation in $r$. The two integration constants are determined exactly the same way as for 
$W_i(r)$.
\end{enumerate}

\subsection{Results for Bulk Fields}\label{resmet}

Without further ado in this subsection we simply present our final results 
for all the fields defined in the previous subsection. 

We have performed all our computations in this section using Mathematica. 
In several instances we have carried out calculations to higher order, in 
the Mathematica file, than we have presented below, mainly to avoid 
burdening the reader with very lengthy expressions.

The solutions presented
in this subsection determine the full first order correction to the gauge 
field, scalar field and metric to the relevant order in an expansion in 
$\epsilon$ and $\frac{1}{e^2}$. Now we choose to  scale $\zeta$ like $\epsilon$. 
We present our results below in terms of the order one field 
$$\chi = \frac{\zeta}{\epsilon} $$
Recall that, according to the results of \S \ref{stability}, our fluid 
becomes unstable whenever $\chi$ exceeds a number of order unity. So while 
$\chi$ can be arbitrarily small, it is unphysical for $\chi$ to be made 
arbitrarily large.

{\it Results for the gauge field and scalar field}
\begin{equation}\label{rrggA}
 \begin{split}
 \delta A_1(r) &=\epsilon\left[\frac{r^2 \left(96 \chi ^2-5\right)+48 \chi ^2+1}{14 r^3}\right]
+ {\cal O}(\epsilon^3)\\ 
 \delta A_2(r) &=\epsilon\left[\frac{\left(2-3 r^2\right) \chi }{7 r^3}\right]
+ {\cal O}(\epsilon^3)\\
 \delta A_3(r) &= -\epsilon\left[ \frac{\left(2 r^2+1\right) \chi }{7 r^3}\right] + {\cal O}(\epsilon^3)\\
 \delta A_4(r) &=\left[ \frac{16 \left(2 r^2+1\right) \chi ^2}{7 r^3}-\frac{2 r}{3
   \left(r^2+1\right)}\right]
+ {\cal O}(\epsilon^2)\\
 \delta A_5(r) &=\epsilon\left[\frac{\left(1-5 r^2\right) \chi }{14 r^3}\right]
+ {\cal O}(\epsilon^3)\\
\delta A_6(r) &=-\frac{2 r}{3 \left(r^2+1\right)}-\frac{8 \left(2 r^2+1\right)
   \chi ^2}{7 r^3}
+ {\cal O}(\epsilon^2)\\
 \end{split}
\end{equation}

\begin{equation}\label{rrggH}
 \begin{split}
 \delta H_1(r) &=\epsilon\left[\frac{r (r+2) \left(96 \chi ^2-5\right)-48 \chi ^2-1}{14 r
   (r+1) \left(r^2+1\right)} \right]+ {\cal O}(\epsilon^3)\\
\delta H_2(r) &=\epsilon\left[ -\frac{\left(3 r^2+6 r+2\right) \chi }{7 r
   \left(r^3+r^2+r+1\right)} \right]+ {\cal O}(\epsilon^3)\\
 \delta H_3(r) &=\epsilon\left[\frac{\left(5 r^2+10 r+1\right) \chi }{7 r
   \left(r^3+r^2+r+1\right)}\right]+ {\cal O}(\epsilon^3)\\
 \delta H_4(r) &=\frac{16 \left(2 r^2+4 r-1\right) \chi ^2}{7 r
   \left(r^3+r^2+r+1\right)}+ {\cal O}(\epsilon^2)\\
 \delta H_5(r) &=\epsilon\left[-\frac{\left(5 r^2+10 r+1\right) \chi }{14 r
   \left(r^3+r^2+r+1\right)}\right]+ {\cal O}(\epsilon^3)\\
 \delta H_6(r) &=-\frac{8 ~[2 r (r+2)-1 ]~\chi ^2}{7 r (r+1) \left(r^2+1\right)}+ {\cal O}(\epsilon^2)\\
 \end{split}
\end{equation}

\begin{equation}\label{rrggL}
 \begin{split}
  \delta L_1(r)& = \epsilon^2\left[\frac{\chi  \left(\log \left(r^2+1\right)-2 \log (r+1)+2 \tan
   ^{-1}(r)-\pi \right)}{4 r^2}\right] + {\cal O}(\epsilon^4)\\
 \delta L_2(r)& = \epsilon^2\left[\frac{\log \left(r^2+1\right)-2 \log (r+1)+2 \tan ^{-1}(r)-\pi
   }{96 r^2}\right] + {\cal O}(\epsilon^4)\\
\delta L_3(r) &= {\cal O}(\epsilon^4)\\
\delta L_4(r) &=-\epsilon\left[\frac{\chi  \left(\log \left(r^2+1\right)-4 \log (r)+2 \log
   (r+1)-2 \tan ^{-1}(r)+\pi \right)}{3 r^2}\right] \\
&~~~~~~~+ {\cal O}(\epsilon^3)\\
\delta L_5(r)&= {\cal O}(\epsilon^4)\\
 \delta L_6(r) &=-\epsilon\left[\frac{\chi  \left(\log \left(r^2+1\right)-4 \log (r)+2 \log
   (r+1)-2 \tan ^{-1}(r)+\pi \right)}{6 r^2}\right]\\
&~~~~ ~~~~+ {\cal O}(\epsilon^3)\\
 \end{split}
\end{equation}

\begin{equation}\label{rrggvec}
 \begin{split}
    X_1(r)&=\epsilon^2\left[\frac{\log \left(r^2+1\right)-2 \log (r+1)+2 \tan ^{-1}(r)-\pi
   }{96 r^2} \right] + {\cal O}(\epsilon^4)\\
 X_2(r)&=\epsilon^2\left[\frac{\chi  \left(\log \left[\frac{r^2+1}{(r+1)^2}\right]+2
   \tan ^{-1}(r)-\pi \right)}{4 r^2} \right] + {\cal O}(\epsilon^4)\\
X_3(r)&= {\cal O}(\epsilon^4)\\
 X_4(r)&= {\cal O}(\epsilon^4)\\
  X_5(r)&=\epsilon\left[-\frac{\chi  \left(\log \left(r^2+1\right)-4 \log (r)+2 \log
   (r+1)-2 \tan ^{-1}(r)+\pi \right)}{4 r^2} \right]\\
&~~~~~~ + {\cal O}(\epsilon^3)\\
 \end{split}
\end{equation}

\begin{equation}\label{rrggp}
 \begin{split}
  \delta \phi_1(r) &=\epsilon^2\bigg[ \frac{3}{14} \left(1-8 \chi ^2\right)\left(\tan ^{-1}(r)  - \log(1 + r) 
- \frac{\pi}{2}\right)\\
&~~~+ \frac{2}{7} \left(36 \chi ^2-1\right) \log (r) +\left(\frac{1}{4}-6 \chi ^2\right) \log \left(r^2+1\right)\bigg]
 +  {\cal O}(\epsilon^4)\\
\delta \phi_2(r) &=\epsilon^2\left[-\frac{1}{28} \chi  \left(-7 \log \left(r^2+1\right)+4 \log
   (r)+10 \log (r+1)-10 \tan ^{-1}(r)+5 \pi \right)\right]\\
&~~~~~ +  {\cal O}(\epsilon^4)\\
 \delta \phi_3(r) &=\epsilon^2\left[ \frac{\chi  \left(-7 \log \left(r^2+1\right)+8 \log (r)+6 \log
   (r+1)-6 \tan ^{-1}(r)+3 \pi \right)}{14 \left(r^2+1\right)}\right]\\
&~~~~~~~~~ +  {\cal O}(\epsilon^4)\\
 \delta \phi_4(r) &=\epsilon\left[ \frac{4 \chi ^2 \left(-7 \log \left(r^2+1\right)+12 \log (r)+2
   \log (r+1)-2 \tan ^{-1}(r)+\pi \right)}{7
   \left(r^2+1\right)}\right] \\
&~~~~~~~~~+  {\cal O}(\epsilon^3)\\
\delta \phi_5(r) &=\epsilon^2\left[-\frac{\chi}{28}   \left(-7 \log \left(r^2+1\right)+8 \log
   (r)+6 \log (r+1)-6 \tan ^{-1}(r)+3 \pi \right)\right]\\
&~~~~~~~~~~~+  {\cal O}(\epsilon^4)\\
\delta \phi_6(r) &=\epsilon\left[-\frac{2}{7} \chi ^2 \left(-7 \log \left(r^2+1\right)+12 \log
   (r)+2 \log (r+1)-2 \tan ^{-1}(r)+\pi \right)\right]\\
&~~~~~~~~~~ +  {\cal O}(\epsilon^3)\\
 \end{split}
\end{equation}

{\it Results for the metric}
\begin{equation}\label{rrmmf}
 \begin{split}
\delta f_1(r) &=\epsilon\left[\frac{-2  \left(80 \chi ^2-3\right)}{7 r^4}\right] + {\cal O}(\epsilon^3)\\
\delta f_2(r) &=\epsilon\left[\frac{16  \chi }{21 r^4}\right] + {\cal O}(\epsilon^3)\\
\delta f_3(r) &=-\epsilon\left[ \frac{12  \chi }{7 r^4}\right] + {\cal O}(\epsilon^3)\\
\\
\delta f_4(r) &=\bigg[-\frac{320 \chi ^2}{21 r^4} -\frac{\left(r^4+1\right) \left(-5 \log \left(r^2+1\right)+8
   \log (r)+2 \log (r+1)+4 \tan ^{-1}(r)\right)}{9 r^2}\\
&~~~~+\frac{2 \left(r^4+1\right) \left(r^2 \left(\pi  r^2-r+\pi
   -3\right)-2\right)}{9 \left(r^6+r^4\right)}\bigg] + {\cal O}(\epsilon^2)\\
\\
 \delta f_5(r) &=\epsilon\left[\frac{6 \chi }{7 r^4} \right] + {\cal O}(\epsilon^3)\\
 \delta f_6(r) &=-\frac{1}{2}\bigg[-\frac{320 \chi ^2}{21 r^4} -\frac{\left(r^4+1\right) \left(-5 \log \left(r^2+1\right)+8
   \log (r)+2 \log (r+1)+4 \tan ^{-1}(r)\right)}{9 r^2}\\
&~~~~+\frac{2 \left(r^4+1\right) \left(r^2 \left(\pi  r^2-r+\pi
   -3\right)-2\right)}{9 \left(r^6+r^4\right)}\bigg] + {\cal O}(\epsilon^2)\\
\end{split}
\end{equation}

\begin{equation}\label{rrmm9}
 \begin{split}
\delta g_1(r) &= {\cal O}(\epsilon^3)\\
\delta g_2(r) &= {\cal O}(\epsilon^3)\\
\delta g_3(r)& = {\cal O}(\epsilon^3)\\
\\
\delta g_4(r) &= \bigg[\frac{1}{18} \left(-5 \log \left(r^2+1\right)+8 \log (r)+2 \log
   (r+1)+4 \tan ^{-1}(r)-2 \pi \right)\\
&~~~~~+\frac{r^6+4 r^5+4 r^4+6 r^3+r^2-2 r-2}{9 (r+1)
   \left(r^3+r\right)^2}\bigg]  + {\cal O}(\epsilon^2)\\
\\
\delta g_5(r) &= {\cal O}(\epsilon^3)\\
\delta g_6(r) &= -\frac{1}{2}\bigg[\frac{1}{18} \left(-5 \log \left(r^2+1\right)+8 \log (r)+2 \log
   (r+1)+4 \tan ^{-1}(r)-2 \pi \right)\\
&~~~~~+\frac{r^6+4 r^5+4 r^4+6 r^3+r^2-2 r-2}{9 (r+1)
   \left(r^3+r\right)^2}\bigg]  + {\cal O}(\epsilon^2)\\
\end{split}
\end{equation}

\begin{equation}\label{rrmmj}
 \begin{split}
\delta K_1(r) &= {\cal O}(\epsilon^3)\\
\delta K_2(r) &= {\cal O}(\epsilon^3)\\
\delta K_3(r)& = {\cal O}(\epsilon^4)\\
\\
\delta K_4(r)& = \bigg[ \frac{r^2}{3}  \left(-5 \log \left(r^2+1\right)+8 \log (r)+2
   \log (r+1)+4 \tan ^{-1}(r)\right)\\
&~~~~~~~+\frac{4-2 r^2 \left(\pi  r^2-r+\pi -3\right)}{3
   \left(r^2+1\right)}\bigg]+ {\cal O}(\epsilon^2)\\
\\
\delta K_5(r)& = {\cal O}(\epsilon^3)\\
\delta K_6(r)& = -\frac{1}{2}\bigg[ \frac{r^2}{3}  \left(-5 \log \left(r^2+1\right)+8 \log (r)+2
   \log (r+1)+4 \tan ^{-1}(r)\right)\\
&~~~~~~~+\frac{4-2 r^2 \left(\pi  r^2-r+\pi -3\right)}{3
   \left(r^2+1\right)}\bigg]+ {\cal O}(\epsilon^2)\\
\end{split}
\end{equation}

\begin{equation}\label{rrmmk}
 \begin{split}
\delta J_1(r)& =-\epsilon^2\bigg[\frac{\left(r^4-1\right) \chi }{4r^2}   \left(\log
   \left(r^2+1\right)-2 \log (r+1)-2 \tan ^{-1}(r)+\pi \right)\\
&~~~~~~~~~+\frac{\chi}{6r^2} (2-3 r)  \bigg]+ {\cal O}(\epsilon^4)\\
\delta J_2(r)& =-\epsilon^2\bigg[-\frac{\left(r^4-1\right)  \left(-\log
   \left(r^2+1\right)+2 \log (r+1)-16 \tan ^{-1}(r)+8 \pi
   \right)}{96 r^2}\\
&~~~~~~~~~+\frac{27 r^4-3 r^3+20 r^2-3 r-19}{144 \left(r^3+r\right)}\bigg]+ {\cal O}(\epsilon^4)\\
\delta J_3(r) &= -\epsilon^2\left(-\frac{r}{6 \left(r^2+1\right)^2}\right) + {\cal O}(\epsilon^3)\\
\delta J_4(r) &= {\cal O}(\epsilon^3)\\
\delta J_5(r)& = {\cal O}(\epsilon^3)\\
\delta J_6(r)& = {\cal O}(\epsilon^3)\\
\end{split}
\end{equation}

\begin{equation}\label{rrmmvec1}
 \begin{split}
Y_1(r)&= -\epsilon^2\bigg[\-\frac{ \left(r^4-1\right)}{96~r^2} \left[-\log
   \left(r^2+1\right)+2 \log (r+1)-16 \tan ^{-1}(r)+8 \pi
   \right]\\
&~~~~~~~~+\frac{27 r^4-3 r^3+20 r^2-3 r-19}{144 r \left(r^2+1\right)} \bigg] + {\cal O}(\epsilon^4)\\
Y_2(r)&= -\epsilon^2\bigg[\frac{\left(r^4-1\right) \chi }{4r^2}  \left(\log
   \left(r^2+1\right)-2 \log (r+1)-2 \tan ^{-1}(r)+\pi \right]\\
&~~~~~~~~~~+\frac{ \chi}{6~r} (2-3 r) \bigg] + {\cal O}(\epsilon^4)\\
 Y_3(r)&=  {\cal O}(\epsilon^4)\\
 Y_4(r)&= \epsilon^2\left[\frac{r}{6 \left(r^2+1\right)^2}\right] + {\cal O}(\epsilon^4)\\
Y_5(r)&= \ {\cal O}(\epsilon^3)\\
\end{split}
\end{equation}

\begin{equation}\label{rrmmvec1}
 \begin{split}
W_1(r)&=\epsilon^3~ \chi  \bigg[\frac{-3 \pi  \left(r^3+r\right)+6 \left(r^3+r\right) \tan
   ^{-1}(r)+6 r^2+4}{16 \left(r^3+r\right)}\bigg] + {\cal O}(\epsilon^5)\\
 W_2(r)&=\epsilon^3  \bigg[\frac{-3 \pi  \left(r^3+r\right)+6 \left(r^3+r\right) \tan
   ^{-1}(r)+6 r^2+4}{32 \left(r^3+r\right)} \bigg] + {\cal O}(\epsilon^5)\\
W_3(r)&= \epsilon^3\left[\frac{\chi  \left[-3 \pi  \left(r^3+r\right)+6
   \left(r^3+r\right) \tan ^{-1}(r)+6 r^2+4\right]}{32
   \left(r^3+r\right)}\right]+ {\cal O}(\epsilon^4)\\
W_4(r)&=  \epsilon^3\left[\frac{\chi  \left[-3 \pi  \left(r^3+r\right)+6
   \left(r^3+r\right) \tan ^{-1}(r)+6 r^2+4\right]}{32
   \left(r^3+r\right)}\right]+ {\cal O}(\epsilon^4)\\
W_5(r)&= -\frac{r^2}{6}  \bigg[-\frac{2 (r+1)}{r^2+1}
-\frac{4}{r^2}+5 \log \left(r^2+1\right)-8 \log (r)-2 \log
   (r+1)\\
&~~~~~~~~~~~~~-4 \tan ^{-1}(r)+2 \pi \bigg] + {\cal O}(\epsilon^2)\\
\end{split}
\end{equation}

\begin{equation}\label{rmten}
 \begin{split}
   Z_1(r)=&~ \frac{r^2}{3} \bigg[\frac{2 (r+1)}{r^2+1}+\frac{4}{r^2}-5 \log
   \left(r^2+1\right)+8 \log (r)\\
&~+2 \log (r+1)+4 \tan ^{-1}(r)-2
   \pi \bigg] + {\cal O}(\epsilon^2)\\
Z_2(r) =&~ \epsilon^3~\chi  \bigg[\frac{-3 \pi  \left(r^3+r\right)+6 \left(r^3+r\right) \tan
   ^{-1}(r)+6 r^2+4}{16 \left(r^3+r\right)}\bigg] +{\cal O}(\epsilon^4) \\
 \end{split}
\end{equation}

\subsection{The $\zeta \to 0$ limit}\label{App:zetazero}

The gravitational solutions presented above are complicated largely because 
they possess very little rotational symmetry. At any given 
spacetime point we have a normal fluid velocity and an independent 
superfluid velocity. These two velocities together break the local 
Lorentz group at a point down to the abelian group $SO(2)$. While we 
have usefully organized the results of our gravitational calculation in 
representations of $SO(2)$, as representations of $SO(2)$ are all one 
dimensional, our solutions admit several different functions of $r$. 

In the special case that $\zeta=0$, however, the residual symmetry group 
about a point is $SO(3)$. $SO(3)$ representation theory is considerably 
more constraining than $SO(2)$ representation theory. This implies that 
the gravitational dual to superfluid dynamics should be considerably simpler
in the special limit $\zeta \to 0$ than in the generic case. 

Let us first present a brief ab initio analysis of the nature of the 
gravitational solution when $\zeta=0$. All independent first derivative
data may be organized into $SO(3)$ scalars, vectors and tensors. These 
may be chosen as follows

{\it Scalar}
\begin{equation*}
 \partial_\mu u^\mu~~\text{and}~~ P^{\mu\nu}\partial_\mu \zeta_\nu
\end{equation*}

{\it Vector}
\begin{equation*}
 u^\mu\partial_\mu u^\nu,~~P^{\mu\nu}\partial_\nu \epsilon~~\text{and}
~~\epsilon^{\mu\nu\lambda\sigma}u_\nu\partial_\lambda\zeta_{\sigma}
\end{equation*}

{\it Tensor}
\begin{equation*}
 \sigma_{\mu\nu}~\text{and} ~~
 \sigma^{(\zeta)}_{\mu\nu}
=P_\mu^\alpha P_\nu^\beta\left(\frac{\partial_\alpha \zeta_\beta +\partial_\beta \zeta_\alpha}{2}
 -\left[ \frac{P^{\theta_1\theta_2} \partial_{\theta_1}\zeta_{\theta_2}}{3}\right] \eta_{\alpha\beta}\right)
\end{equation*}

Note of course that an $SO(3)$ vector or an $SO(3)$ may be decomposed 
into an $SO(2)$ vector and a scalar, while an $SO(3)$ tensor is composed 
of an $SO(2)$ tensor, vector and scalar. In $SO(2)$ terms, therefore, 
the data listed above totals to 7 scalars, 5 vectors and two tensor.

It follows from symmetry considerations (and the fact that our parity 
conserving gravitational system will never generate a parity violating 
vector term, so we can ignore the third vector above) that it must be 
possible to write the 
metric and gauge field, in the $\zeta \to 0$ limit, in the form 
\begin{equation}\label{metuni}
 \begin{split}
ds^2 &= -2g\left(\frac{r}{r_c}\right) u^\mu dx^\mu dr
+\bigg[  -r_c^2f\left(\frac{r}{r_c}\right) u_\mu u_\nu  + r^2 P_{\mu\nu}\bigg]dx^\mu dx^\nu\\
&+r_c^2 F\left(\frac{r}{r_c}\right) \sigma_{\mu\nu}dx^\mu dx^\nu\\
&+\frac{1}{e^2}\bigg\{-2\left[{\mathcal G}_1\left(\frac{r}{r_c}\right)\left( \partial_\mu u^\mu\right)
+{\mathcal G}_2\left(\frac{r}{r_c}\right
)\left(  P^{\mu\nu}\partial_\mu \zeta_\nu\right)\right]u_\mu dx^\mu dr\\
&+r_c^2\left[{\mathcal F}_1\left(\frac{r}{r_c}\right)\left( \partial_\mu u^\mu\right)
+{\mathcal F}_2\left(\frac{r}{r_c}\right
)\left(  P^{\mu\nu}\partial_\mu \zeta_\nu\right)\right]u_\mu u_\nu dx^\mu dx^\nu \\
&+ r_c^2\left[{\cal V}_1\left(\frac{r}{r_c}\right) \left(u.\partial\right) u_\nu 
+{\cal V}_2\left(\frac{r}{r_c}\right)  P_{\nu}^\alpha\partial_\alpha \epsilon \right]u_\mu dx^\mu dx^\nu\\
&+r_c^2\left[ {\cal T}_1\left(\frac{r}{r_c}\right)\sigma_{\mu\nu}
 +
{\cal T}_2\left(\frac{r}{r_c}\right)\sigma^{(\zeta)}_{\mu\nu} \right] dx^{\mu}dx^{\nu}\bigg\}
+{\cal O}\left(\frac{1}{e^4}\right)\\
\\
A &= \frac{1}{r_c}H\left(\frac{r}{r_c}\right) u^\mu\partial_\mu \\
&+ \left[{\mathcal A}_1\left(\frac{r}{r_c}\right)\left( \partial_\mu u^\mu\right)
+{\mathcal A}_2\left(\frac{r}{r_c}\right
)\left(  P^{\mu\nu}\partial_\mu \zeta_\nu\right)\right]\partial_r\\
&+\frac{1}{r_c^2} \left[{\mathcal H}_1\left(\frac{r}{r_c}\right)\left( \partial_\mu u^\mu\right)
+{\mathcal H}_2\left(\frac{r}{r_c}\right
)\left(  P^{\mu\nu}\partial_\mu \zeta_\nu\right)\right]u^\mu\partial_\mu\\
&+\frac{1}{r_c^2} {\cal L}_1\left(\frac{r}{r_c}\right) \left(u.\partial\right)u^\mu\partial_\mu
+\frac{1}{r_c^2} {\cal L}_2\left(\frac{r}{r_c}\right)P^{\mu\nu}\partial_\nu\epsilon\partial_\mu
+{\cal O}\left(\frac{1}{e^2}\right)
 \end{split}
\end{equation}

The results of the previous subsection must obey several relations in the 
limit $\zeta \to 0$ for them to agree with the form 
presented in \eqref{metuni}. \footnote{
A direct comparison
between these two forms is complicated by an irritating feature; the 
coordinate choice of the previous subsection differs from the one above 
(it breaks manifest $SO(3)$ invariance) even in the limit $\zeta \to 0$. 
In Appendix \ref{rot} we have explicitly performed the coordinate change 
that allows one to transform the results between coordinates.}. 
In Appendix \ref{rot} we have explicitly verified that each required 
relation is indeed obeyed. The results presented in the previous subsection 
are consistent with the form \eqref{metuni} once we make the identifications

\begin{equation}\label{somfix}
 \begin{split}
  &{\cal V}_1(r) =\epsilon^2\left[\frac{r}{3 \left(r^2+1\right)^2}\right] + {\cal O}(\epsilon^4)\\
  &{\cal V}_2(r)=   {\cal O}(\epsilon^4)\\
  &{\cal T}_1(r)=-\frac{r^2}{3}  \bigg[-\frac{2 (r+1)}{r^2+1}
-\frac{4}{r^2}+5 \log \left(r^2+1\right)-8 \log (r)-2 \log
   (r+1)\\
&~~~~~~~~~~~~~-4 \tan ^{-1}(r)+2 \pi \bigg] + {\cal O}(\epsilon^2)\\
 &{\cal T}_2(r)=   {\cal O}(\epsilon^3)\\
 &{\cal G}_1(r)=   0,~~ \text{(Required by Weyl invariance)}\\
&{\cal F}_1(r)=  {\cal O}(\epsilon^2)\\
&{\cal G}_2(r)=  {\cal O}(\epsilon^3)\\
&{\cal F}_2(r)=  \frac{6\epsilon}{7r^4}+{\cal O}(\epsilon^3)\\
 \end{split}
\end{equation}
and
\begin{equation}\label{gafnr}
 \begin{split}
  &{\cal A}_1(r) =-\frac{2r}{3(r^2 +1)} + {\cal O}(\epsilon^2)\\
  &{\cal A}_2(r)= \frac{\epsilon}{14 r^3}+  {\cal O}(\epsilon^3)\\
 &{\cal H}_1(r) = {\cal O}(\epsilon^2)\\
  &{\cal H}_2(r)=  \epsilon\left[\frac{-5r (r+2) -1}{14 r
   (r+1) \left(r^2+1\right)} \right]+  {\cal O}(\epsilon^3)\\
 &{\cal L}_1(r) = {\cal O}(\epsilon^4)\\
  &{\cal L}_2(r)= \epsilon^2\left[\frac{\log \left(r^2+1\right)-2 \log (r+1)+2 \tan ^{-1}(r)-\pi
   }{96 r^2}\right] + {\cal O}(\epsilon^4)
   \end{split}
\end{equation}

\subsection{Stress Tensor, Charge current and the Josephson Equation}
\label{resstress}

The results of the previous subsection may be used to read off the values 
of the boundary stress tensor, the boundary current and the correction 
to the Josephson equation at first order in the derivative expansion. 
Like all the calculations in this paper our results are obtained 
in a power series expansion in $\epsilon$ and $\frac{1}{e^2}$. 

We parameterize our boundary stress tensor and current as 
\begin{equation}
 \begin{split}
  T^{\mu\nu} &=  \frac{1}{16 \pi G}\bigg[A u^\mu u^\nu + B n^\mu n^\nu  
+ C \left( n^\mu u^\nu +  u^\mu n^\nu \right)
+\left( \frac{A-B}{2}\right) {\tilde P}^{\mu\nu}\bigg] + {\tilde \pi}^{\mu\nu}\\
 J^{\mu}& =\frac{1}{16 \pi G }\left[Q_1 u^\mu + Q_2 n^\mu\right]  + 
{\tilde J}^\mu_{diss}\\
 \end{split}
\end{equation}
where $A$, $B$, $C$, $Q_1$ and $Q_2$ are functions of $\epsilon(x)$, $\zeta(x)$ and $r_c(x)$ as given 
in equations \eqref{abc} and \eqref{q1q2}. We further expand the 
corrections to the perfect fluid stress tensor and current as 

\begin{equation}\label{covstfi}
 \begin{split}
  16 \pi G {\tilde \pi}_{\mu\nu} &= -2r_c^3~\sigma_{\mu\nu}+\frac{1}{e^2}\bigg[r_c^3\sum_{i =1}^7 S_i P_i\left(~n_\mu n_\nu  -\frac{1}{2}{\tilde P}_{\mu\nu}\right)\\
&+ r_c^3 \sum_{i = 1}^5 v_i \bigg(n_\mu [V_i]_\nu + n_\nu [V_i]_\mu \bigg) 
+  r_c^3 \sum_{i = 1}^2 t_i ~ [T_i]_{\mu\nu} \bigg] + {\cal O}\left(\frac{1}{e^4}\right)\\
\\
 16 \pi G  {\tilde J}^\mu_{diss} &= \frac{r_c^2}{e^2}\sum_{i=1}^7 S_i\left( a_i u^\mu + b_i n^\mu\right)+ \frac{r_c^2}{e^2} \sum_{i = 1}^5 c_i~ [V_i]^\mu
+ {\cal O}\left(\frac{1}{e^4}\right)\\
\\
\mu_{diss} & =\sum_{i = 1}^7 \delta\mu_i ~S_i + {\cal O}\left(\frac{1}{e^2}\right) 
 \end{split}
\end{equation}

Our results are given as follows. 

{\it Results for stress tensor:}
\begin{equation}\label{rrst}
 \begin{split}
 P_1 &= {\cal O}(\epsilon ^4),
~~~ P_2 = {\cal O}(\epsilon ^4),
~~~P_3=  {\cal O}(\epsilon^4),
~~~ P_4 = {\cal O}(\epsilon^3),
~~~ P_5 ={\cal O}(\epsilon^4),
 ~~~P_6 = {\cal O}(\epsilon^3)\\
v_1 &= {\cal O}(\epsilon^5),~~~
v_2 =  {\cal O}(\epsilon^5),~~~
v_3 =  {\cal O}(\epsilon^5),~~~
v_4 =  {\cal O}(\epsilon^5),~~~
v_5 ={\cal O}(\epsilon^4) \\
t_1&=  {\cal O}(\epsilon^4),~~~~t_2={\cal O}(\epsilon^4) \\
\end{split}
\end{equation}

{\it Results for current:}
\begin{equation}\label{taritfal}
 \begin{split}
a_1& =\epsilon \left[\frac{3}{7} \left(3-80~ \chi ^2\right)\right]    + {\cal O}(\epsilon^3), ~~~
b_1 =  \epsilon^2 ~\chi +{\cal O}(\epsilon^4)\\
a_2& =\epsilon \left(\frac{8~ \chi }{7}\right)    + {\cal O}(\epsilon^3), ~~~
b_2 =  -\epsilon^2\left[-\frac{1}{24}\right] +{\cal O}(\epsilon^4)\\
a_3& =  -\epsilon\left(\frac{18\chi }{7}\right)    + {\cal O}(\epsilon^3), ~~b_3 =  {\cal O}(\epsilon^4)\\
a_4& = -\left(\frac{160~\chi^2 }{7}\right)    + {\cal O}(\epsilon^2), ~~b_4 =  {\cal O}(\epsilon^4)\\
a_5& =\epsilon \left(\frac{9~ \chi }{7}\right)    + {\cal O}(\epsilon^3), ~~~b_5 =  {\cal O}(\epsilon^4)\\
a_6& = \left(\frac{80~\chi^2 }{7}\right)    + {\cal O}(\epsilon^2), ~~b_6 =  {\cal O}(\epsilon^4)\\
\\
c_1 &= \frac{\epsilon^2}{24}+{\cal O}(\epsilon^4) \\
c_2 &= \epsilon^2~\chi +{\cal O}(\epsilon^4)\\
c_3 &= {\cal O}(\epsilon^4)\\
c_4 &= {\cal O}(\epsilon^4)\\
c_5 &= \epsilon^3~\chi\left(\frac{-1 + 2\log(2)}{16}\right) + {\cal O}(\epsilon^4)
 \end{split}
\end{equation}

{\it Results for the correction to the Josephson equation:}
\begin{equation}\label{dasafal}
 \begin{split}
\delta\mu_1&=\epsilon \left[\frac{1}{14} \left(5-96~ \chi ^2\right)\right]    + {\cal O}(\epsilon^3)\\
\delta\mu_2&=\epsilon \left(\frac{3~ \chi }{7}\right)    + {\cal O}(\epsilon^3)\\
\delta\mu_3&= -\epsilon\left(\frac{5 \chi}{7} \right)+ {\cal O}(\epsilon^3)\\
\delta\mu_4&= -\left(\frac{32~ \chi^2}{7} \right) + {\cal O}(\epsilon^2)\\
\delta\mu_5&=\epsilon \left(\frac{5~ \chi }{14}\right)    + {\cal O}(\epsilon^3)\\
\delta\mu_6&= \left(\frac{16~ \chi^2}{7} \right) + {\cal O}(\epsilon^2)
 \end{split}
\end{equation}

In $\zeta\rightarrow0$ limit this derivative corrections to 
stress tensor, charge current and the  phase equation take the following
form
\begin{equation}\label{sunyazeta1}
 \begin{split}
 \lim_{\zeta\rightarrow0}\tilde \pi^{\mu\nu}&= -2r_c^3\sigma^{\mu\nu} +{\cal O}(\epsilon^3)\\
 \lim_{\zeta\rightarrow0}\tilde J^\mu_{diss}&=
 r_c^2\left\{ \alpha_1u^\mu\left[ P^{ab}\partial_a\zeta_b\right] 
+ \alpha_2 P^{\mu\nu}\partial_\nu \epsilon\right\}\\
 \lim_{\zeta\rightarrow0}\tilde \mu_{diss}&=  \alpha_3\left[ P^{ab}\partial_a\zeta_b\right] 
 \end{split}
\end{equation}
where 
$$\alpha_1 = \frac{9}{7} + {\cal O}(\epsilon^3),
~~\alpha_2 = \frac{\epsilon}{24}+ {\cal O}(\epsilon^4),
~~\alpha_3 = \frac{5}{14} + {\cal O}(\epsilon^3)$$

\subsection{Weyl Covariance of our bulk fields and boundary currents}

In this section we will demonstrate that our fluid dynamical solutions 
must, on general grounds, obey certain constraints that follow 
from the requirement of Weyl invariance. We then verify that our 
explicit solution does indeed obey these constraints, providing a nontrivial 
check on these solutions. 

All computations reported in this paper 
have been performed for superfluid motion on 
a flat boundary metric. However our final 
results must be the restriction to a flat boundary of results that 
apply in a general weakly curved space. The (boundary) generally covariant
version of our final bulk metric, stress tensor etc are all given simply
by promoting all derivatives to covariant derivatives (ambiguities in this 
procedure and boundary curvature terms all show up only at second order 
in the derivative expansion). 

Given these results in a general boundary spacetime, it follows 
on general grounds (see \cite{Bhattacharyya:2008ji}) that our bulk metric, 
gauge field and scalar fields must enjoy invariance under the following 
spacetime dependent Weyl transformations and coordinate redefinitions.
\begin{equation*}\label{wt}
\begin{split}
&\tilde r = r  \text{e}^{ \psi(v,x_i)},~~\tilde g_{\mu\nu} = g_{\mu\nu} \text{e}^{-2 \psi(v,x_i)}\\
&\tilde u_\mu = u_\mu \text{e}^{-\psi(v,x_i)},
~~\tilde n_\mu =n_\mu \text{e}^{- \psi(v,x_i)},~~\tilde\zeta = \zeta,~~\tilde\epsilon = \epsilon
\end{split}
\end{equation*}
Note that the Weyl transformed metric $\tilde g_{\mu\nu}$ is, in general, 
not flat even if the original metric is. Let us work in the special case 
that the original metric $g_{\mu\nu}$ is taken to be flat. The 
boundary connection with respect to  $\tilde g_{\mu\nu}$ is non zero and is given by
$${{\tilde \Gamma}^\sigma}_{\mu\nu} = - \left(\delta^\sigma_\mu \partial_\nu \psi +\delta^\sigma_\nu \partial_\mu \psi
- \eta_{\mu\nu}\partial^\sigma\psi\right)$$
The new frame covariant derivatives of $u_\mu$ and $n_\mu$ are given by 
\begin{equation*}
\begin{split}
\tilde\nabla_\mu \tilde u_\nu & = \text{e}^{- \psi}\left[\partial_\mu u_\nu + u_\mu\partial_\nu\psi 
- \eta_{\mu\nu}(u.\partial)\psi\right]\\
\tilde\nabla_\mu \tilde n_\nu & = \text{e}^{-\psi}\left[\partial_\mu n_\nu + n_\mu\partial_\nu\psi
 - \eta_{\mu\nu}(n.\partial)\psi\right]\\
\end{split}
\end{equation*}

Using these expressions one can deduce the transformation properties of the scalar,  vector and the tensor
 forms appearing in the bulk solution

\begin{equation}\label{poribartan}
 \begin{split}
  \tilde S_1 &= \text{e}^{\psi} S_1,~~\tilde S_2 = \text{e}^{\psi} S_2\\
 \tilde S_3 &= \text{e}^{\psi}\left[S_3 - (n.\partial)\psi\right],~~\tilde S_4 = \text{e}^{\psi}\left[S_4 - (u.\partial)\psi\right],\\
 \tilde S_5 &= \text{e}^{\psi}\left[S_5 -2 (n.\partial)\psi\right],~~\tilde S_6 = \text{e}^{\psi}\left[S_6 -2 (u.\partial)\psi\right],\\
\\
  \left[\tilde V_1\right]_\mu &=[ V_1]_\mu,~~\left[\tilde V_2\right]_\mu = [ V_2]_\mu\\
  \left[\tilde V_3\right]_\mu &= [V_3]_\mu +\tilde P^\nu_\mu\partial_\nu\psi,
~~ \left[\tilde V_4\right]_\mu = [V_4]_\mu -\tilde P^\nu_\mu\partial_\nu\psi,~~\left[\tilde V_5\right]_\mu = [ V_5]_\mu\\
\\
  \left[\tilde T_1\right]_{\mu\nu} &=\text{e}^{-\psi}~[ T_1]_{\mu\nu},~~
  \left[\tilde T_2\right]_{\mu\nu} =\text{e}^{-\psi}~[ T_2]_{\mu\nu}\
 \end{split}
\end{equation}
If we transform the gauge field and the metric from the new Weyl frame (the frame with tilde variables) 
to the old Weyl frame (the frame where the variables are denoted without tilde), the equilibrium solution 
itself generates some new terms due to the $r$ coordinate redefinition.
In the new frame the coordinates are $\tilde r = r  \text{e}^{ \psi(v,x_i)}$ and $\tilde x^\mu = x^\mu$.
This implies the following transformation rule for the differentials.
\begin{equation}\label{ruletrans}
\begin{split}
 d\tilde r &=  \text{e}^{ \psi(v,x_i)}\left(dr  + r dx^\mu\partial_\mu \psi\right)\\
d\tilde x^\mu &= d x^\mu\\
\tilde \partial_\mu &= \frac{\partial r}{\partial \tilde x_\mu}\partial_r + \partial_\mu\\
&=\left[ r\text{e}^{\psi}\partial_\mu\psi \right]\partial_r + \partial_\mu
\end{split}
\end{equation}

This induces the following transformations on 
gauge field
\begin{equation}\label{gaugewtr}
 \begin{split}
\tilde A &=
\frac{1}{\tilde r_c}\left[ H\left(\frac{\tilde r}{\tilde r_c}\right)(\tilde u.\tilde\partial) 
+ L\left(\frac{\tilde r}{\tilde r_c}\right)(\tilde n.\tilde\partial)\right]\\
&=- \frac{r\text{e}^{\psi}}{r_c}
 \left[H\left(\frac{r}{r_c}\right)( u.\partial \psi)
+ L\left(\frac{ r}{ r_c}\right)( n.\partial\psi)\right] \partial_r \\
&+\frac{1}{r_c}\left[ H\left(\frac{r}{r_c}\right)( u.\partial) 
+ L\left(\frac{ r}{ r_c}\right)( n.\partial)\right]\\
\\
&= -r\text{e}^{\psi}
 \left[H\left(r\right)( u.\partial \psi)
+ L\left(r\right)( n.\partial\psi)\right] \partial_r 
+r\left[ H\left(r\right)( u.\partial) 
+ L\left(r\right)( n.\partial)\right]\\
\end{split}
\end{equation}
In the last line we have used the scaling symmetry to set $r_c  =1$.

Similarly the equilibrium metric also transforms and  the nontrivial transformation is generated due
to the term $dx^\mu dr$. 
\begin{equation}\label{metwtr}
\begin{split}
 &-2 g\left(\frac{\tilde r}{\tilde r_c}\right) \tilde u_\mu dx^\mu d\tilde r
 =-2 g\left(\frac{ r}{ r_c}\right)  u_\mu ~dx^\mu ~\left(d r + r\partial_\nu \psi dx^\nu\right)\\
&= -2g\left(\frac{ r}{ r_c}\right) u_\mu dx^\mu dr 
-2r g\left(\frac{ r}{ r_c}\right) u_\mu u_\nu(u.\partial)\psi dx^\mu dx^\nu\\
& -r g\left(\frac{ r}{ r_c}\right)\bigg[ (u_\mu n_\nu + u_\nu n_\mu)  (n.\partial)\psi
+\left( u_\mu \tilde P^\sigma_\nu + u_\nu \tilde P^\sigma_\mu\right) \partial_\sigma\psi\bigg]
dx^\mu dx^\nu\\
\\
&= -2g\left(r\right) u_\mu dx^\mu dr 
-2r g\left(r\right) u_\mu u_\nu(u.\partial)\psi dx^\mu dx^\nu\\
& -r g\left(r\right)\bigg[ (u_\mu n_\nu + u_\nu n_\mu)  (n.\partial)\psi
+\left( u_\mu \tilde P^\sigma_\nu + u_\nu \tilde P^\sigma_\mu\right) \partial_\sigma\psi\bigg]
dx^\mu dx^\nu\\
&\text{Here also in the last step  the scaling symmetry is used to set $r_c$ =1 }
 \end{split}
\end{equation}
Combining these transformations we find the transformed metric, gauge field 
and scalar have the expected form (expected according
to \eqref{wt} ) together with some additional pieces 
that multiply a single derivative of $\psi$. The coefficients of these 
unwanted pieces themselves have no derivatives, and must vanish in order 
that our result respect Weyl invariance. This requirement imposes the following 
simple algebraic conditions on the fields in the metric, scalar and gauge field:
\begin{equation}\label{weylconstgauge}
 \begin{split}
&\delta A_3(r) +2~ \delta A_5(r) -r L(r) = 0\\
&\delta A_4(r) +2~\delta A_6(r) -rH(r) = 0\\
&\delta H_3(r) + 2~\delta H_5(r) = 0,~~~
\delta H_4(r) + 2~\delta H_6(r) = 0\\
&\delta L_3(r) + 2~\delta L_5(r) = 0,~~~
\delta L_4(r) + 2~\delta L_6(r) = 0\\
&\delta\phi_3(r) + 2~\delta\phi_5(r) = 0,~~~
\delta \phi_4(r) + 2~\delta\phi_6(r) = 0\\
&X_3(r) -X_4(r) =0\\
 \end{split}
\end{equation}

\begin{equation}\label{weylconstmet}
 \begin{split}
&\delta f_3(r) +2~ \delta f_5(r)  = 0\\
&\delta f_4(r) +2~\delta f_6(r) +2rg(r) = 0\\
&\delta J_3(r) + 2~\delta J_5(r) + rg(r) = 0,~~~
\delta J_4(r) + 2~\delta J_6(r) = 0\\
&\delta K_3(r) + 2~\delta K_5(r) = 0,~~~
\delta K_4(r) + 2~\delta K_6(r) = 0\\
&Y_3(r) -Y_4(r) - rg(r) = 0,~~~
W_3(r) - W_4(r) = 0\\
 \end{split}
\end{equation}

We also require that the stress tensor, charge current and Josephson equation 
in our model are invariant under Weyl transformations. As these boundary 
quantities are all independent of $r$, the redefinition of $r$ is irrelevant
to the study of Weyl transformations of these quantities. Using only the 
equations \eqref{poribartan} we find the following constraints on the 
coefficients in \eqref{covstfi} 
\begin{equation}\label{weylconststc}
 \begin{split}
&P_3 +2~ P_5  = P_4 +2~ P_6= 0\\
&v_3 - v_4 = 0\\
\\
&a_3 +2~ a_5  = a_4 +2~ a_6= 0\\
&b_3 +2~ b_5  = b_4 +2~b_6= 0\\
&c_3 - c_4 = 0\\
&\delta\mu_3 +2~ \delta\mu_5  = \delta\mu_4 +2~ \delta\mu_6= 0\\
 \end{split}
\end{equation}

The equations \eqref{weylconstmet}, \eqref{weylconstgauge} and 
\eqref{weylconststc} must apply to any consistent asymptotically AdS 
solution of gravitational equations. In particular these equations must 
apply to the results of this paper, and constitute a nontrivial consistency
check on our algebra. We have explicitly checked that the results 
of \S \ref{resmet} and \S \ref{resstress} obey these 
constraints, to the calculated order in $\epsilon$ and $\frac{1}{e^2}$.

\subsection{Entropy Current from Gravity}

Fluid flows obtained from the fluid gravity correspondence are automatically
equipped with families of local entropy currents of positive divergence. 
A particularly natural choice for this entropy current was presented in 
equation 3.11 of 
\cite{Bhattacharyya:2008xc}.
Using this formula for our solution we have computed the entropy current dual 
to our fluid flow. This entropy current has a piece at ${\cal O}(1)$ and
a piece at ${\cal O}(1/e^2)$, and takes the form 
\begin{equation}\label{entcurp}
 \begin{split}
  4 G J^\mu_s = r_c^3 u^\mu + \frac{ r_c^2}{e^2}\sum_{i=1}^7 S_i\left( \kappa^{(u)}_i u_\mu + \kappa^{(n)}_i n_\mu\right)
+ \frac{r_c^2}{e^2}\sum_{i = 1}^5 \kappa^{(v)}_i~ [V_i]_\mu + {\cal O}\left(\frac{1}{e^4}\right)\\
 \end{split}
\end{equation}
where 
\begin{equation}\label{expent}
 \begin{split}
\kappa^{(u)}_1& =\frac{1}{2}\epsilon \left[\frac{3}{7} \left(3-80~ \chi ^2\right)\right]    + {\cal O}(\epsilon^3), ~~~
\kappa^{(n)}_1 =  \frac{1}{2}\epsilon^2 ~\chi +{\cal O}(\epsilon^3)\\
\kappa^{(u)}_2& =\frac{1}{2}\epsilon \left(\frac{8~ \chi }{7}\right)    + {\cal O}(\epsilon^3), ~~~
\kappa^{(n)}_2 = -\frac{1}{2} \epsilon^2\left[-\frac{1}{24}\right] +{\cal O}(\epsilon^3)\\
\kappa^{(u)}_3& = - \frac{1}{2}\epsilon\left(\frac{18\chi }{7}\right)    + {\cal O}(\epsilon^3), ~~\kappa^{(n)}_3 =  {\cal O}(\epsilon^3)\\
\kappa^{(u)}_4& = -\frac{1}{2}\left(\frac{160~\chi^2 }{7}\right)    + {\cal O}(\epsilon^2), ~~\kappa^{(n)}_4 =  {\cal O}(\epsilon^3)\\
\kappa^{(u)}_5& =\frac{1}{2}\epsilon \left(\frac{9~ \chi }{7}\right)    + {\cal O}(\epsilon^3), ~~~\kappa^{(n)}_5 =  {\cal O}(\epsilon^3)\\
\kappa^{(u)}_6& =\frac{1}{2} \left(\frac{80~\chi^2 }{7}\right)    + {\cal O}(\epsilon^2), ~~\kappa^{(n)}_6 =  {\cal O}(\epsilon^3)\\
\end{split}
\end{equation}
\begin{equation}\label{expent2}
\begin{split}
\kappa^{(v)}_1 &= \frac{1}{2}\frac{\epsilon^2}{24}+{\cal O}(\epsilon^3) \\
\kappa^{(v)}_2 &= \frac{1}{2}\epsilon^2~\chi +{\cal O}(\epsilon^3)\\
\kappa^{(v)}_3 &= {\cal O}(\epsilon^3)\\
\kappa^{(v)}_4 &= {\cal O}(\epsilon^3)\\
\kappa^{(v)}_5 &= {\cal O}(\epsilon^3)
 \end{split}
\end{equation}

Quite remarkably this gravitational entropy current agrees exactly with 
the simple fluid dynamical current described in \eqref{mfent} (to the 
order at which we have done the calculation). 



Actually, all but the first two terms on the RHS of \eqref{mfent} are
${\cal O}(\epsilon^3)$ or higher. It follows that to ${\cal O}(\epsilon^2)$ 

\begin{equation}
J^\mu _S = s u^\mu - \frac{\mu}{T} {\tilde J}^\mu_{diss} + {\cal O}(\epsilon^3)
\end{equation}
We have checked that the gravitational entropy current, presented in this 
subsection, exactly agrees with this form to  this order.

\section{Transformation to standard frames and identification 
of the dissipative parameters} \label{resinDifframe}

The equations of gravitational super fluid dynamics, derived in the 
previous section are presented in a modified phase frame that is 
adapted to the expectation value of the operatore $\epsilon(x)$, and 
is not particularly natural from a fluid dynamical point of view. 

In this section we will transform our results to the 
$\mu_{diss}=0$ fluid frame and the transverse fluid frames. We will 
then compare these results with the general `theory' of 
dissipative dynamics presented in \S \ref{dissrev}. We will find perfect 
agreement with the general structures predicted in \S \ref{dissrev}, and 
so be able to read off the values of all 10 nonzero dissipative fluid 
parameters. 

To begin this subsection we first recall the structure of the boundary 
equations that emerged from our gravitational computations of the 
previous section. In the previous section our gravitational solutions
were parameterized by the eight fields $u^\mu(x)$, $\epsilon(x)$, $r_c(x)$ 
and $\zeta^\mu(x)$. In this section we will find it convenient to use the 
first of \eqref{mudef} to define a new field 
$\mu_0(x)$ and to eliminate $\epsilon(x)$ in favour of the new field 
$\mu_0(x)$. We will always suppose that this has been done in what follows. 

Our gravitational results for the stress tensor, current and superfluid phase 
are of the form 
\begin{equation} \begin{split} \label{gravstruc}
 T^{\mu\nu} &= \left[\rho_n(r_c,\mu_0,\xi_0) + P(r_c,\mu_0 , \xi_0)\right] u^\mu u^\nu 
+ P(r_c,\mu_0,\xi_0)\eta^{\mu\nu} + f(\xi_0,\mu_0,r_c)\xi^\mu_0\xi^\nu_0 + \tilde \pi^{\mu\nu}\\
J^\mu &= q_n(r_c,\xi_0,\mu_0) u^\mu - f(r_c,\xi_0,\mu_0) \xi_0^\mu + \tilde J^\mu_{diss}\\
\xi^\mu &
 = -(\mu_0 + \mu_{diss})u^\mu + r_c\zeta^\mu \\
\end{split}
\end{equation}
where the functions $ \rho_n$, $P$ etc are the thermodynamical functions 
derived in \S \ref{static}. Our results are presented in a modified 
phase frame. We remind the reader that this means that we have presented 
our answers in terms of a 
new auxiliary phase field $\xi_0^\mu$ and its modulus $\xi_0$ defined as 
\begin{equation} \begin{split} \label{xinotdef}
\xi_0^\mu &= -\mu_0 u^\mu +r_c \zeta^\mu\\
\xi_0 &= \sqrt{\mu_0^2 -r_c^2\zeta^2}\\
\end{split}
\end{equation}
Note that $\xi_0^\mu$ is not equal to the phase field $\xi^\mu$ (because 
$\mu_0 \neq u\xi$ ). Instead the relationship between these two fields
is given by 
\begin{equation} \begin{split} \label {xirel}
\xi^\mu &= \xi_0^\mu - \mu_{diss}u^\mu\\
\xi & = \xi_0 - \mu_{diss}\frac{\mu_0}{\xi_0}\\
\end{split}
\end{equation}

In this section we wish to transform our gravitational results into 
the $\mu_{diss}=0$ fluid frame and the transverse fluid frame.
As we have explained above, \eqref{gravstruc} is in a modified phase 
frame not a fluid frame. This means that \eqref{gravstruc} 
differs from the form \eqref{ltconst} in two ways. 
First thermodynamical functions, like the pressure, in 
\eqref{gravstruc}, are functions of $(\mu_0, r_c, \xi_0)$. According to the
form specified in \eqref{ltconst} the choice and definition of two of the three 
thermodynamical variables- e.g. the chemical potential and temperature - 
is upto the user, and can reasonably chosen in a fluid frame 
 to be $\mu_o$ and $r_c$. If we work in a fluid frame, however, the 
third variable has to be $\xi$, the magnitude of the phase field. 
As we have seen from \eqref{xirel} $\xi_0 \neq \xi$. 

The second way in which \eqref{gravstruc} differs from the from 
the standard fluid frame form \eqref{ltconst}
is that the fourth term in the expression for $T^{\mu\nu}$ in \eqref{gravstruc}
is proportional to $\xi_0^\mu \xi_0^\nu$. Similarly the second term in the 
expression for $J^\mu_{diss}$ in \eqref{gravstruc} is proportional to $\xi_0^\mu$. 
The corresponding terms in \eqref{ltconst}, however, are proportional to 
$\xi^\mu \xi^\nu$ and $\xi^\mu$ respectively. 

These discrepancies are easily cured. In order to move to a fluid 
frame, all one needs to do is substitute the expressions
for $\xi_0$ and $\xi_0^\mu$ as functions of $\xi$ and $\xi^\mu$ 
(see \eqref{xinotdef}) into \eqref{gravstruc}. That is we must perform 
a {\it prescribed} field redefinition on $\xi^\mu_0$ to take it to the 
field $\xi^\mu$. In addition we are free to perform any additional 
field redefinitions
\begin{equation} \begin{split} \label{fd}
\mu_0&=\mu+\delta \mu\\
r_c&= {\tilde r_c} + \delta r_c\\
u^\mu&= {\tilde u}^\mu + \delta u^\mu
\end{split}
\end{equation}
in order to transform to any particular fluid dynamical frame. Of course 
$\delta \mu$, $\delta r_c$ and $\delta u^\mu$ above are all necessarily of 
first or higher order in derivatives.
 
We will now describe how these general ideas can be used in practice 
to transform our gravitational results into the $\mu_{diss}=0$ frame and the 
transverse frames respectively.

\subsection{Transformation to the $\mu_{diss}=0$ frame}

In order to transform to the $\mu_{diss}=0$ frame we must take 
$\delta \mu=-\mu_{diss}$ in \eqref{fd}. We can immediately work out 
the effect of this field redefinition, combined with the change of variables 
from $\xi^\mu_0$ to $\xi^\mu$, on the current and stress tensor. We find
\begin{equation}\label{gravframecur}
 \begin{split}
  J^\mu &= q_n(r_c,\xi_0,\mu_0) u^\mu - f(r_c,\xi_0,\mu_0) \xi_0^\mu + \tilde J^\mu_{diss}\\
&=q_n(r_c,\xi,\mu) u^\mu - f(r_c,\xi,\mu) \xi^\mu \\
&+( d q_n + \mu_0 df) u^\mu  + \mu_{diss} f(r_c,\xi_0,\mu_0)u^\mu
-df~\zeta^\mu + \tilde J^\mu_{diss}\\
 \end{split}
\end{equation}
and 
\begin{equation}\label{gravframstress}
 \begin{split}
 T^{\mu\nu} &= \left[\rho_n(r_c,\mu_0,\xi_0) + P(r_c,\mu_0 , \xi_0)\right] u^\mu u^\nu 
+ P(r_c,\mu_0,\xi_0)\eta^{\mu\nu} + f(\xi_0,\mu_0,r_c)\xi^\mu_0\xi^\nu_0 + \tilde \pi^{\mu\nu}\\
&= \left[\rho_n(r_c,\mu,\xi) + P(r_c,\mu , \xi)\right] u^\mu u^\nu 
+ P(r_c,\mu,\xi)\eta^{\mu\nu} +f(r_c,\xi,\mu)\xi^\mu\xi^\nu \\
&+\left[d\rho +\mu_0^2 ~df\right] u^\mu u^\nu 
+
\mu_{diss} f(r_c,\xi_0 ,\mu_0)(u^\mu\xi_0^\nu+u^\nu\xi_0^\mu) \\
&+df~\zeta^\mu\zeta^\nu+dP~ P^{\mu\nu} + \tilde \pi^{\mu\nu}\\
 \end{split}
\end{equation}
where $f(r_c,\xi,\mu) = \frac{q_s(r_c,\xi,\mu)}{\xi} = \frac{\rho_s(r_c,\xi,\mu)}{\xi^2}$ and the operation $`d'$ acting on any function of $r_c$, $\xi$ and 
$\mu$ is given by 
$$ dA =-\mu_{diss}\frac{\mu_0}{\xi_0}\left[ \frac{\partial A(r_c,\xi,\mu)}{\partial\xi}\right]
-\mu_{diss}\left[ \frac{\partial A(r_c,\xi,\mu)}{\partial\mu}\right]$$
The gravitational stress tensor and charge current have now been recast 
into the fluid dynamical form with  
\begin{equation}\begin{split}\label{npnj}
\pi^{\mu\nu}&=\left[d\rho +\mu_0^2 ~df\right] u^\mu u^\nu 
+
\mu_{diss} f(r_c,\xi_0 ,\mu_0)(u^\mu\xi_0^\nu+u^\nu\xi_0^\mu) 
+df~\zeta^\mu\zeta^\nu+dP~ P^{\mu\nu} + \tilde \pi^{\mu\nu} \\
J^\mu_{diss}&=( d q_n + \mu_0 df) u^\mu  + \mu_{diss} f(r_c,\xi_0,\mu_0)u^\mu
-df~\zeta^\mu + \tilde J^\mu_{diss}\\
\mu_{diss}&=0
\end{split}
\end{equation}
While ${\tilde \pi}^{\mu\nu}$ was orthogonal to the velocity field $u^\mu$, 
the same is not true of $\pi^{\mu\nu}$ in \eqref{npnj}. 
In order to enforce the transversality of $\pi^{\mu\nu}$ (a defining condition
for the $\mu_{diss}=0$ frame), we must now redefine $r_c$ and $u^\mu$. 
The determination of $\delta r_c$ and $\delta u^\mu$ is particularly 
simple at leading order in $\frac{1}{e^2}$ (the order to which our 
gravitational results have been obtained). Recall that, to leading (unit) order
in $\frac{1}{e^2}$, $\tilde\pi^{\mu\nu}$ is already transverse to $u_\mu$. 
The piece of $T^{\mu\nu}$ that is not transverse to $u_\mu$ starts out at 
${\cal O}\left(\frac{1}{e^2}\right)$. It follows that the $\delta r_c$ and $\delta u^\mu$ 
must both be of order ${\cal O}(\frac{1}{e^2})$. It is important 
to implement this field redefinition only in the ${\cal O}(1)$ part of 
the equilibrium or perfect fluid part of $T^{\mu\nu}$ (as we are keeping 
track of the final answer only ${\cal O}(\frac{1}{e^2})$ and only to first 
order in fluid derivatives). This piece and its transformations are 
given by 
\begin{equation}\label{zeerr}
\begin{split}
& T^{\mu\nu}|_{\text{equilibrium at ${\cal O}\left(\frac{1}{e^2}\right)^0 $}}\\
 = &r_c^4(4 u^\mu u^\nu + \eta^{\mu\nu}) \\
= &\tilde r_c^4(4 \tilde u^\mu \tilde u^\nu + \eta^{\mu\nu}) 
+ 4\tilde r_c^4\left[\frac{\delta r_c}{r_c}\left(3u^\mu u^\nu + P^{\mu\nu}\right)+
\left(\delta u^\mu u^\nu + \delta u^\nu u^\mu\right)\right]
\end{split}
\end{equation}
Adding \eqref{zeerr} to \eqref{gravframstress} gives the complete 
expression of  $\pi^{\mu\nu}$ after this variable redefinition. 
$\pi^{\mu\nu}$ is transverse if we choose 
 \begin{equation}\label{ffssf}
 \begin{split}
  12 r_c^3\delta r_c &=-\left(d\rho_n  + \mu_0^2~df - 2 \mu_) f~\mu_{diss}\right)\\
4r_c\delta u^\mu& =\left( df - f~ \mu_{diss}\right)\zeta^\mu 
 \end{split}
\end{equation}
(as can be verified by dotting $\pi^{\mu\nu}$ with $u^\mu u^\nu$ and with 
$u^\mu n^\nu$). With $\delta r_c$ and $\delta u^\mu$ chosen as above
\begin{equation}\label{beginpi}
\pi^{\mu\nu} = df~\zeta^\mu\zeta^\nu+\left(dP + 4 r_c^3 \delta r_c\right)~ P^{\mu\nu}
+ \tilde \pi^{\mu\nu}
\end{equation}
(all expressions involving $\delta u^\mu$ cancel). 

Note that the corrections to $J^\mu_{diss}$ that arise from the field 
redefinitions of $r_c$ and $u^\mu$ are all at ${\cal O}(\frac{1}{e^2})$. 
As we have not kept track of the current to this order in our gravitational 
computation, we will ignore all such terms; $J^\mu_{diss}$ continues to be 
given by \eqref{npnj}.

Note that the final expression \eqref{beginpi} for $\pi^{\mu\nu}$ 
depends on $\delta r_c$ but not on $\delta u^\mu$ (this is, of course, true
only to first order in the derivative expansion). While $\delta r_c$ is 
given by \eqref{ffssf}, it is equally well given by solving the equation
\begin{equation}\label{andr}
df ~\zeta^2 + 3 \left(dP + 4 r_c^3 \delta r_c\right) = 0
\end{equation}
That \eqref{andr} must be true follows from the trace of 
 \eqref{beginpi} and 
the observation that 
$\pi^{\mu\nu}$ and $\tilde \pi^{\mu\nu}$ are both traceless. That 
the expressions for $\delta r_c$ obtained from \eqref{andr} and
\eqref{ffssf} agree follows from acting the operator $`d'$ on the 
equation 
$$(-\rho_n + 3P -f \xi^2 )=0$$
which itself is simply an assertion of the tracelessness of the perfect 
fluid stress tensor 
\footnote{
\begin{equation}\label{footeq}
 \begin{split}
0&=d(-\rho_n + 3 P - f\xi^2)
=- d\rho_n + 3 dP - \xi^2 df - 2\xi f d\xi\\
&=- d\rho_n + 3 dP - \xi^2 df + 2 \xi f\left[ \frac{\mu}{\xi}\mu_{diss}\right]
=\zeta^2 df + 3 dP -[d\rho_n + \mu^2 df - 2 \mu f~ \mu_{diss}]\\
&= \zeta^2 df + 3 dP  + 12 r_c^3\delta r_c
 \end{split}
\end{equation}
}.

Inserting the expression for $\delta r_c$ from \eqref{andr} into 
the expression for $\pi^{\mu\nu}$ in \eqref{beginpi} we conclude that
the gravitational expressions for $\pi^{\mu\nu}$ and $J^\mu_{diss}$ in the 
$\mu_{diss}=0$ frame are given by 
\begin{equation}\label{finalexct}
 \begin{split}
  J^\mu_{diss} &= ( d q_n + \mu_0~ df) u^\mu  + \mu_{diss} f(\xi_0,\mu_0)u^\mu
-df~\zeta^\mu + \tilde J^\mu_{diss}\\
\pi^{\mu\nu} &= \frac{df~ \zeta^2}{3}\left(2n^\mu n^\nu -\tilde P^{\mu\nu}\right)
+{\tilde \pi}^{\mu\nu}
 \end{split}
\end{equation}

In the rest of this subsection we will now compare this gravitational result 
to the `standard form' for the dissipative corrections to the stress 
tensor and charge current predicted by our `theory' of Weyl Invariant 
fluid dynamics in \S \ref{dissrev}. This standard form was given in 
\eqref{nfstcr} and is reproduced here for the convenience of the reader

\begin{equation}\label{nfstcr2}
\begin{split}
 \pi^{\mu\nu}&= T^3\bigg[\left[ P_{{\cal S}} {\cal S} + P_b S_b + P_{w}S_w\right]
\left(n^\mu n^\nu - \frac{ P^{\mu\nu}}{3}\right)\\
 &~~~~~~~~+E_a \left(V_a^\mu n^\nu + V_a^\nu n^\mu\right) 
+E_b \left(V_b^\mu n^\nu + V_b^\nu n^\mu\right) \\
&~~~~~~~~+\tau {\cal T}^{\mu\nu}\bigg]\\
\\
J^\mu_{diss}&= T^2\bigg[\left[ Q_{{\cal S}} {\cal S} + Q_b S_b  + Q_{w}S_w\right]
u^\mu\\
&~~~~~~~~~+\left[ R_{\cal S} {\cal S} + R_b S_b + R_{w}S_w\right]n^\mu\\
&~~~~~~~~~+ C_a V_a^\mu + C_b V_b^\mu\bigg]
\end{split}
 \end{equation}

where 
$${\cal S} = (u. \partial)\left(\frac{\mu}{T}\right), ~S_b = (n. \partial)\left(\frac{\mu}{T}\right),
~S_w  =  n^\mu n^\nu \sigma_{\mu\nu}=\frac{2 S_4 - S_6}{3}$$ and 
$$V_a^\mu  = \tilde P^{\mu\nu}\partial_\nu \left(\frac{\mu}{T}\right),
~V_b^\mu = \tilde P^{\mu\alpha}\sigma_{\alpha\beta}n^\beta $$
The standard form described above uses a basis of independent first derivative
terms that differs from our choice of independent first derivative 
terms in our gravitational solution. In order to compare the two forms 
we simply reexpress $S_a$, $S_b$ and $V_a$ in terms of the independent 
first derivative forms employed in the gravity calculation, utilizing the 
equations of motion. We get
\begin{equation}\label{exgegfg}
 \begin{split}
{\cal S} &= \epsilon^2 \frac{2 \chi^2}{7} (2 S_4 - S_6) + \epsilon^3\left( \frac{1 + 48 \chi^2}{56}\right) S_1
+ \epsilon^3 \frac{\chi}{14}S_2 \\
&- \epsilon^3 \frac{\chi}{56}(2 S_3 - S_5) + {\cal O}(\epsilon^4)\\
\\
S_b &= - \epsilon^2 \chi S_1 - \frac{\epsilon^2}{24} S_2 + {\cal O}(\epsilon^4)\\
V_a^\mu &= - \epsilon^2 \chi V_2^\mu - \frac{\epsilon^2}{24} V_1^\mu + {\cal O}(\epsilon^4)\\
 \end{split}
\end{equation}

We then compare the expressions so obtained with our gravitational 
results \eqref{finalexct}. We find that the two expressions match perfectly 
if we choose
\begin{equation}\label{lastcoef}
 \begin{split}
16\pi G Q_{\cal S} &=\frac{\pi}{e^2}\left[ -\frac{208}{ \epsilon^2} + {\cal O}(\epsilon)^0\right] 
+ {\cal O}\left(\frac{1}{e^4}\right)\\
16\pi G Q_b &=\frac{\pi}{e^2}\left[ -240 \frac{\chi}{ \epsilon} + {\cal O}(\epsilon)^0\right] 
+ {\cal O}\left(\frac{1}{e^4}\right)\\
16\pi G Q_w &=\frac{\pi^2}{e^2}\left[240 \chi^2 + {\cal O}(\epsilon)\right] 
+ {\cal O}\left(\frac{1}{e^4}\right)\\
\\
16\pi G R_{\cal S} &= \frac{\pi}{e^2}\left[-240 \frac{\chi}{\epsilon} + {\cal O}(\epsilon)^0\right] 
+ {\cal O}\left(\frac{1}{e^4}\right)\\
16\pi G R_b &=\frac{\pi}{e^2}\left[- 1 - 288 \chi^2 + {\cal O}(\epsilon)\right] 
+ {\cal O}\left(\frac{1}{e^4}\right)\\
16\pi G R_w &= \frac{\pi^2}{e^2}\left[ 288~\epsilon \chi^3+ {\cal O}(\epsilon)^2\right] 
+ {\cal O}\left(\frac{1}{e^4}\right)\\
\\
16\pi G P_{\cal S} &= \frac{\pi^2}{e^2}\left[240\chi^2+ {\cal O}(\epsilon)\right] 
+ {\cal O}\left(\frac{1}{e^4}\right)\\
16\pi G P_b &=\frac{\pi^2}{e^2}\left[ 288~\epsilon \chi^3+ {\cal O}(\epsilon)^2\right] 
+ {\cal O}\left(\frac{1}{e^4}\right)\\
16\pi G P_w &= -3\pi^3 +\frac{\pi^3}{e^2}\left[-6
-\left(\frac{1}{4}-3\chi^2- 288~ \chi^4\right)\epsilon^2+ {\cal O}(\epsilon)^3\right] 
+ {\cal O}\left(\frac{1}{e^4}\right)\\
\end{split}
\end{equation}

\begin{equation}\label{lastcoef4}
\begin{split}
16\pi G C_a &= \frac{\pi}{e^2}\left[-1 + {\cal O}(\epsilon)\right] 
+ {\cal O}\left(\frac{1}{e^4}\right) \\
16\pi G C_b &= \frac{\pi^2}{e^2}\left[\epsilon^3 \chi \left(\frac{-1 + \log(4)}{8}\right) + {\cal O}(\epsilon)^4\right] 
+ {\cal O}\left(\frac{1}{e^4}\right) \\
\\
16\pi G E_a &= {\cal O}(\epsilon) ^3
+ {\cal O}\left(\frac{1}{e^4}\right)\\
16\pi G E_b &= -2\pi^3 +\frac{\pi^3}{e^2}\left[-4 +\left(\frac{1}{6}-2\chi^2\right)\epsilon^2
 + {\cal O}(\epsilon)^4\right]
+ {\cal O}\left(\frac{1}{e^4}\right)\\
16\pi G \tau & =-2\pi^3 +\frac{\pi^3}{e^2}\left[-4 +\left(\frac{1}{6}-2\chi^2\right)\epsilon^2
 + {\cal O}(\epsilon)^4\right]
+ {\cal O}\left(\frac{1}{e^4}\right)\\
\end{split}
\end{equation}

Note that $Q_w=P_a$, $R_w=P_b$, $R_a=Q_b$ and $C_b=E_a$, so that our 
results obey the Onsager relations listed in \S \ref{mdweyl}. It may 
also be verified that these results obey all the positivity constraints 
required on general grounds in \S \ref{mdweyl}.

\subsection{Transformation to the Transverse Frame}

Following the discussion of the previous subsection, it is also possible 
to cast our gravitational results into the transverse fluid frame. 
As mentioned in \S \ref{ep}, the basis of first derivative quantities 
with non zero coefficients most suitable for this frame are 

\begin{equation}
\begin{split}
&S_a= \partial_\mu\left(\frac{q_s}{\xi}\xi^\mu\right);~~
S_b=\left( n^{\mu}\partial_{\mu}\right)\left(\frac{\mu}{T}\right);~~
S_{w}=  n^\mu n^\nu \sigma_{\mu\nu};\\
&V_a^{\mu} = \tilde{P}^{\sigma\mu}\partial_{\sigma}\left( \frac{\mu}{T}\right);
~~V_b^{\mu} = \tilde{P}^{\alpha\mu}\sigma_{\alpha\nu} n^\nu;
~~{\cal T}^{\mu\nu} = \tilde{P}^{\alpha\mu}\tilde{P}^{\beta\nu}\sigma_{\mu\nu}\\
\end{split}
\end{equation}

These quantities may be expressed in term of the quantities (defined in 
\eqref{covsc}) used for the gravity calculation as follows

\begin{equation}
  \begin{split}
S_a &=\frac{1}{16\pi G~e^2} \left[\epsilon^2 \frac{8 \chi^2}{7} (2 S_4 - S_6) 
+ \epsilon^3\left( \frac{-5 + 96 \chi^2}{28}\right) S_1
- \epsilon^3 \frac{3\chi}{14}S_2 + \epsilon^3 \frac{5\chi}{28}(2 S_3 - S_5) 
+ {\cal O}(\epsilon^4) \right];\\
S_b &= - \epsilon^2 \chi S_1 - \frac{\epsilon^2}{24} S_2 + {\cal O}(\epsilon^4);~~
S_w = \frac{2 S_4 -S_6}{3};\\
V_a^\mu &= - \epsilon^2 \chi V_2^\mu - \frac{\epsilon^2}{24} V_1^\mu + {\cal O}(\epsilon^4);
~~V_b^\mu  =\frac{V_5^\mu}{2} ;~~{\cal T}^{\mu\nu} = \frac{T_1^{\mu\nu}}{2} \\
 \end{split}
\end{equation}
Let us rewrite the first derivative corrections to 
charge current obtained from gravity (given in \eqref{covstfi} and \eqref{taritfal})
in the following schematic form
\begin{equation}
 \tilde J^\mu _{diss} = \left(\frac{1}{16 \pi G}\right) 
\frac{r_c^2}{e^2}\left( \tilde j_u u^\mu +\tilde  j_n n^\mu + \sum_i c_i [V_i]^{\mu} \right).
\end{equation}
In the gravity solution the  stress tensor ( $\tilde \pi^{\mu\nu}$) is given as the following
(see \eqref{rrst}).
$$16\pi G~ \tilde \pi^{\mu\nu} = \left[-2 r_c^3 \sigma^{\mu\nu} + {\cal O}(\epsilon^3)\right] 
+ {\cal O}\left(\frac{1}{e^4}\right)$$

Then using the procedure outlined in the previous section and 
in \S \eqref{transf} we can compute first derivative 
corrections to stress tensor, charge current and chemical potential
in the transverse frame (which we denote by $\pi^{(T)}_{\mu \nu}$,
$(J^{(T)}_{\text{diss}})_{\mu}$ and $\mu^{(T)}_{\text{diss}}$ respectively).
We find 
\begin{equation}\label{transfrTJmu}
 \begin{split}
  16 \pi G~(\pi^{(T)})^{\mu \nu} &= \frac{6}{5}\left[\tilde j_u 
- 4  \mu_{diss}\right]  \zeta^2\left(n^\mu n^\nu -\frac{P^{\mu\nu}}{3}\right)
+{\tilde \pi}^{\mu\nu},\\
16\pi G~(J^{(T)}_{\text{diss}})^{\mu} &=
\frac{r_c^2}{e^2}\left( \left(\tilde{j}_n -\frac{6\zeta}{5}\left[\tilde j_u 
- 4  \mu_{diss}\right] \right)n^{\mu} + \sum_i c_i [V_i]^{\mu} \right),\\
\mu^{(T)}_{\text{diss}} &=  \frac{1}{10}\left[\tilde j_u - 14 ~\mu_{diss}\right],
 \end{split}
\end{equation}
where $\mu_{diss}$ is the first derivative correction to the chemical potential
obtained from gravity given in \eqref{covstfi} and in \eqref{dasafal}. 
Here the formulas presented in \eqref{transfrTJmu} are valid only at the leading order in $\epsilon$
for each independent data at one  derivative order.

As in the previous subsection, we then consider the expected standard form 
fluid expression which is given by
\begin{equation}\label{weystrcr2}
\begin{split}
 \pi^{\mu\nu}&= T^3\bigg[\left( P_a S_a + P_b S_b + P_w S_w
 \right)
\left( n_{\mu} n_{\nu} -\frac{P_{\mu\nu}}{3}\right)\\
 &~~~~~~~~+E_a \left(V_a^\mu n^\nu + V_a^\nu n^\mu\right) 
+E_b \left(V_b^\mu n^\nu + V_b^\nu n^\mu\right)
 \\
&~~~~~~~~+\tau {\cal T}^{\mu\nu} \bigg]\\
J^\mu_{diss}&=
T^2 \bigg[\left( R_a S_a + R_b S_b + R_w S_w  \right)n^\mu\\
&~~~~~~~~~+ C_a V_a^\mu  + C_b V_b^\mu
 \bigg]\\
\mu_{diss}& = -\left[ Q_a S_a + Q_b S_b + Q_w S_w \right] 
\end{split}
 \end{equation}

Just as in the previous section the gravity result after 
the frame transformation \eqref{transfrTJmu} perfectly fits into the
above form provided we identify
{\it equation 8.22 }
\begin{equation}\label{coeftrans}
  \begin{split}
Q_a & =16\pi G \left(\frac{ e^2}{\pi^3}\right)\left[-\frac{52}{25\epsilon^2} + {\cal O}(\epsilon)^0\right] 
+ {\cal O}\left(\frac{1}{e^2}\right)^0\\
R_a &= \frac{1}{\pi}\left[- \frac{24\chi}{25\epsilon} + {\cal O}(\epsilon)^0\right] 
+ {\cal O}\left(\frac{1}{e^2}\right)\\
P_a& =  \left[\frac{24}{25 }\chi^2+ {\cal O}(\epsilon)\right] 
+ {\cal O}\left(\frac{1}{e^2}\right)\\
Q_b& =\frac{1}{\pi}\left[- \frac{24\chi}{25\epsilon} + {\cal O}(\epsilon)^0\right] 
+ {\cal O}\left(\frac{1}{e^2}\right)\\
R_b &=\frac{\pi}{(16\pi G)e^2}\left[\left(- 1 - \frac{288}{25} \chi^2 \right) + {\cal O}(\epsilon)\right] 
+ {\cal O}\left(\frac{1}{e^4}\right)\\
P_b& =\frac{\pi^2}{(16\pi G)e^2}\left[ \frac{288}{25 }~\epsilon \chi^3+ {\cal O}(\epsilon)^2\right] 
+ {\cal O}\left(\frac{1}{e^4}\right)\\
Q_w& =\left[ \frac{24}{25} \chi^2 + {\cal O}(\epsilon)\right] 
+ {\cal O}\left(\frac{1}{e^4}\right)\\
R_w &= \frac{\pi^2}{(16\pi G)e^2}\left[ \frac{288}{25}~\epsilon \chi^3+ {\cal O}(\epsilon)^2\right] 
+ {\cal O}\left(\frac{1}{e^4}\right)\\
P_w &= -\frac{3\pi^3}{16\pi G}
 +\frac{\pi^3}{(16\pi G)e^2}\left[-6
-\left(\frac{1}{4}-3\chi^2+\frac{288}{25 }~ \chi^4\right)\epsilon^2+ {\cal O}(\epsilon)^3\right] \\
&~~~~~+ {\cal O}\left(\frac{1}{e^4}\right)\\
\end{split}
\end{equation}
and
\begin{equation}\label{coeftrans4}
  \begin{split}
16\pi G~E_a ~~ &= \frac{\pi^2}{e^2}\left[{\cal O}(\epsilon) ^3\right] + {\cal O}\left(\frac{1}{e^4}\right) \\
16\pi G ~C_a ~~ &=\frac{\pi}{e^2}\left[ -1 + {\cal O}(\epsilon)\right] 
+ {\cal O}\left(\frac{1}{e^4}\right) \\
16\pi G~E_b &= -2\pi^3 +\frac{\pi^3}{e^2}\left[-4 +\left(\frac{1}{6}-2\chi^2\right)\epsilon^2
 + {\cal O}(\epsilon)^4\right]
+ {\cal O}\left(\frac{1}{e^4}\right)\\
16\pi G~C_b &=\frac{\pi^2}{e^2}\left[ \epsilon^3 \chi \left(\frac{-1 + \log(4)}{8}\right) 
+ {\cal O}(\epsilon)^4\right]+ {\cal O}\left(\frac{1}{e^4}\right) \\
16\pi G~\tau& =-2\pi^3 +\frac{\pi^3}{e^2}\left[-4 +\left(\frac{1}{6}-2\chi^2\right)\epsilon^2
 + {\cal O}(\epsilon)^4\right]
+ {\cal O}\left(\frac{1}{e^4}\right)\\
\end{split}
\end{equation}
Note that in this transverse frame also we have
 $Q_w=P_a$, $R_w=P_b$, $R_a=Q_b$ and $C_b=E_a$, which constitutes
the expected Onsager relations. All the positivity constraints 
given in \S \ref{weyl} are also obeyed in this frame.

In $\zeta\rightarrow0$ limit  derivative corrections to 
stress tensor, charge current and the  phase equation in transeverse frame take the following
form
\begin{equation}\label{sunyazeta2}
 \begin{split}
 \lim_{\zeta\rightarrow0} \left[\pi^{(T)}\right]^{\mu\nu}&=T^3\beta_1\sigma^{\mu\nu}\\
 \lim_{\zeta\rightarrow0} \left[J_{diss}^{(T)}\right]^\mu&=
T^2\beta_2P^{\mu\nu}\partial_\nu\left(\frac{\mu}{T}\right)\\
 \lim_{\zeta\rightarrow0}\mu^{(T)}_{diss}&=  \beta_3\partial_\mu\left(\frac{q_s}{\xi}\xi^\mu \right)
 \end{split}
\end{equation}
where 
\begin{equation}\label{coeffhh}
 \begin{split}
 \beta_1 &= -2\pi^3 +\frac{\pi^3}{e^2}\left[-4 +\frac{\epsilon^2}{6}
 + {\cal O}(\epsilon)^4\right]
+ {\cal O}\left(\frac{1}{e^4}\right)\\
\beta_2 &=\frac{1}{16\pi G}\left[ -\frac{\pi}{e^2}+ {\cal O}(\epsilon^3)\right] + {\cal O}\left(\frac{1}{e^4}\right)\\
\beta_3 &=16\pi G \left(\frac{ e^2}{\pi^3T^3}\right)\left[-\frac{52}{25\epsilon^2} + {\cal O}(\epsilon)^0\right] 
+ {\cal O}\left(\frac{1}{e^2}\right)^0\\
 \end{split}
\end{equation}

\subsection{Transformation to the Landau-Lifshitz-Clark-Putterman Frame}

Our results may also be transformed to the Landau-Lifshitz-Clark-Putterman
frame. We have not fully worked through the complicated algebra needed 
for this process. However we have verified that this transformation yields 
$$16\pi G~Q_2^{(s)}=\frac{\pi^3}{e^2}\left[-\frac{6}{7}\zeta^3 + {\cal O}(\epsilon^4)\right] 
+ {\cal O}\left(\frac{1}{e^4}\right) $$
Note in particular that $Q_2^{(s)}$ is not zero, predicted by the formulas
of Clark and Putterman.

\section{Discussion}

Our paper suggests many directions of for future research. 
In this paper we have studied the bulk dual to Einstein-Maxwell systems in the 
absence
of a bulk Chern Simons term . A Chern Simon's term has already been shown to 
lead to yield qualitatively new contributions to charged fluid dynamics even 
in the absence of charged scalar fields \cite{Bhattacharyya:2007vs, 
Erdmenger:2008rm, Banerjee:2008th, Son:2009tf} . It seems certain to contribute 
new and interesting terms to charged superfluid dynamics as well. We leave a 
study of the effect of this Chern Simons term on holographic 
superfluid flows to future work.

Relatedly, we should emphasize that the framework for dissipative fluid 
dynamics presented in this paper crucially assumes that the 
superfluid entropy current takes the canonical form discussed in 
\S \ref{CEntrop}. While this assumption turned out to be true of the 
gravitational system studied in this paper we do not know of any physical 
reason for it always to be true. In particular we think it very likely that 
the entropy current will be modified in superfluids dual to a gravitational 
system with a nonzero Chern Simon's term as described in the previous 
paragraph. It would be very interesting to understand the general rules for 
the entropy current at first order in the derivative expansion, and 
thereby to construct the most general framework for first order dissipative
superfluid dynamics.

It would of course be interesting to study dissipative corrections to 
holographic superfluid dynamics away from the rather strange limit (of very 
large $e$ and perturbatively in the superfluid condensate) employed in this 
paper. Infact it is interesting that a continuation of the model studied in 
this paper to a particular order one value of $e$ and supplemented with 
a particular Chern Simons term and a more complicated scalar potential 
(however with squared mass =-4 as in this paper)
 is a consistent truncation of IIB supergravity on $AdS_5 \times S^5$ 
\cite{Bhattacharyya:2010yg}. It would be fascinating to be able to 
work out the map from gravity to fluid dynamics in this system, though
that might require numerical work. It would also be interesting to 
follow the investigation of \cite{Bhattacharyya:2007vs} to study whether 
hairy black holes in global $AdS_5$ (see e.g. \cite{Basu:2010uz, 
Bhattacharyya:2010yg, Sonner:2009fk}) can be thought of as stationary 
superfluid flows.

The task described in the previous paragraphs
is hampered by the lack of explicit analytic 
solutions for equilibrium configurations. One could imagine circumventing this 
lack in two different ways. First it would be very interesting to 
attempt an abstract analysis of dissipative fluid flows, along the lines 
described in Appendix \ref{SgravThermo} (following Sonner and Withers)
for equilibrium flows. The aim of such an abstract analysis could be 
to prove that the gravitational entropy current 
always agrees with the canonical form presented in \S \ref{dissrev}
in the absence of bulk Chern Simons terms, and to prove that all 
Onsager relations are obeyed. This would constitute a proof that 
gravitational super fluid dynamics falls within the framework of 
dissipative superfluid dynamics described in \S \ref{dissrev} above (in the 
absence of bulk Chern Simons terms). A second possible approach to the 
same problem is numerical; a clever use of numerical techniques should 
permit the numerical generation of the gravitational solutions dual 
to arbitrary first order superfluid flows at finite values of $e$.

\acknowledgments

{\bf Acknowledgments}

This paper has benefited enormously from comments by C. Herzog and 
A. Yarom on the first version of our draft.
We would like to acknowledge useful discussions and correspondences with 
P. Basu, K. Damle, B. Ezuthachan, R. Gopakumar, S. Gutti, R. Loganayagam, C. Herzog, 
M. Rangamani, A. Saha, A. Sen, D.T. Son, V. Tripathi, S. Wadia and 
A. Yarom. The work 
of S.M. was supported in part by a Swarnajayanti Fellowship.
We would all also like to acknowledge our debt to the people of India for 
their generous and steady support to research in the basic sciences.

\appendix

\section{More about the canonical entropy current}\label{entropy}

\subsection{Frame invariance of the fluid frame entropy current}
In this subsection we will demonstrate that the entropy current 
 $$J^\mu_s = s u^\mu - \frac{\mu}{T} J^\mu_{diss} - \frac{u_\nu \pi^{\mu\nu}}{T}$$
is frame invariant. 

Let us perform the following change of variables (we work accurately 
only to first order in the derivatives) 
\begin{equation}\label{nutannaam}
 \begin{split}
u^\mu &= {\bar u}^\mu + \delta u^\mu\\
T &= \bar T + \delta T\\
\mu &= \bar\mu + \delta\mu
 \end{split}
\end{equation}
Under this redefinition $J^\mu_{diss}$ and $\pi^{\mu\nu}$ change according to 
the formulae given in \eqref{channge}.
\begin{equation}\label{chrpt}
 \begin{split}
 \delta \pi^{\mu\nu} & =
( u^\mu \delta u^\nu +  u^\nu \delta u^\mu) (P+\rho_n) +
 u^\mu  u^\nu d(P+\rho_n) + \frac{\xi^\mu \xi^\nu}{\xi^2} d(\rho_s)
+ \eta^{\mu\nu} dP\\
\delta J^{\mu}_{diss}&=q_n \delta  u^\mu + d q_n  u^\mu - d q_s
\frac{\xi^\mu}{\xi} \\
 \end{split}
\end{equation}
In the new frame the entropy current is given by

\begin{equation}\label{sahaj}
\begin{split}
\bar J^\mu_s =& \bar s(\bar\mu,\bar T)\bar u^\mu - \frac{\mu}{T} \left(J^\mu_{diss} + \delta J^{\mu}_{diss}\right) - \frac{u_\nu \left(\pi^{\mu\nu}+ \delta \pi^{\mu\nu}\right)}{T}\\
=& J^\mu_s- ds~  u^\mu- s\delta u^\mu - \frac{\mu}{T}\left(q_n \delta  u^\mu + d q_n  u^\mu - d q_s
\frac{\xi^\mu}{\xi}\right)\\
&+ \frac{P + \rho_n}{T}\delta u^\mu + \frac{u^\mu}{T} d(P + \rho_n) - \frac{\mu d\rho_s}{T\xi}\frac{\xi^\mu}{\xi} - \frac{u^\mu}{T} dP\\
=&J^\mu_s + u^\mu \left(\frac{d\rho_n}{T} -ds - \frac{\mu}{T}dq_n\right) + \delta u^\mu \left(\frac{P + \rho_n}{T} - s - \frac{\mu}{T} q_n\right)\\
&- \frac{\xi^\mu}{\xi}\frac{\mu}{T}\left( \frac{ d\rho_s}{\xi}- d q_s\right)
\end{split}
\end{equation}
In the above expression each of the terms inside the bracket vanishes because of thermodynamic identities
 (recall also that $\xi = \mu_s$). \footnote{
Recall that
the symbol $d$ denotes the change of a quantity under a frame change field 
redefinition, and that the microscopically defined field $\xi^\mu(x)$ is taken
to be the same in all frames so that, in particular, $d\xi =0$. }

It follows that 
 $$\bar J^\mu_s = J^\mu_s$$

\subsection{The Divergence of the fluid frame entropy current}

Using the relation that $\mu_s =\xi =\sqrt{-\xi_\mu\xi^\mu}$  the stress tensor 
current and the phase as given in
\eqref{ltconst} can be rewritten as
 \begin{equation} \label{ltconstr}
\begin{split}
T^{\mu\nu}&=(\rho_n+P) u^\mu u^\nu + P \eta^{\mu\nu}
+\frac{\rho_s}{\mu_s^2} \xi^\mu \xi^\nu + \pi^{\mu\nu} \\
J^\mu &=q_n u^\mu - \frac{q_s}{\mu_s} \xi^{\mu}+ J^\mu_{diss}\\
u^\mu \xi_\mu&=\mu + \mu_{diss}\\
\end{split}
\end{equation}
We will find use, below, for the following thermodynamical relationships:
\begin{equation}\label{usetherm}
 \begin{split}
  s&= \frac{\rho_n + P - \mu q_n}{T}\\
d\rho_n &= \mu dq_n + Tds - q_s d\mu_s
\end{split}
\end{equation}
Now divergence of $su^\mu$ gives the following  expression.
\begin{equation}\label{divergent}
 \begin{split}
  &\partial_\mu \left[s u^\mu\right] \\
=& u.\partial s + s (\partial.u)\\
= &\frac{1}{T} \left[u.\partial \rho_n + \left(\rho_n + P\right)(\partial.u)\right]
-\frac{\mu}{T}\left[u.\partial q_n + q_n (\partial.u)\right] + \frac{q_s}{T} u.\partial \mu_s
 \end{split}
\end{equation}
Using conservation of stress tensor (the equation to be used is $ u_\nu \partial_\mu T^{\mu\nu} =0$) one can evaluate the expression $\left[u.\partial \rho_n + \left(\rho_n + P\right)(\partial.u)\right] $.
\begin{equation}\label{strcon}
 \begin{split}
&u.\partial \rho_n + \left(\rho_n + P\right)(\partial.u)\\
=& \frac{\rho_s}{\mu_s^2}\left[(u.\xi) (\partial.\xi)  + u^\nu(\xi.\partial)\xi_\nu \right]+(u.\xi) (\xi.\partial)\left(\frac{\rho_s}{\mu_s^2}\right) + u_\nu \partial_\mu \pi^{\mu\nu}\\
=&\frac{\rho_s}{\mu_s^2}\left[\mu(\partial.\xi)  + u^\nu(\xi.\partial)\xi_\nu \right]+\mu (\xi.\partial)\left(\frac{\rho_s}{\mu_s^2}\right)
+\mu_{diss}\partial_\mu\left(\frac{\rho_s}{\mu_s^2}\xi^\mu\right) +u_\nu \partial_\mu \pi^{\mu\nu}\\
=&\frac{\rho_s}{\mu_s^2}\left[\mu(\partial.\xi)  - \mu_s(\xi.\partial)\mu_s\right]+\mu (\xi.\partial)\left(\frac{\rho_s}{\mu_s^2}\right)
+\mu_{diss}\partial_\mu\left(\frac{\rho_s}{\mu_s^2}\xi^\mu\right) +u_\nu \partial_\mu \pi^{\mu\nu}
\end{split}
\end{equation}
In the last line we have used the following identity derived from the curl-free condition.
\begin{equation}\label{curlfreeid}
 \begin{split}
 & u^\mu\left(\partial_\mu \xi_\nu - \partial_\nu\xi_\mu\right)\xi^\nu =0\\
\Rightarrow &u^\nu (\xi.\partial)\xi_\nu = \frac{1}{2}(u.\partial )(\xi_\mu \xi^\mu) = -\mu_s (u.\partial)\mu_s
\end{split}
\end{equation}
Similarly using the current conservation equation one can evaluate the expression $\left[u.\partial q_n + q_n (\partial.u)\right]$
\begin{equation}\label{curcon}
 \begin{split}
u.\partial q_n + q_n (\partial.u) = \frac{q_s}{\mu_s} (\partial.\xi) + (\xi.\partial)\left(\frac{q_s}{\mu_s}\right) - \partial.J_{diss}
\end{split}
\end{equation}
Adding all these equations one finds the following expression for the divergence of the entropy current
\begin{equation}\label{divergentf0}
 \begin{split}
  &\partial_\mu \left[s u^\mu\right] \\
=& \left(\frac{\rho_s}{\mu_s^2} - \frac{q_s}{\mu_s}\right)\left[\frac{\partial.\xi - \mu_s (\xi.\partial)\mu_s}{T}\right]
+\frac{\mu}{T}(\xi.\partial)\left(\frac{\rho_s}{\mu_s^2} - \frac{q_s}{\mu_s}\right)\\
& + \frac{1}{T}\left[\mu_{diss}\partial_\mu\left(\frac{\rho_s}{\mu_s^2}\xi^\mu\right) +u_\nu \partial_\mu \pi^{\mu\nu}\right] + \frac{\mu}{T}\left(\partial.J_{diss}\right)
 \end{split}
\end{equation}

The first line is zero if $\rho_s = \mu_s q_s$ and the second line can be written as
\begin{equation}\label{divergentf}
 \begin{split}
  &\partial_\mu \left[s u^\mu - \frac{\mu}{T} J^\mu_{diss} - \frac{u_\nu \pi^{\mu\nu}}{T}\right]\\
 =& -\partial_\mu\left[\frac{u_\nu}{T} \right]\pi^{\mu\nu}  - \partial_\mu\left[\frac{\mu}{T} \right]J_{diss}^{\mu} +\frac{\mu_{diss}}{T}\partial_\mu\left(\frac{\rho_s}{\mu_s^2}\xi^\mu\right)
 \end{split}
\end{equation}

\subsection{Direct Thermodynamical determination of the modified 
phase entropy curent and its divergence} \label{cb}

As we have explained in the main text, a modified phase fluid description 
works in terms of a gradient vector $\xi_o^\mu$ defined by 
$$\xi^\mu = \xi_0^\mu -\mu_{diss} u^\mu$$
where $\xi_0^\mu = -\mu u^\mu+ \zeta^\mu=-\mu u^\mu +\zeta n^\mu $

In a modified phase frame, the stress tensor and charge current are taken 
to have the form
\begin{equation}
 \begin{split}
  T^{\mu\nu} &= (\rho_s + P) u^\mu u^\nu + P \eta^{\mu\nu} + f \xi_0^\mu \xi_0^\nu + \tilde \pi^{\mu\nu}\\
J^\mu &= q_s u^\mu - f\xi_0^\mu  + \tilde J^\mu_{diss}\\
 \end{split}
\end{equation}
where  $ f = \frac{\rho_s}{\xi_0^2} = \frac{q_s}{\xi_0}$, and all 
thermodynamical quantities are functions of $T, \mu, \xi_0$. 

We will find use, below, for the following thermodynamical relationships:
\begin{equation}\label{usethermp}
 \begin{split}
  s&= \frac{\rho_n + P - \mu q_n}{T}\\
d\rho_n &= \mu dq_n + Tds - f \xi_0 d\xi_0
\end{split}
\end{equation}
\
We now compute the divergence of the vector 
$su^\mu$ 
\begin{equation}\label{divergent}
 \begin{split}
  &\partial_\mu \left[s u^\mu\right] \\
=& u.\partial s + s (\partial.u)\\
= &\frac{1}{T} \left[u.\partial \rho_n + \left(\rho_n + P\right)(\partial.u)\right]
-\frac{\mu}{T}\left[u.\partial q_n + q_n (\partial.u)\right] + \frac{q_s}{T} u.\partial \mu_s
 \end{split}
\end{equation}
Using conservation of stress tensor (the equation to be used 
is $ u_\nu \partial_\mu T^{\mu\nu} =0$) one can evaluate the expression 
$\left[u.\partial \rho_n + \left(\rho_n + P\right)(\partial.u)\right] $ as
\begin{equation}\label{strcon}
 \begin{split}
&u.\partial \rho_n + \left(\rho_n + P\right)(\partial.u)\\
=& \mu \partial_\mu (f \xi_0^\mu) + f u_\nu (\xi_0.\partial)\xi_0^\nu + u_\nu\partial_\mu \tilde \pi^{\mu\nu}\\
\end{split}
\end{equation}

Similarly using the current conservation equation one can evaluate the 
expression $\left[u.\partial q_n + q_n (\partial.u)\right]$
\begin{equation}\label{curconp}
 \begin{split}
u.\partial q_n + q_n (\partial.u) =\partial_\mu (f \xi_0^\mu) - \partial_\mu \tilde J^\mu_{diss}
\end{split}
\end{equation}
The fact that the vector $\xi_0$ is curl free gives the following identity
\begin{equation}\label{curlfreep}
 \begin{split}
\xi_0 (u.\partial)\xi_0 + u_\mu (\xi_0.\partial)\xi^\mu_0 & = - \mu_{diss} \xi_0^\nu(u.\partial)u_\nu
 + (\zeta.\partial)\mu_{diss}\\
\end{split}
\end{equation}

Using all these equations one finds the following expression for the 
divergence of the entropy current
\begin{equation}\label{divergentf0}
 \begin{split}
\partial_\mu \left[s u^\mu\right] 
=& \frac{\mu}{T}\partial_\mu \tilde J^\mu_{diss} + \frac{u_\nu \partial_\mu \tilde \pi^{\mu\nu}}{T}
-\frac{f}{T} \mu_{diss} \xi^\nu_0 (u.\partial)u_\nu - \frac{f}{T}(\zeta.\partial)\mu_{diss}
 \end{split}
\end{equation}

It follows that if we define an entropy current by 
\begin{equation}\label{pentdef}
 \tilde J^\mu_S =s(\xi_0) u^\mu - \frac{\mu}{T} \tilde J^\mu_{diss} - \frac{u_\nu \tilde \pi^{\mu\nu}}{T}
+\frac{f}{T}\mu_{diss} \zeta^\mu
\end{equation}
then its divergence is given by
\begin{equation}\label{divergentfpa}
 \begin{split}
\partial_\mu\tilde J^\mu_S
 =& -\partial_\mu\left[\frac{u_\nu}{T} \right]\tilde \pi^{\mu\nu} 
 - \partial_\mu\left[\frac{\mu}{T} \right]\tilde J_{diss}^{\mu} 
+\mu_{diss}P^{\mu\nu}\partial_\mu\left(\frac{f\zeta_\nu}{T}\right)
 \end{split}
\end{equation}

In deriving this expression we have used the following equations.
\begin{equation}\label{simplify}
 \begin{split}
 &\frac{f}{T}\left[ \xi_0 (u.\partial)\xi_0 + u_\mu (\xi_0.\partial)\xi^\mu_0\right] \\
& = \frac{f}{T}\left[- \mu_{diss} \xi_0^\nu(u.\partial)u_\nu
 + (\zeta.\partial)\mu_{diss}\right]\\
 =&- \partial_\mu \left(\frac{f}{T} \zeta^\mu \mu_{diss}\right)
 + \mu_{diss}\left[\partial_\mu \left(\frac{f~\zeta^\mu}{T} \right)
 - \frac{f}{T} \xi_0^\nu(u.\partial)u_\nu\right] \\
 =&- \partial_\mu \left(\frac{f}{T} \zeta^\mu \mu_{diss}\right)
 + \mu_{diss}\left[\partial_\mu \left(\frac{f~\zeta^\mu}{T} \right)
 - \frac{f}{T} \zeta^\nu(u.\partial)u_\nu\right] \\
=&- \partial_\mu \left(\frac{f}{T} \zeta^\mu \mu_{diss}\right)
 + \mu_{diss}\left[\partial_\mu \left(\frac{f~\zeta^\mu}{T} \right)
 + \frac{f}{T} u^\nu(u.\partial)\zeta_\nu\right] \\
=&- \partial_\mu \left(\frac{f}{T} \zeta^\mu \mu_{diss}\right)
 + \mu_{diss}P^{\mu\nu}\left[\partial_\mu \left(\frac{f~\zeta_\nu}{T} \right)\right] \\
 \end{split}
\end{equation}

\subsubsection{Fluid frame entropy current transformed to the modified 
phase frame}\label{cc}

In this subsection we will check that the modified frame entropy current 
\eqref{pentdef} is precisely the current what we get by transforming the 
general fluid frame entropy current to modified phase variables. 
As we have explained in the main text, the relation between dissipative 
parameters in a general fluid frame and the corresponding modified 
phase frame is given by 
\begin{equation}\label{eqtocopy}
 \begin{split}
  \pi^{\mu\nu} &= \tilde \pi^{\mu\nu} +d(\rho_n + P) u^\mu u^\nu + dP\eta^{\mu\nu}
+ df \xi_0^\mu\xi_0^\nu \\
&+ f ~\mu_{diss} (u^\mu\xi_0^\nu + u^\nu \xi_0^\mu)\\
J^\mu_{diss} &= \tilde J^\mu_{diss} + dq_n u^\mu - d f \xi_0^\mu -f~ \mu_{diss} u^\mu
 \end{split}
\end{equation}
where $`d'$ of any function denotes
$$dA = A(\xi_0) - A(\xi) = -\mu_{diss}\left(\frac{\mu}{\xi_0}\right)\frac{\partial A}{\partial \xi}$$

The transformation of the canonical fluid frame entropy current into
the modified phase frame is given by 
\begin{equation}\label{transent}
 \begin{split}
  J_S^\mu\\
 =& s(\xi) u^\mu - \frac{u_\nu\pi^{\mu\nu}}{T} - \frac{\mu}{T} J^\mu_{diss}\\
          =& s(\xi_0) u^\mu - \frac{u_\nu\tilde\pi^{\mu\nu}}{T} - \frac{\mu}{T}\tilde J^\mu_{diss}
-\left(\frac{\mu}{T} d q_n - \frac{d\rho_n}{T} + ds\right)u^\mu + \frac{f~\mu_{diss}}{T} \xi_0^\mu\\
&= s(\xi_0) u^\mu - \frac{u_\nu\tilde\pi^{\mu\nu}}{T} - \frac{\mu}{T}\tilde J^\mu_{diss}
-\frac{f\xi_0}{T} ~d\xi_0 u^\mu + \frac{f~\mu_{diss}}{T} \xi_0^\mu\\
&= s(\xi_0) u^\mu - \frac{u_\nu\tilde\pi^{\mu\nu}}{T} - \frac{\mu}{T}\tilde J^\mu_{diss}
+\frac{f ~\mu_{diss}}{T}\left(\mu u^\mu+ \xi_0^\mu\right)\\
&= s(\xi_0) u^\mu - \frac{u_\nu\tilde\pi^{\mu\nu}}{T} - \frac{\mu}{T}\tilde J^\mu_{diss}
+\frac{f ~\mu_{diss}}{T}\zeta^\mu
 \end{split}
\end{equation}
In going from the second to third line we have used the following 
thermodynamic relation
$$d\rho_n = \mu dq_n + T ds - f\xi d\xi$$
In going from the third line to fourth line we have used the fact that
$$d\xi_0 = -\mu_{diss}\left(\frac{\mu}{\xi_0}\right)$$

Using equation \eqref{transent} and \eqref{pentdef} it follows that 
 $$\tilde J^\mu_S = J^\mu_S$$

\section{Details on Superfluid Thermodynamics from Gravity} \label{SgravThermo}

In this section we will use the bulk equations of motion to demonstrate 
that our solution obeys two Gibbs Duhem type relations. We will also 
review the demonstration of \cite{Sonner:2010yx} that the onshell action 
for the solution is the negative of its pressure, and that the equation 
for the infinitesimal variation of this pressure obeys a thermodynamical 
first law.

\subsection{Bulk equations, symmetries, conventions etc}

The Lagrangian we study is given by
\begin{equation}\label{genlag}
\begin{split}
 {\cal L} &=\frac{\sqrt{g}}{16\pi G}\bigg( R +12\\
&~~~~~~~~~~~~- \frac{1}{e^2}\left[V_1(\phi \phi^*) F_{ab}F^{ab} 
+ V_2(\phi\phi^*) D_a\phi \bar{D}^a \phi^* + V_3(\phi\phi^*)\right]\bigg)
\end{split}
\end{equation}
where $V_3(x)$ is taken to vanish at its minimum, 
$D_a = \nabla - i A_a$ and $\bar D_a = \nabla + i A_a$ and all the potentials 
$V_1,~V_2,~V_3$ are real.

The bulk equations that follow by extremizing this Lagrangian are given by
\begin{equation}\label{eqm}
 \begin{split}
&\bar D_a \left[V_2(\phi\phi^*)\bar D^a \phi^*\right] = \frac{\partial V_1}{\partial\phi} F_{ab}F^{ab}
  +\frac{\partial V_2}{\partial\phi}D_a\phi \bar{D}^a \phi^*
 + \frac{\partial V_3}{\partial\phi}\\
& D_a \left[V_2(\phi\phi^*) D^a \phi\right] = \frac{\partial V_1}{\partial\phi^*} F_{ab}F^{ab}
  +\frac{\partial V_2}{\partial\phi^*}D_a\phi \bar{D}^a \phi^*
 + \frac{\partial V_3}{\partial\phi^*}\\
&\nabla_a\left[V_1(\phi\phi^*)F^{ab}\right] =
- \frac{i V_2(\phi\phi^*)}{4}\left[\phi{\bar D}^b \phi^* - \phi^* D^b\phi\right]\\
&R_{ab} - \frac{1}{2} R g_{ab} - 6g_{ab} = {\cal T}_{ab}
 ={\cal T}^\text{em}_{ab} + {\cal T}^\text{sc}_{ab} \\
&\text{where}\\
&{\cal T}^\text{em}_{ab}  = -\frac{2V_1(\phi\phi^*)}{e^2} \left(F_{ac}{F^c}_b 
+ \frac{1}{4}g_{ab}F^{c_1 c_2}F_{c_1c_2}\right)\\
&{\cal T}^\text{sc}_{ab}  = \frac{V_2(\phi\phi^*)}{e^2}
 \left(\frac{D_a\phi{\bar D}_b\phi^* + { \bar D}_a\phi^* D_b\phi}{2}
- \frac{1}{2}g_{ab}D_c\phi{\bar D}^c\phi^*\right)-\left[\frac{V_3(\phi\phi^*)}{2 e^2}\right]g_{ab}\\
 \end{split}
\end{equation}
Here small Latin letters denote the bulk coordinates $(\{r,v,x,y,z\})$
 and Greek letters denote the boundary coordinates $(\{v,x,y,z\})$. Greek indices are
 lowered or raised using the metric $\eta$.

In this section we study solutions that preserve translational invariance 
in the field theory 
directions. By scaling our solution we can always ensure that its horizon 
is located at $r=1$; we make this choice in what follows. 
Our spacetime has a $d$ parameter set of killing vectors that 
generate translations in the field theory directions. We now identify two 
special killing vectors among this set. Let  $k^a = \{0,u^\mu\}$, 
be the unique killing vector that is null on the horizon. Further let 
$l^a = \{0,n^\mu\}$.
\newline
Where $n^\mu$ is defined as
$$n^\mu =\frac{ P^{\mu\nu}\xi_\nu}{\sqrt{\xi_\nu P^{\mu\nu}\xi_\nu}},~~~
\xi_\nu = \lim_{r\rightarrow\infty}g_{\mu a}A^a$$
where $A = A^a\partial_a$ is the gauge field and $P^{\mu\nu} = \eta^{\mu\nu} + u^\mu u^\nu$.
\newline
Below we will identify the vector $u^\mu$ with the normal fluid four velocity, 
while the vector $n^\mu$ points in the spatial direction of the superfluid 
velocity, in a frame in which the normal fluid is at rest.

As we have explained, the most general ansatz for our solution is given by 
\begin{equation}\label{ansatz}
 \begin{split}
  ds^2 &= -2 g(r)~u_\mu dx^\mu dr  - f(r)~ u_\mu u_\nu dx^\mu dx^\nu 
+j(r)\left( u_\mu n_\nu + u_\nu n_\mu\right) dx^\mu dx^\nu \\
&+k(r) n_\mu n_\nu  dx^\mu dx^\nu 
+r^2\tilde P_{\mu\nu}dx^\mu dx^\nu \\
A &= A^r(r)\partial_r + H(r)~u^\mu\partial_\mu + L(r)~ n^\mu\partial_\mu\\
 \end{split}
\end{equation}
where
$$\tilde P_{\mu\nu}=\eta_{\mu\nu} +u_\mu u_\nu - n_\mu n_\nu $$

As we have mentioned above, we work in a gauge in which the scalar field is 
set to be real. This will be consistent with the equation of motion
if the difference between the first and the second equation in \eqref{eqm} 
vanishes when $\phi^*(r)$ is set
 equal to $\phi(r)$. One can check that when all potentials are real and there are no explicit
 dependence on the boundary coordinates the scalar field can be consistently chosen to be real
 if $A^r(r)$ is set identically to zero. Consequently, on our solutions, 
$$A^r = 0~~\text{and}~~\phi(r) = \phi^*(r)$$

We now make the following convenient coordinate choices. Let $v$ represent 
a flat normalized boundary coordinate in in the direction of constant vector 
$u^\mu$ and let $x$ represent the coordinate is in the direction of constant 
vector $n^\mu$. In other words, in our new coordinate system,  
$$k^a = \{0,1,0,0,\cdots\}~~\text{and}~~l^a = \{0,0,1,0,\cdots\}$$
where we list a vector by $(k^r, k^v, k^x \ldots )$. 
In this coordinate system the metric and the gauge field take the form 
\begin{equation}\label{ansatzcc}
 \begin{split}
  ds^2 &= 2g(r)~dv~ dr  - f(r)dv^2 -2j(r)~dv~dx +k(r)~ dx^2
+r^2 \left(\sum dy_i^2\right)\\
A^r& = 0,~~A^v = H(r),~~A^x = L(r)
 \end{split}
\end{equation}
 
Recall that in our coordinates the horizon occurs at $r_H = 1$. Because 
the horizon is a null surface, the norm of the normal vector vanishes. As the 
oneform dual to the normal is simply $dr$ we conclude that 
\begin{equation}\label{grr}
g^{rr}(r = 1) =\left[ \frac{j^2(r ) + f(r) k(r)}{g^2(r) k(r)}\right]_{r=1} = 0 
\end{equation}
Recall also that, by definition, $k^a$ is a generator of the horizon, and 
so should be the dual to the normal $dr$ at the horizon. This requires 
that $g_{v v}$ vanish at the horizon, so we conclude that $f(r= 1) = 0$. 
It follows then from \eqref{grr} that $j(r=1) = 0$. None of the other 
functions that appear in equation \eqref{ansatzcc} are restricted at the 
horizon, except that they should be regular.
\newline
At infinity we will impose the condition that our space is 
asymptotically AdS which implies that 
\begin{equation}\label{bndexp2}
 \begin{split}
  \lim_{r\rightarrow\infty}g(r)&= 1 + \frac{g_4}{r^4} + {\cal O}(r^{-6})\\
  \lim_{r\rightarrow\infty}f(r)&=r^2 + \frac{f_2}{r^2} + {\cal O}(r^{-4})\\
  \lim_{r\rightarrow\infty}k(r)&= r^2 + \frac{k_2}{r^2} + {\cal O}(r^{-4})\\
  \lim_{r\rightarrow\infty}j(r)&=  \frac{j_2}{r^2} + {\cal O}(r^{-4})\\
\end{split}
\end{equation}
and that 
\footnote{The existence of such solution will impose some constraints 
on the potential 
such that the contribution of gauge field and
 matter field to the bulk stress tensor vanishes sufficiently rapidly as $r\rightarrow\infty$. 
}
\begin{equation}\label{gagecg}
 \begin{split}
&  \lim_{r\rightarrow\infty}A_v = \xi_v = ~\text{finite}\\
 & \lim_{r\rightarrow\infty}A_x = \xi_x = ~\text{finite}\\
 & \lim_{r\rightarrow\infty}\left[\sqrt{g}~V_1(\phi\phi^*) F^{vr}\right]=  ~\text{finite}\\
 & \lim_{r\rightarrow\infty}\left[\sqrt{g}~V_1(\phi\phi^*) F^{xr}\right]=  ~\text{finite}\\
 \end{split}
\end{equation}

The first two equations in  \eqref{gagecg} may be regarded as definitions 
of the constants $\xi_v$ and $\xi_x$. The last two equations assert the 
finiteness of the boundary charge density and the boundary 
current in the $x$ direction. 

\subsection{Derivation of the Smarr Gibbs Duhem Relations}

\subsubsection{Basic Strategy}

It is a geometrical fact (see e.g. section 5.3 of \cite{Townsend:1997ku}) that if $\zeta^a$
is a killing vector 
$$R_{ab}\zeta^b = \nabla_b\nabla_a\zeta^b.$$
It follows that, in our background,  
\begin{equation}\label{lemma}
 \begin{split}
 R^a_b\zeta^b&= \nabla_b\nabla^a\zeta^b \\
&= -  \nabla_b\nabla^b\zeta^a \\
& = -\frac{1}{\sqrt{g}}\partial_b\left[\sqrt{g}g^{bc_1}g^{ac_2}\nabla_{c_1}\zeta_{c_2}\right]\\
&=  -\frac{1}{\sqrt{g}}\partial_r\left[\sqrt{g}g^{rc_1}g^{ac_2}\nabla_{c_1}\zeta_{c_2}\right]
 \end{split}
\end{equation}
where we have used the Killing equation in the second line, the fact that 
$\nabla^b\zeta^a$ is antisymmetric in the third line  and 
the fact that all functions in our metric and gauge field depend only on $r$
in the last line.  Plugging in the explicit form of our metric and making 
different choices for the free indices we find 
\begin{equation}\label{identity}
 \begin{split}
  \sqrt{g} R^v_v &= -\partial_r\left[\sqrt{g}g^{ra}g^{vb}\nabla_ak_b\right] 
= -\partial_r\left(\frac{r^2[k(r)f'(r) + j(r)j'(r)]}{2 g(r)\sqrt{k(r)}}\right)\\
  \sqrt{g} R^x_x &= -\partial_r\left[\sqrt{g}g^{ra}g^{xb}\nabla_al_b\right] 
= -\partial_r\left(\frac{r^2[f(r)k'(r) + j(r)j'(r)]}{2 g(r)\sqrt{k(r)}}\right)\\
  \sqrt{g} R^{y}_{y} &= -\partial_r\left[\sqrt{g}g^{ra}g^{yb}\nabla_aq_b\right] 
= -\partial_r\left(\frac{r[k(r)f(r) + j^2(r)]}{ g(r)\sqrt{k(r)}}\right)\\
  \sqrt{g} R^x_v &= -\partial_r\left[\sqrt{g}g^{ra}g^{vb}\nabla_al_b\right] 
= -\partial_r\left(\frac{r^2[j(r)f'(r) - f(r)j'(r)]}{2 g(r)\sqrt{k(r)}}\right)\\
\text{where} ~~~q^a &= \{0,0,0,1,0,\cdots\}\\
 \end{split}
\end{equation}

The RHS of each of \eqref{identity} are total derivatives. Our basic strategy
is to find appropriate linear combinations of the four relations above so that
the LHS of these equations also reduce to total derivatives onshell (i.e. upon
using Einstein's equations). We will find three such linear combinations 
and so deduce the vanishing of three different expressions, each of which 
is a total derivative. We will then integrate these three equations to obtain 
three relations between quantities at infinity (i.e. conserved charges, 
currents, and superfluid velocities) and quantities at the horizon 
(entropy and temperature). Using these relationships we will be able to 
identify the vector $u^\mu$ (defined in terms of the generator of the 
horizon) with the normal velocity of the fluid, and also deduce two 
distinct Smarr Gibbs Duhem relations. In the next subsubsection we 
present the algebra involved in identifying the relevant total derivatives. 
In the subsequent subsection we will integrate these total derivatives to 
deduce physical conclusions. 

\subsubsection{Total derivatives that vanish onshell}

We use Einstein equations to simplify the LHS of the equations above.
\begin{equation}\label{einsami}
R^a_b = {\cal T}^a_b -\left( \frac{{\cal T}^a_a}{3}\right)\delta^a_b - 4\delta^a_b 
\end{equation}
Here ${\cal T}_{ab}$ is the bulk stress tensor.

 In order to do this we will now 
present and manipulate explicit expressions for the relevant bulk stress
tensors.

The contribution of electromagnetic and matter fields to the stress tensor
above is given by 
\begin{equation}\label{step1}
 \begin{split}
{\cal S}^\text{em}_{ab} &=
{\cal T}^\text{em}_{ab}  -\left( \frac{[{\cal T}^\text{em}]^a_a}{3}\right)g_{ab}\\
&=
 -\frac{2V_1(\phi\phi^*)}{e^2} \left(F_{ac}{F^c}_b 
+ \frac{g_{ab}}{6}F^{c_1 c_2}F_{c_1c_2}\right)\\
{\cal S}^\text{sc}_{ab} &=
{\cal T}^\text{sc}_{ab}  -\left( \frac{[{\cal T}^\text{sc}]^a_a}{3}\right)g_{ab}\\
&=
\left[\frac{V_2(\phi)}{e^2}\right]\left(\partial_a\phi\partial_b\phi +A_a A_b \phi^2\right)
+ \left[\frac{V_3(\phi)}{3e^2}\right]g_{ab}
 \end{split}
\end{equation}
where we have used the fact that the scalar field is real in the last equation.

Using the fact that all components of the gauge field, metric and scalar field 
are functions of $r$ only,
one can simplify those components of the Maxwell equations where the free index is the boundary direction.
\begin{equation}\label{step2}
 \begin{split}
 & \nabla_a\left[V_1(\phi\phi^*)F^{a\mu}\right] =
-  \frac{iV_2(\phi\phi^*)}{4}\left[\phi{\bar D}^\mu \phi^* - \phi^* D^\mu\phi\right]\\
\Rightarrow&\partial_r\left[\sqrt{g}~V_1(\phi)F^{r\mu}\right] = \frac{\sqrt{g}}{2}A^\mu V_2(\phi)\phi^2
 \end{split}
\end{equation}

We now find explicit expressions (in terms of the functions that appear in our
ansatz for the metric and gauge field) for the components of 
${\cal S}^\text{em}_{ab}$ that we will need in the sequel.
Using the equation \eqref{step2}  .
\begin{equation}\label{stepmid1}
 \begin{split}
\sqrt{g}\left[{\cal S}^\text{em}\right]^v_v =& -\frac{2}{e^2} \sqrt{g}V_1(\phi)
\left[\frac{2}{3}(\partial_r A_v)F^{vr} - \frac{1}{3}(\partial_r A_x)F^{xr}\right]\\
=& -\frac{2}{e^2} \partial_r\left[\sqrt{g}V_1(\phi)
\left(\frac{2}{3} A_vF^{vr} - \frac{1}{3}A_xF^{xr}\right)\right]\\
&+\frac{2}{e^2}\left[\frac{2}{3}A_v\partial_r\left(\sqrt{g} V_1(\phi)F^{vr}\right)
-\frac{A_x}{3}\partial_r\left(\sqrt{g} V_1(\phi)F^{xr}\right)\right] \\
=& -\frac{2}{e^2} \partial_r\left[\sqrt{g}V_1(\phi)
\left(\frac{2}{3} A_vF^{vr} - \frac{1}{3}A_xF^{xr}\right)\right]\\
&-\frac{\sqrt{g}\phi^2V_2(\phi)}{e^2}\left[\frac{2}{3}A_vA^v
-\frac{A_x A^x}{3}\right] \\
 \end{split}
\end{equation}
Similarly
\begin{equation}\label{stepmid2}
 \begin{split}
\sqrt{g}\left[{\cal S}^\text{em}\right]^x_x =& -\frac{2}{e^2} \sqrt{g}V_1(\phi)
\left[\frac{2}{3}(\partial_r A_x)F^{xr} - \frac{1}{3}(\partial_r A_v)F^{vr}\right]\\
=& -\frac{2}{e^2} \partial_r\left[\sqrt{g}V_1(\phi)
\left(\frac{2}{3} A_xF^{xr} - \frac{1}{3}A_vF^{vr}\right)\right]\\
&-\frac{\sqrt{g}\phi^2V_2(\phi)}{e^2}\left[\frac{2}{3}A_xA^x
-\frac{A_v A^v}{3}\right] \\
 \end{split}
\end{equation}

\begin{equation}\label{stepmid3}
 \begin{split}
\sqrt{g}\left[{\cal S}^\text{em}\right]^{y}_{y} =& \frac{2\sqrt{g}V_1(\phi)}{3e^2} 
\left[(\partial_r A_x)F^{xr}+(\partial_r A_v)F^{vr}\right]\\
=& \frac{2}{(3)e^2} \partial_r\left[\sqrt{g}V_1(\phi)
\left( A_xF^{xr}+A_vF^{vr}\right)\right]\\
&+\sqrt{g}\frac{\phi^2V_2(\phi)}{3e^2}\left[A_xA^x
+ A_v A^v\right] \\
 \end{split}
\end{equation}

\begin{equation}\label{stepmid4}
 \begin{split}
\sqrt{g}\left[{\cal S}^\text{em}\right]^x_v =& -\frac{2}{e^2} \sqrt{g}V_1(\phi)
\left[(\partial_r A_v)F^{xr}\right]\\
=& -\frac{2}{e^2} \partial_r\left[\sqrt{g}V_1(\phi)
 A_vF^{xr}\right] - \frac{\sqrt{g}V_2(\phi)\phi^2}{e^2}A^xA_v\\
 \end{split}
\end{equation}

In a similar manner we now find explicit expressions for the same components 
of ${\cal S}^\text{sc}_{ab}$
\begin{equation}\label{stepmid5}
 \begin{split}
\sqrt{g}\left[{\cal S}^\text{sc}\right]^v_v =& \frac{\sqrt{g}}{e^2} \left(V_2(\phi)A^v A_v\phi^2 
+\frac{V_3(\phi)}{3}\right)\\
\sqrt{g}\left[{\cal S}^\text{sc}\right]^x_x =& \frac{\sqrt{g}}{e^2} \left(V_2(\phi)A^x A_x\phi^2 
+\frac{V_3(\phi)}{3}\right)\\
\sqrt{g}\left[{\cal S}^\text{sc}\right]^{y}_{y}
 =& \frac{\sqrt{g}V_3(\phi)}{(3)e^2} \\
\sqrt{g}\left[{\cal S}^\text{sc}\right]^x_v =& \frac{\sqrt{g}\phi^2 V_2(\phi)}{e^2} A^x A_v
\\
 \end{split}
\end{equation}
With these expressions in hand it is not difficult to  determine three 
linear combinations of the equations appearing in  \eqref{einsami}  so that the 
RHS of of these equations is a total derivative. We find 

\begin{equation}\label{comb1}
 \begin{split}
  &\sqrt{g} \left(R^v_v - R^{y}_{y}\right) \\
=& \sqrt{g}\bigg(\left[{\cal S}^\text{em}\right]^v_v +\left[{\cal S}^\text{sc}\right]^v_v
 -\left[{\cal S}^\text{em}\right]^{y}_{y}-\left[{\cal S}^\text{sc}\right]^{y}_{y}\bigg) \\
=&-\frac{2}{e^2}\partial_r\left[A_v \left(\sqrt{g} V_1(\phi) F^{vr}\right)\right]\\
\\
&\sqrt{g} \left(R^x_x - R^{y}_{y}\right)\\
=& \sqrt{g}\bigg(\left[{\cal S}^\text{em}\right]^x_x +\left[{\cal S}^\text{sc}\right]^x_x
 -\left[{\cal S}^\text{em}\right]^{y}_{y}-\left[{\cal S}^\text{sc}\right]^{y}_{y}\bigg) \\
=&-\frac{2}{e^2} \partial_r\left(A_x \left(\sqrt{g} V_1(\phi) F^{xr}\right)\right)\\
\\
&\sqrt{g}R^x_v\\
=& \sqrt{g}\bigg(\left[{\cal S}^\text{em}\right]^x_v +\left[{\cal S}^\text{sc}\right]^x_v
 \bigg) \\
=& -\frac{2}{e^2}
\partial_r \left(A_v \left(\sqrt{g} V_1(\phi) F^{xr}\right)\right)\\
\end{split}
\end{equation}

\subsubsection{Integrating the Total Derivatives}

We now plug \eqref{comb1} into 
\eqref{identity} to obtain
three total derivatives that vanish onshell. We now integrate these expressions 
from the horizon to infinity to obtain
\begin{equation}\label{intcomb1}
 \begin{split}
& \left[\frac{r[k(r)f(r) + j^2(r)]}{ g(r)\sqrt{k(r)}}
-\frac{r^2[k(r)f'(r) + j(r)j'(r)]}{2 g(r)\sqrt{k(r)}}\right]^{r=\infty}_{r = 1}\\
 =& -\frac{2}{e^2}\left[A_v \left(\sqrt{g} V_1(\phi) F^{vr}\right)\right]^{r=\infty}_{r = 1}\\
 \end{split}
\end{equation}
and
\begin{equation}\label{intcomb2}
 \begin{split}
 & \left[\frac{r[k(r)f(r) + j^2(r)]}{ g(r)\sqrt{k(r)}}
-\frac{r^2[f(r)k'(r) + j(r)j'(r)]}{2 g(r)\sqrt{k(r)}}\right]^{r=\infty}_{r = 1}\\
 =& -\frac{2}{e^2}\left[A_x \left(\sqrt{g} V_1(\phi) F^{xr}\right)\right]^{r=\infty}_{r = 1}\\
 \end{split}
\end{equation}
and
\begin{equation}\label{intcomb3}
 \begin{split}
 & \left[\frac{r^2[j(r)f'(r) - f(r)j'(r)]}{2 g(r)\sqrt{k(r)}}\right]^{r=\infty}_{r = 1}\\
 =& \frac{2}{e^2}\left[A_v \left(\sqrt{g} V_1(\phi) F^{xr}\right)\right]^{r=\infty}_{r = 1}\\
 \end{split}
\end{equation}
The asymptotic expansion \eqref{bndexp2}  and \eqref{gagecg} at infinity and 
the fact that $$f(r = 1) = j(r=1)=0$$ at the horizon allow us to simplify 
the LHSs of \eqref{intcomb1}, \eqref{intcomb2} and 
\eqref{intcomb3} as follows:
\begin{equation}\label{lhsend}
 \begin{split}
& \left[\frac{r[k(r)f(r) + j^2(r)]}{ g(r)\sqrt{k(r)}}
-\frac{r^2[k(r)f'(r) + j(r)j'(r)]}{2 g(r)\sqrt{k(r)}}\right]^{r=\infty}_{r = 1}   =2 f_2
 + \frac{\sqrt{k(1)}f'(1)}{2g(1)}\\
\\
&\left[\frac{r[k(r)f(r) + j^2(r)]}{ g(r)\sqrt{k(r)}}
-\frac{r^2[f(r)k'(r) + j(r)j'(r)]}{2 g(r)\sqrt{k(r)}}\right]^{r=\infty}_{r = 1} = 2k_2\\
\\
& \left[\frac{r^2[j(r)f'(r) - f(r)j'(r)]}{2 g(r)\sqrt{k(r)}}\right]^{r=\infty}_{r = 1}=2j_2
 \end{split}
\end{equation}
The RHSs of \eqref{intcomb1}, \eqref{intcomb2} and \eqref{intcomb3} may 
also be simplified upon noting that according to the prescription of 
AdS/CFT the 
boundary current $J^\mu$ is given by 
$$J^\mu = \frac{1}{16\pi G}\lim_{r\rightarrow\infty}\left(\frac{4}{e^2}\sqrt{g}V_1(\phi)F^{\mu r}\right),~~
\xi_\mu = \lim_{r\rightarrow\infty}g_{\mu a}A^a$$
and also that 
$$A_v(r = 1) = -\left[f(r)H(r) + j(r) L(r)\right]_{r=1} = 0$$
(because $f$ and $j$ vanish at the horizon) and that 
$$F^{xr}(r=1) =\left[ \nabla^r A^x\right]_{r = 1} =\left[ g^{rr}\nabla_r A^x\right]_{r=1} = 0$$
because $g^{rr}(r=1) = 0$
Using these relations the RHSs of those three equations may be simplified 
as follows
\begin{equation}\label{rhsend}
 \begin{split}
  & -\frac{2}{e^2}\left[A_v \left(\sqrt{g} V_1(\phi) F^{vr}\right)\right]^{r=\infty}_{r = 1}
 = -\frac{16 \pi G}{2} \xi_v J^v
 = \frac{16 \pi G}{2} \xi_v J_v\\
\\
& -\frac{2}{e^2}\left[A_x \left(\sqrt{g} V_1(\phi) F^{xr}\right)\right]^{r=\infty}_{r = 1} 
= -\frac{16 \pi G}{2} \xi_x J^x
= -\frac{16 \pi G}{2} \xi_x J_x \\
\\
& -\frac{2}{e^2}\left[A_v \left(\sqrt{g} V_1(\phi) F^{xr}\right)\right]^{r=\infty}_{r = 1}
 = -\frac{16 \pi G}{2} \xi_v J^x
= -\frac{16 \pi G}{2} \xi_v J_x \\
 \end{split}
\end{equation}

Finally plugging the simplifications \eqref{lhsend} and \eqref{rhsend} 
into \eqref{intcomb1}, \eqref{intcomb2} and \eqref{intcomb3} obtain the 
final result of this subsubsection
\begin{equation}\label{finalrel}
 \begin{split}
  f_2 &= - \frac{\sqrt{k(1)}f'(1)}{4g(1)} + \frac{16 \pi G}{4}\xi_v J_v\\
k_2 &=-\frac{16 \pi G}{4}\xi_x  J_x\\
j_2 &= \frac{16 \pi G}{4}\xi_v J_x
 \end{split}
\end{equation}

 \subsubsection{Constraints on boundary stress tensor and current etc} 
\label{GDSrel}

Let us first use the identities in \eqref{finalrel} of the previous 
subsubsection to demonstrate that the boundary stress tensor dual to 
our gravitational solution takes the form 
\begin{equation}\label{stressdecomp}
T_{\mu\nu} = (\rho_n + P)~ u_\mu u_\nu + \rho_s~ (u_s)_\mu (u_s)_\nu 
+P\eta_{\mu\nu}
\end{equation}
In other words, $u^\mu$, defined in this section as the killing vector that 
reduces to the black hole horizon, is also the normal fluid velocity, 
according to our definition of the normal fluid. In order to check the 
absence of a normal fluid super fluid cross term in \eqref{stressdecomp}
it is necessary and sufficient, in our coordinates, to check that 
\begin{equation}\label{condition}
 \frac{T_{xx} - T_{yy}}{T_{vx}} =\frac{\xi_x}{\xi_v}
\end{equation}
Now using the usual definition of the stress tensor we find 
\footnote{In order to obtain these relations we worked, for concreteness, 
with a scalar field is dual to an operator of dimension $2$, i.e. one whose 
normalizable solutions look like
 $$  \lim_{r\rightarrow\infty}\phi(r)=  \frac{\phi_2}{r^2} + {\cal O}(r^{-4}).$$}
\begin{equation}\label{bstexpl}
 \begin{split}
  T_{vv}&= \frac{1}{16\pi G}\left(-3f_2 + 6 g_4 + 4k_2  + \frac{\phi_2^2}{e^2}\right)\\
T_{vx} &= \frac{1}{16\pi G}\left(- 4j_2\right)\\
T_{xx} &= \frac{1}{16\pi G}\left(-f_2 - 6 g_4 - \frac{\phi_2^2}{e^2}\right)\\
T_{y y} &= T_{z z} =  \frac{1}{16\pi G}\left(-f_2 - 6 g_4 - 4k_2  - \frac{\phi_2^2}{e^2}\right)\\
 \end{split}
\end{equation}

Using these relations, the LHS of \eqref{condition} reduces to
$\frac{k_2}{j_2}$. Using the last two equations in \eqref{finalrel} then 
immediately yields \eqref{condition}.

Let us now demonstrate the first relation in \eqref{fconst}. In order to 
do this we note that the horizon area of our  solution is equal to 
$\sqrt{k(1)}$ and so that the entropy density of our solution is given by 
$\frac{\sqrt{k(1)}}{4 G}$. The periodicity of the Euclidean time circle in our
circle is equal to $\beta = \frac{1}{T} =4\pi\left( \frac{ g(1)}{f'(1)}\right)$.
It follows that 
\begin{equation}\label{ent}
T s = \frac{1}{16 \pi G} \frac{\sqrt{k(1)}f'(1)}{g(1)}
\end{equation}

All other thermodynamical charges may be evaluated from the stress tensor 
as follows. In our coordinates
\begin{equation}\label{compexp1}
 \begin{split}
  \rho_s& =T_{vx}\left(\frac{\xi^2}{\xi_v~\xi_x}\right)\\
\rho_n + P &= T_{vv} + T_{y y}- T_{vx}\left(\frac{\xi_v}{\xi_x}\right)\\
q_s &= -J_x \left(\frac{\xi}{\xi_x}\right)\\ 
q_n &= -\left[ J_v - J_x \left(\frac{\xi_v}{\xi_x}\right)\right]
 \end{split}
\end{equation}

Substituting \eqref{bstexpl} in the first two equations of \eqref{compexp1} and then using \eqref{finalrel}
and the last two equations of  \eqref{compexp1} one finds the following relations
\begin{equation}\label{endrel}
 \begin{split}
  16 \pi G (P + \rho_n) &= \frac{\sqrt{k(1)}f'(1)}{g(1)} +  16 \pi G\xi_v q_n =  \frac{\sqrt{k(1)}f'(1)}{g(1)} 
+ 16 \pi G(\xi_\mu u^\mu)q_n\\
16 \pi G  \rho_s &= 16 \pi G \xi q_s
 \end{split}
\end{equation}
Finally using \eqref{ent} these two equations turn into 
\begin{equation}\label{endrel}
 \begin{split}
  P + \rho_n &= Ts+  \xi_\mu u^\mu q_n \\
\rho_s &= \mu_s q_s
 \end{split}
\end{equation}
the two Smarr-Gibbs-Duhem relations we set out to prove.

\subsection{The on-shell action}
In this subsection we will demonstrate that the onshell bulk action of 
our solution is the negative of its pressure.

The bulk stress-tensor appearing in the Einstein equation that follows from the 
action in \eqref{sysg} is given by
\begin{equation}\label{stressten}
T_{a b} = {\cal T}^\text{em}_{ab} + {\cal T}^\text{sc}_{ab} +6g_{ab},
\end{equation}
where 
\begin{equation}
 \begin{split}\label{tmattmax}
{\cal T}^\text{em}_{ab}  =& -\frac{2V_1(\phi\phi^*)}{e^2} \left(F_{ac}{F^c}_b 
+ \frac{1}{4}g_{ab}F^{c_1 c_2}F_{c_1c_2}\right)\\
{\cal T}^\text{sc}_{ab}  =& \frac{V_2(\phi\phi^*)}{e^2}
 \left(\frac{D_a\phi{\bar D}_b\phi^* + { \bar D}_a\phi^* D_b\phi}{2}
- \frac{1}{2}g_{ab}D_c\phi{\bar D}^c\phi^*\right)-\frac{V_3(\phi\phi^*)}{2 e^2}g_{ab} \\
 \end{split}
\end{equation}

We consider solution of the form \eqref{metgaganza}. Then it follows that the 
$yy$-component of the stress-tensor is given by
\begin{equation}\label{Tyy}
 T_{yy} = \frac{1}{2} ~g_{yy} ~\left(\mathcal{L} - \mathcal{R}\right)
\end{equation}
In deriving this relation we have crucially used the fact that
the term $D_{y}\phi D_{y} \phi$ in $T_{mat}$ and $F_{y c} F^{c}_{\ y}$ in 
$T_{max}$ is zero. This is because $\phi$ and the non-zero components of the
gauge field is not a function of the coordinate $y$ and the $y$-component 
of the gauge field is zero in our chosen form of the solution.
Also we recall the identity for the Einstein Tensor $E_{\mu \nu}$,
$$ E^{\mu}_{~\mu} = -\frac{3}{2}\mathcal{R}.$$
Using the above identity and the Einstein equation $T_{yy} = E_{yy}$, we
have from \eqref{Tyy}
\begin{equation}
 \mathcal{L} = 2 \frac{E_{yy}}{g_{yy}} -\frac{2}{3} E^{\mu}_{~\mu}.
\end{equation}
Now for the metric ansatz in \eqref{metgaganza} this expression reduces to 
\begin{equation}
 \mathcal{L} = \frac{1}{ \sqrt{-g}} \frac{d}{dr} \left( \frac{2}{r} \sqrt{-g} ~g^{rr} \right). 
\end{equation}
With the help of this relation our onshell action reduces to an integration of a total derivative
\begin{equation}
 \mathcal{S}_\text{OS} 
= \frac{1}{16 \pi G} \int d^5x \frac{d}{dr} \left( \frac{2}{r} \sqrt{-g} ~g^{rr} \right)
\end{equation}
Hence, after performing the above integration we get surface terms from the horizon and 
boundary. Since $g^{rr}$ is zero at the horizon only the boundary term contributes
and we have 
\begin{equation}
 \mathcal{S}_\text{OS} 
= \frac{\text{Vol}_4}{16 \pi G}  \left( \frac{2}{r} \sqrt{-g} ~g^{rr} \right)\bigg|_{bdy}.
\end{equation}
In order to avoid the singularities near the boundary we have to add the 
required counterterms to the on shell action 
\begin{equation}
 S_{CT} = \frac{2}{16 \pi G} \int_{bdy} \sqrt{-\gamma} \left( \mathcal{K} - 3 + \lambda \phi^2 \right),
\end{equation}
where $\gamma$ and $\mathcal{K}$ are respectively the induced metric and extrinsic curvature of
a constant $r$ surface and $\nu \phi^2$ is the mass counterterm for the scalar field. This mass counterterm 
is used to cancel any boundary divergence of the scalar field. In this subsection we keep it arbitrary 
as our argument here is independent of the specific value of $\nu$; we assume the scalar field to be norsmalizable.

Let us now consider the following asymptotic expansion of the 
functions that appear in  the metric and the scalar field
\begin{equation}\label{bndexp2}
\begin{split}
 \lim_{r\rightarrow\infty}g(r)&= 1 + \frac{g_4}{r^4} + {\cal O}(r^{-6})\\
  \lim_{r\rightarrow\infty}f(r)&= r^2 + \frac{f_2}{r^2} + {\cal O}(r^{-4})\\
  \lim_{r\rightarrow\infty}k(r)&= r^2 + \frac{k_2}{r^2} + {\cal O}(r^{-4})\\
  \lim_{r\rightarrow\infty}j(r)&=  \frac{j_2}{r^2} + {\cal O}(r^{-4})\\
  \lim_{r\rightarrow\infty} \phi(r) &=\frac{\phi_2}{r^2} + {\cal O}(r^{-3}) 
\end{split}
\end{equation}

With these expansions we find that renormalized on shell action 
evaluates to 
\begin{equation}\label{onsh}
 S_{ROS} = S_{OS}-S_{CT} = \frac{1}{16 \pi G} \left(f_2 + 6 ~g_4 + 4 ~k_2 - 2 \ \lambda \ \phi_2^2 \right)
\end{equation}
Now the boundary stress-tensor is given by 
\begin{equation}\label{bdyst}
 T^{(\text{bdy})}_{\mu \nu} =-\frac{1}{8 \pi G} 
  \lim_{r \rightarrow \infty} r^4 \bigg(
 K^{\mu}_{\nu} - K\delta^\mu_\nu 
+ 3 \delta^{\mu}_{\nu}-\frac{\lambda \phi^2}{2}\delta^\mu_\nu\bigg)
\end{equation}

In the fluid dynamic limit the $yy$-component of this boundary stress tensor yields the
pressure of the fluid. Plugging in the metric ansatz \eqref{metgaganza} into \eqref{bdyst}
we have 
\begin{equation}\label{pres}
 T^{(\text{bdy})}_{yy} \equiv P = - \frac{1}{16 \pi G} \left(f_2 + 6 ~g_4 + 4 ~k_2 - 2 \ \lambda\ \phi_2^2 \right)
\end{equation}
Hence, comparing \eqref{pres} and \eqref{onsh} we have 
$$ S_{ROS} = -(\text{Vol}_4) ~P.$$
which implies that the renormalized on-shell action density is given by $-P$, 
as we set out to prove \footnote{Here $\lambda$ is some constant fixed by the requirement that the limit in 
\eqref{bdyst} is finite
even when the non normalizable mode of the bulk scalar field is turned on. But when the scalar field 
is normalizable the limit  exists for any constant
$\lambda$ and in such cases the identity, proved here, is  true irrespective of the value of $\lambda$}.

\section{Instabilities at large superfluid velocities}\label{stab}

In this section we study the linearized equations of superfluid dynamics 
in Herzog's model in a very simple context. First we restrict attention 
to the strict probe limit $e \to \infty$. In this limit the only dynamical 
variables are $\xi_\mu$ and $\epsilon$. 
Next we restrict attention to perfect fluids. Finally we work at leading 
order in the $\epsilon$ expansion. 

We assume that 
$\partial_0 \phi=\mu(\epsilon)$ and $\partial_x \phi = \zeta$ where 
$\phi$ is the superfluid phase. It is consistent, to linear order, 
 to truncate to this special form (where we retain only two of the 
variables of fluid dynamics) if we also assume that 
$\epsilon=\epsilon(x, t)$ and $\zeta=\zeta(x,t)$. More specifically we set

\begin{equation} \label{solform}
\begin{split}
 \epsilon &= \epsilon_0 + \delta\epsilon \text{e}^{i \omega v + i k x}\\
\zeta &= \zeta_0 + \delta\zeta \text{e}^{i \omega v + i k x}\\
\end{split}
\end{equation}

In order to obtain the dispersion relation for our fluctuations we need to 
solve the equations 
\begin{equation}
 \begin{split}
  \partial_\mu  J^\mu =0,~~\partial_\mu \xi_\nu - \partial_\nu\xi_\mu =0
 \end{split}
\end{equation}
where the second equation is nontrivial only when we choose $(\mu, \nu)$ 
to be $(0,1)$. In the first equation $J^\mu$ is given by 
\eqref{fromst} retaining only terms of quadratic order in smallness of 
$\epsilon$ and $\zeta$, assuming that the two quantities scale in the same 
way. In other words the terms in the current that we have 
neglected are of the order 
${\cal O}(\epsilon^3,~\zeta^3,~\epsilon^2\zeta,~\zeta^2\epsilon)$.

The equations of motion have a solution of the form \eqref{solform} only 
when the following `characteristic' equation is obeyed
\begin{equation}
 \begin{split}
  \left(
\begin{array}{cc}
 \frac{\epsilon_0 k}{24} & k\zeta_0+\omega  \\
 \frac{6 \epsilon_0 k\zeta_0+7 \epsilon_0 \omega}{12} & \frac{3
  \epsilon_0^2 k+24 \zeta_0 \omega }{12}
\end{array}
\right)
\left(
\begin{array}{c}
 \delta \epsilon \\
 \delta\zeta
\end{array}
\right) = 0.
 \end{split}
\end{equation}
This equation yields the dispersion relation 
$$\omega = -\frac{6}{7} k \zeta_0 \pm \frac{k}{28}\sqrt{14\epsilon_0^2 - 96 \zeta_0^2}$$
Notice that $\omega$ develops an imaginary piece when 
$\frac{\zeta_0}{\epsilon_0}  \geq \sqrt{\frac{7}{48}} = \frac{1}{4}\sqrt{\frac{7}{3}}  $
demonstrating that the system we study has an instability when this 
inequality is satisfied. 

At the point of onset of instability
the eigenvector that corresponds to the zero eigenvalue is given by 
\begin{equation}
\begin{split}
\left(
\begin{array}{c}
 \delta \epsilon \\
 \delta\zeta
\end{array}
\right) =\left(
\begin{array}{c}
1 \\
-\frac{1}{2}\sqrt{\frac{7}{3}}
\end{array}
\right)
 \end{split}
\end{equation}




\section{Limit $\zeta\rightarrow0$ } \label{rot}

As we have explained in the main text, the metric and gauge field 
dual to a super fluid flow must take the simplified form 
\eqref{metuni} in the limit $\zeta \to 0$. In this Appendix we will 
explicitly verify that the results for our metric and gauge field 
at nonzero $\zeta$ reduce to this rotationally invariant form 
when $\zeta$ is set to zero, and read off all the unknown functions 
of radius in \eqref{metuni}

Let us start with a metric and gauge field of the form \eqref{metuni} 
and rewrite it in the language of our metric and gauge field at nonzero 
$\zeta$. That is we set $\zeta^\mu = \zeta n^\mu$ where $n^\mu$ is a 
unit vector orthogonal to $u^\mu$, and eventually set $\zeta$ to zero. 
The scalar vector and the tensor quantities 
defined above \eqref{metuni} become

\begin{equation}\label{inourfr}
 \begin{split}
 & \partial_\mu\zeta_\nu = n_\nu \partial_\mu \zeta\\
& P^{\mu\nu}\partial_\mu \zeta_\nu=n^\mu \partial_\mu \zeta = S_1\\
&\sigma^{(\zeta)}_{\mu\nu} = \frac{2 S_1}{3}\left(n_\mu n_\nu -\frac{\tilde P_{\mu\nu}}{2}\right)
+ \frac{n_\nu [V_2]_\mu + n_\mu [V_2]_\nu}{2}\\
&P^{\mu\nu}\partial_\nu \epsilon = n^\mu S_2 + [V_1]^\mu\\
&\partial_\mu u^\mu = S_4 + S_6\\
&u.\partial u_\mu =  S_3 ~n_\mu + [V_4]_\mu\\
&\sigma_{\mu\nu} = \frac{2 S_4 - S_6}{3}\left(n_\mu n_\nu -\frac{\tilde P_{\mu\nu}}{2}\right)
+ \frac{n_\nu [V_5]_\mu + n_\mu [V_5]_\nu}{2} + \frac{[T_1]_{\mu\nu}}{2}\\
 \end{split}
\end{equation}

Plugging these expressions into \eqref{metuni} we find 
\begin{equation}\label{metunichb}
 \begin{split}
ds^2 &= -2g\left(\frac{r}{r_c}\right) u^\mu dx^\mu dr
+\bigg[  -r_c^2f\left(\frac{r}{r_c}\right) u_\mu u_\nu  + r^2 P_{\mu\nu}\bigg]dx^\mu dx^\nu\\
&+r_c F\left(\frac{r}{r_c}\right) \sigma_{\mu\nu}dx^\mu dx^\nu\\
&+\frac{1}{e^2}\bigg\{-\frac{2}{r_c}\left[{\mathcal G}_1\left(\frac{r}{r_c}\right)\left( S_4 + S_6\right)
+{\mathcal G}_2\left(\frac{r}{r_c}\right
)S_1\right]u_\mu dx^\mu dr\\
&+r_c\left[{\mathcal F}_1\left(\frac{r}{r_c}\right)\left( S_4 + S_6\right)
+{\mathcal F}_2\left(\frac{r}{r_c}\right
)S_1\right]u_\mu u_\nu dx^\mu dx^\nu \\
&+ r_c\left[{\cal V}_1\left(\frac{r}{r_c}\right)~S_3 +{\cal V}_2\left(\frac{r}{r_c}\right)~S_2\right]
u_\mu  n_\nu dx^\mu dx^\nu\\
&+r_c\left[\frac{2{\cal T}_2\left(\frac{r}{r_c}\right)}{3}S_1 
+  \frac{{\cal T}_1\left(\frac{r}{r_c}\right)}{3}(2S_4 - S_6)\right]
\left[n_\mu n_\nu -\frac{\tilde P_{\mu\nu}}{2}\right]dx^{\mu}dx^{\nu}\\
&+r_cu_\mu\left[{\cal V}_1\left(\frac{r}{r_c}\right) [V_4]_\nu+
{\cal V}_2\left(\frac{r}{r_c}\right) [V_1]_\nu\right]dx^{\mu}dx^{\nu}\\
&+r_c n_\mu \left[{\cal T}_1\left(\frac{r}{r_c}\right) [V_5]_\nu
 +
{\cal T}_2\left(\frac{r}{r_c}\right) [V_2]_\nu \right] dx^{\mu}dx^{\nu}
+\frac{r_c}{2} {\cal T}_1\left(\frac{r}{r_c}\right) [T_1]_{\mu\nu} dx^\mu dx ^\nu\bigg\}
+{\cal O}\left(\frac{1}{e^4}\right)\\
\\
A &= \frac{1}{r_c}H\left(\frac{r}{r_c}\right) u^\mu\partial_\mu \\
&+ \left[{\mathcal A}_1\left(\frac{r}{r_c}\right)\left( S_4 + S_6\right)
+{\mathcal A}_2\left(\frac{r}{r_c}\right
)S_1\right]\partial_r\\
&+\frac{1}{r_c^2} \left[{\mathcal H}_1\left(\frac{r}{r_c}\right)\left(S_4 + S_6\right)
+{\mathcal H}_2\left(\frac{r}{r_c}\right
)S_1\right]u^\mu\partial_\mu\\
&+\frac{1}{r_c^2} \left[{\cal L}_1\left(\frac{r}{r_c}\right)S_3 
+{\cal L}_2\left(\frac{r}{r_c}\right)S_2\right]~n^\mu\partial_\mu
+\frac{1}{r_c^2} \left[{\cal L}_1\left(\frac{r}{r_c}\right)[V_4]^\mu 
+ {\cal L}_2\left(\frac{r}{r_c}\right)[V_1]^\mu\right]\partial_\mu\\
&+{\cal O}\left(\frac{1}{e^2}\right)
 \end{split}
\end{equation}

We would now like to compare these expressions with the $\zeta \to 0$ 
limit of the metric and gauge field at nonzero $\zeta$. There is, however, 
an immediate complication. The metric in \eqref{metunichb} is not in the 
gauge employed in our computation at nonzero $\zeta$. In that more 
general calculation we have chosen a gauge such that the coefficient of 
$\tilde P_{\mu\nu} dx^\mu dx^\nu$ 
term in the metric is $r^2$. But in \eqref{metunichb}  the coefficient of
 $\tilde P_{\mu\nu} dx^\mu dx^\nu$ 
term is given by $r^2 +  \frac{h(r)}{e^2}$ where 
$$h(r) = - \frac{r_c}{2} \left[\frac{2{\cal T}_2\left(\frac{r}{r_c}\right)}{3}S_1 
+  \frac{{\cal T}_1\left(\frac{r}{r_c}\right)}{3}(2S_4 - S_6)\right]$$

So in order to compare \eqref{metunichb} with the results at nonzero $\zeta$ 
we must first perform a coordinate redefinition upto first order in derivative.
Define 
\begin{equation}\label{rredef}
 \begin{split}
  &\tilde r^2 = r^2 + \frac{h(r)}{e^2}\\
&\text{such that}\\
& r \approx \tilde r - \frac{h(\tilde r)}{2 e^2\tilde r}\\
&d r \approx d\tilde r \left[1- \left(\frac{h(\tilde r)}{2e^2\tilde r}\right)'\right]
 \end{split}
\end{equation}

Since this coordinate redefinition has a factor of $\frac{1}{e^2}$ it does not affect the gauge field as
the gauge field is computed only upto order ${\cal O}\left(\frac{1}{e^2}\right)^0$.
 However the coordinate transformation does affect the metric, though 
only in the scalar sector (i.e. in the coefficients of scalar terms), 
basically because the coordinate transformation performed above is 
a scalar operation. Once this shift of $r$ coordinate is implemented new 
terms are generated from the expansion of the uncharged black-brane metric
 (the metric at order ${\cal O}\left(\frac{1}{e^2}\right)^0$).

After performing the relevant coordinate transformation, the 
metric takes the following form
\begin{equation}\label{metunifg}
 \begin{split}
ds^2 &= -2g\left(\frac{\tilde r}{r_c}\right) u^\mu dx^\mu dr
+\bigg[  -r_c^2f\left(\frac{\tilde r}{r_c}\right) u_\mu u_\nu  +\tilde r^2 P_{\mu\nu}\bigg]dx^\mu dx^\nu\\
&-\left[2r u_\mu\left(\left(u.\partial\right)u_\nu  - 
\frac{1}{3}\left(\partial.u\right) u_\nu\right) +r_c F\left(\frac{r}{r_c}\right)\sigma_{\mu\nu}\right]
dx^\mu dx^\nu\\
&+\frac{1}{e^2}\bigg\{-\frac{2}{r_c}\left[{\mathcal G}_1\left(\frac{\tilde r}{r_c}\right)\left( S_4 + S_6\right)
+{\mathcal G}_2\left(\frac{\tilde r}{r_c}\right
)S_1- r_c\left(\frac{h(\tilde r)}{2\tilde r}\right)'\right]u_\mu dx^\mu dr\\
&+r_c\left[{\mathcal F}_1\left(\frac{\tilde r}{r_c}\right)\left( S_4 + S_6\right)
+{\mathcal F}_2\left(\frac{\tilde r}{r_c}\right
)S_1+ \frac{h(\tilde r)}{r_c}\left(1+\frac{r_c^4}{\tilde r^4}\right)\right]u_\mu u_\nu dx^\mu dx^\nu \\
&+ r_c\left[{\cal V}_1\left(\frac{\tilde r}{r_c}\right)~S_3 +{\cal V}_2\left(\frac{\tilde r}{r_c}\right)~S_2\right]
u_\mu  n_\nu dx^\mu dx^\nu\\
&+r_c\left[\frac{2{\cal T}_2\left(\frac{\tilde r}{r_c}\right)}{3}S_1 
+  \frac{{\cal T}_1\left(\frac{\tilde r}{r_c}\right)}{3}(2S_4 - S_6)-\frac{h(\tilde r)}{r_c}\right]
n_\mu n_\nu dx^{\mu}dx^{\nu}\\
&+r_cu_\mu\left[{\cal V}_1\left(\frac{\tilde r}{r_c}\right) [V_4]_\nu+
{\cal V}_2\left(\frac{\tilde r}{r_c}\right) [V_1]_\nu\right]dx^{\mu}dx^{\nu}\\
&+r_c n_\mu \left[{\cal T}_1\left(\frac{\tilde r}{r_c}\right) [V_5]_\nu
 +
{\cal T}_2\left(\frac{\tilde r}{r_c}\right) [V_2]_\nu \right] dx^{\mu}dx^{\nu}
+\frac{r_c}{2} {\cal T}_1\left(\frac{\tilde r}{r_c}\right) [T_1]_{\mu\nu} dx^\mu dx ^\nu\bigg\}
+{\cal O}\left(\frac{1}{e^4}\right)\\
\end{split}
\end{equation}

Now this metric matches the $\zeta\rightarrow 0$ limit of the metric and 
gauge field in this paper provided, among other conditions, that 

{\it The constraints for the metric:}
\begin{equation}\label{zetazero}
 \begin{split}
 & \lim_{\chi\rightarrow0}\delta g_2(r)
 =  \lim_{\chi\rightarrow0}\delta g_3(r) =  \lim_{\chi\rightarrow0}\delta g_5(r) = 0\\
& \lim_{\chi\rightarrow0}\delta f_2(r)
 =  \lim_{\chi\rightarrow0}\delta f_3(r) =  \lim_{\chi\rightarrow0}\delta f_5(r) = 0\\
& \lim_{\chi\rightarrow0}\delta K_2(r)
 =  \lim_{\chi\rightarrow0}\delta K_3(r) =  \lim_{\chi\rightarrow0}\delta K_5(r) = 0\\
& \lim_{\chi\rightarrow0}\delta J_1(r)
 =  \lim_{\chi\rightarrow0}\delta J_4(r) =  \lim_{\chi\rightarrow0}\delta J_5(r) = 
\lim_{\chi\rightarrow0}\delta J_6(r) =0\\
& \lim_{\chi\rightarrow0}\delta J_3(r) =  \lim_{\chi\rightarrow0}Y_4(r) = \frac{1}{2}{\cal V}_1(r)\\
& \lim_{\chi\rightarrow0}\delta J_2(r) =  \lim_{\chi\rightarrow0}Y_1(r) = \frac{1}{2}{\cal V}_2(r)\\
&\lim_{\chi\rightarrow0}\delta K_6(r) = - \lim_{\chi\rightarrow0}\frac{\delta K_4(r)}{2}
 =-\lim_{\chi\rightarrow0}\delta Z_1(r)=-\lim_{\chi\rightarrow0}\delta W_5(r)=-\frac{1}{2}{\cal T}_1(r)\\
&\lim_{\chi\rightarrow0}\delta K_1(r) =2\lim_{\chi\rightarrow0}\delta W_2(r) = {\cal T}_2(r)
 \end{split}
\end{equation}

{\it The constraints for the gauge field}
\begin{equation}\label{gagacoo}
 \begin{split}
 & \lim_{\chi\rightarrow0}\delta A_4(r)
 =  \lim_{\chi\rightarrow0}\delta A_6(r) =  {\cal A}_1(r)\\
& \lim_{\chi\rightarrow0}\delta H_4(r)
 =  \lim_{\chi\rightarrow0}\delta H_6(r) =  {\cal H}_1(r)\\
& \lim_{\chi\rightarrow0}\delta L_3(r)
 =  \lim_{\chi\rightarrow0}\delta X_4(r) =  {\cal L}_1(r)\\
& \lim_{\chi\rightarrow0}\delta L_2(r)
 =  \lim_{\chi\rightarrow0}\delta X_1(r) =  {\cal L}_2(r)\\
 \end{split}
\end{equation}

It turns out all these constraints (and all others that are required 
to ensure that the bulk metric is an analytic function of the vector 
field $w^\mu$ in the $\zeta \to 0$ limit) 
are true upto the order the metric is calculated and using them 
one can read off ${\cal V}_1(r)$, ${\cal V}_2(r)$,  ${\cal T}_1(r)$ and ${\cal T}_2(r)$ and 
hence the gauge transformation parameter $h(r)$.
Similarly the functions  ${\cal G}_1(r)$, ${\cal G}_2(r)$,  ${\cal F}_1(r)$ and ${\cal F}_2(r)$ appearing in
$u_\mu dx^\mu dr$ and $u_\mu u_\nu dx^\mu dx^\nu$ terms are related to the
 corresponding  functions appearing in the metric at nonzero $\zeta$.  
These relations also imply some
constraint on the $\chi\rightarrow0$ limit. But they are a bit complicated to write because they
 involve the function $h(r)$ and its derivative with respect to $r$.
 Once  ${\cal V}_1(r)$, ${\cal V}_2(r)$,  ${\cal T}_1(r)$ and ${\cal T}_2(r)$ are all fixed, using them
 it becomes easier read off  ${\cal G}_1(r)$, ${\cal G}_2(r)$,  ${\cal F}_1(r)$ and ${\cal F}_2(r)$.

\bibliographystyle{utphys}
\bibliography{superfluids}

\end{document}